\documentclass[12pt, letterpaper]{article} 

\usepackage[top=1in, bottom=1in, left=1in, right=1in]{geometry} 
\usepackage{setspace}
\usepackage{graphicx}
\renewcommand{\thefigure}{\arabic{section}.\arabic{figure}}

\usepackage{amsmath,amsthm,amsfonts,amssymb,amscd, fancyhdr, color, comment, graphicx, environ}
\usepackage{float}
\usepackage{mathrsfs}
\usepackage{lastpage}
\usepackage[dvipsnames]{xcolor}
\usepackage[framemethod=TikZ]{mdframed}
\usepackage{enumerate}
\usepackage[shortlabels]{enumitem}
\usepackage{fancyhdr}
\usepackage{indentfirst}
\usepackage{listings}
\usepackage{sectsty}
\usepackage{thmtools}
\usepackage{shadethm}
\usepackage{hyperref}
\usepackage{setspace}
\hypersetup{
    colorlinks=true,
    linkcolor=black,
    filecolor=black,
    urlcolor=black,
    citecolor=black
}
\usepackage{import}
\usepackage{tikz-feynman}
\usepackage{subcaption}
\usepackage{tikz}
\usepackage{quantikz}

\mdfsetup{skipabove=\topskip,skipbelow=\topskip}
\newrobustcmd\ExampleText{%
An \textit{inhomogeneous linear} differential equation has the form
\begin{align}
L[v ] = f,
\end{align}
where $L$ is a linear differential operator, $v$ is the dependent
variable, and $f$ is a given non−zero function of the independent
variables alone.
}
\mdfdefinestyle{theoremstyle}{%
linecolor=black,linewidth=1pt,%
frametitlerule=true,%
frametitlebackgroundcolor=gray!20,
innertopmargin=\topskip,
}
\mdtheorem[style=theoremstyle]{Problem}{Problem}
\definecolor{codegreen}{rgb}{0,0.6,0}
\definecolor{codegray}{rgb}{0.5,0.5,0.5}
\definecolor{codepurple}{rgb}{0.58,0,0.82}
\definecolor{backcolour}{rgb}{0.95,0.95,0.92}

\lstdefinestyle{mystyle}{
    backgroundcolor=\color{backcolour},   
    commentstyle=\color{codegreen},
    keywordstyle=\color{magenta},
    numberstyle=\tiny\color{codegray},
    stringstyle=\color{codepurple},
    basicstyle=\ttfamily\footnotesize,
    breakatwhitespace=false,         
    breaklines=true,                 
    captionpos=b,                    
    keepspaces=true,                 
    numbers=left,                    
    numbersep=5pt,                  
    showspaces=false,                
    showstringspaces=false,
    showtabs=false,                  
    tabsize=2
}

\usepackage[utf8]{inputenc}

\usepackage{wrapfig}

\usepackage[justification=centering]{caption}

\usepackage{xcolor}

\usepackage{titlesec}

\titleformat{\section}
  {\normalfont\Large\bfseries}{}{0pt}{}

\renewcommand\thesubsection{\arabic{subsection}.}

\renewcommand\thesubsubsection{\thesubsection\arabic{subsubsection}}

\newcounter{subsubsubsection}[subsubsection]
\renewcommand\thesubsubsubsection{\thesubsubsection.\arabic{subsubsubsection}}
\newcommand{\subsubsubsection}[1]{%
  \refstepcounter{subsubsubsection}
  \par\medskip 
  \noindent{\bfseries \thesubsubsubsection\quad #1}
  \par\smallskip 
  \addcontentsline{toc}{paragraph}{\protect\numberline{\thesubsubsubsection}#1}%
}

\usepackage{tocloft}

\usepackage{multirow}

\catcode`\<=\active \def<{
\fontencoding{T1}\selectfont\symbol{60}\fontencoding{\encodingdefault}}
\catcode`\>=\active \def>{
\fontencoding{T1}\selectfont\symbol{62}\fontencoding{\encodingdefault}}
\catcode`\<=\active \def<{
\fontencoding{T1}\selectfont\symbol{60}\fontencoding{\encodingdefault}}
\usepackage{amsmath}
\usepackage[makeroom]{cancel}

\usepackage[T1]{fontenc}
\usepackage[utf8]{inputenc}      
\usepackage{newcent} 

\usepackage{pifont}
\newcommand{\cmark}{\ding{51}}%
\newcommand{\xmark}{\ding{55}}%

\begin{document}

\providecommand{\keywords}[1]
{
  \small	
  \textbf{\textit{Keywords---}} #1
}
\begin{titlepage}
    \begin{center}
            
        \huge
        {Design and FPGA Implementation of WOMBAT: A Deep Neural Network Level-1 Trigger System for Jet Substructure Identification and Boosted $H \rightarrow b\bar{b}$ Tagging at the CMS Experiment}
            
        \vfill
        \Large
            
        \textbf{Mila Bileska}       \\               
         \vspace{0.2cm}
        Advisor: Isobel Ojalvo\\
        Second Reader: James Olsen\\

        \vfill\large

        \vspace{1cm}
            
        \includegraphics[width=0.4\textwidth]{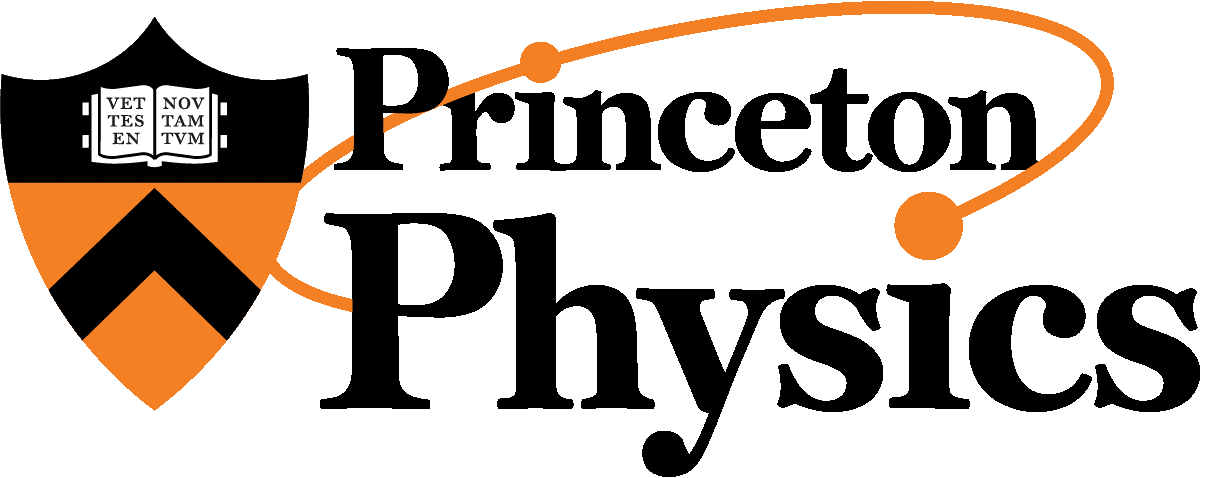}
        \\
        
        \large
        
        \today

\newpage
\clearpage
\pagenumbering{gobble}
        \begin{abstract}

This thesis investigates the physics performance, trigger efficiency, and Field Programmable Gate Array (FPGA) implementation of machine learning (ML)-based algorithms for Lorentz-boosted $H\rightarrow b\bar{b}$ tagging within the CMS Level-1 Trigger (L1T) under Phase-1 conditions. The proposed algorithm, WOMBAT (Wide Object ML Boosted Algorithm Trigger), comprises a high-performance Master Model (W-MM) and a quantized, FPGA-synthesizable Apprentice Model (W-AM), benchmarked against the standard Single Jet 180 and the custom rule-based JEDI (Jet Event Deterministic Identifier) triggers.

All algorithms process calorimeter trigger primitive data to localize boosted $H\rightarrow b\bar{b}$ jets. Outputs are post-processed minimally to yield real-valued $(\eta, \phi)$ jet coordinates at trigger tower granularity.

Trigger rates are evaluated using 2023 CMS ZeroBias data (0.64 fb$^{-1}$), with efficiency assessed via a Monte Carlo sample of $H\rightarrow b\bar{b}$ offline reconstructed AK8 jets. W-MM achieves a $1$ kHz rate at an offline jet $p_T$ threshold of $146.8$ GeV, $40.6$ GeV lower than Single Jet 180, while maintaining comparable signal efficiency. W-AM reduces the threshold further to $140.4$ GeV, with reduced efficiency due to fixed-output constraints and limited multi-jet handling.

FPGA implementation targeting the Xilinx Virtex-7 XC7VX690T confirms that W-AM meets resource constraints with a pre-place-and-route latency of 22 clock cycles ($137.5$ ns). In contrast, JEDI requires excessive resource usage and a 56-cycle latency, surpassing the 14-cycle L1T budget.

These results underscore trade-offs between physics performance and hardware constraints: W-MM offers the highest tagging performance but exceeds current FPGA capacity; W-AM is deployable with reduced efficiency; JEDI remains deployable with moderate efficiency but increased latency. Originally developed for Run-3 CMS L1T, WOMBAT serves as a proof-of-concept for Phase-2 triggers, where hardware advances will enable online deployment of more sophisticated ML-based L1T systems.

        \end{abstract}


    \end{center}
\end{titlepage}

\newpage\clearpage\pagenumbering{arabic}  

\newpage

{\footnotesize \tableofcontents \newpage
\listoffigures\newpage \listoftables 
\begin{figure}[H]
    \centering
    \includegraphics[width=0.8\linewidth]{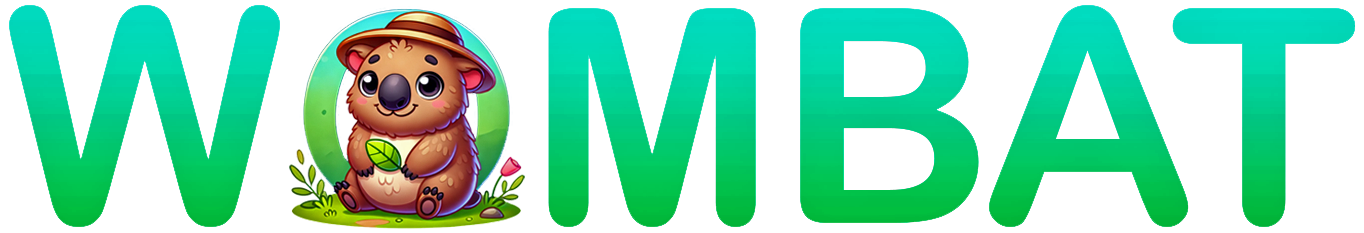}
    \caption{WOMBAT Logo Design by M. Bileska}
    \label{fig:enter-label}
\end{figure}
\newpage}

\stepcounter{section}
\section*{Chapter {I}: The Large Hadron Collider, CMS Experiment, and Level-1 Trigger}
\addcontentsline{toc}{section}{Chapter {I}: The Large Hadron Collider, CMS Experiment, and Level-1 Trigger}
\setcounter{figure}{0}
\markboth{Chapter {I}: The Large Hadron Collider, CMS Experiment, and Level-1 Trigger}{}

\subsection{Physics Goals Driving Trigger Development}

As the operating conditions of hadron colliders become more extreme --- characterized by unprecedented event rates, pileup densities, and detector occupancies --- the task of real-time event selection becomes central to the pursuit of new physics. In high-energy experiments, the trigger system serves as the earliest stage of online event selection, employing low-latency, hardware-implemented algorithms that perform rapid reconstruction of detector signals. By suppressing dominant backgrounds, trigger systems are designed to maintain sensitivity to signatures consistent with target processes, such as high transverse momentum (boosted) decays or rare topologies indicative of physics beyond the Standard Model (BSM).

Designed to cope with the extreme data rates and event complexities at collider experiments, trigger systems must operate under strict constraints on latency, bandwidth, and hardware resources. This imposes a limit on the expressiveness of algorithms that can be deployed in real-time. Traditionally, trigger logic has relied on heuristic or rule-based approaches optimized for speed rather than flexibility. However, recent advances in machine learning (ML), combined with the increasing programmability of modern Field Programmable Gate Arrays (FPGAs), have opened new avenues for implementing data-driven, high-performance decision-making within the tight operational constraints of trigger systems.

At the Large Hadron Collider (LHC), the highest-energy particle accelerator currently in operation, one of the key targets for precision measurements and new physics searches is the study of boosted Higgs bosons decaying to bottom quark-antiquark pairs ($H\rightarrow b\bar{b}$). This decay channel dominates the Higgs boson's branching ratios and provides access to the bottom-quark Yukawa coupling --- a fundamental parameter that determines the interaction strength between fermions and the Higgs field, which underlies the mechanism of mass generation in the Standard Model. Furthermore, $H\rightarrow b\bar{b}$ decays constitute the most common final state in Higgs boson pair production, which offers a direct probe of the Higgs self-coupling and the shape of the Higgs potential. However, isolating these decays in a hadronic environment presents a formidable challenge due to overwhelming quantum chromodynamics (QCD) multijet backgrounds and the limited angular separation of decay products in the boosted regime. These challenges are expected to intensify significantly during the High-Luminosity LHC (HL-LHC) era, where pileup and event rates will increase substantially. At the same time, efficiently capturing Higgs boson pair production events remains a key objective, with boosted topologies offering a powerful probe of this process, making real-time identification of $H\rightarrow b\bar{b}$ decays a high-priority target for triggering strategies.

This thesis presents the development and evaluation of WOMBAT (Wide Object ML Boosted Algorithm Trigger), a machine learning-based trigger system designed for the identification of boosted $H\rightarrow b\bar{b}$ decays at the Compact Muon Solenoid (CMS) Level-1 Trigger (L1T). Intended to operate within the constraints of hardware-based trigger systems, WOMBAT leverages calorimetric information to identify spatial and kinematic features of Higgs decays in high-density hadronic environments. By applying custom deep learning techniques to low-latency data streams, WOMBAT aims to enhance the sensitivity to boosted $H\rightarrow b\bar{b}$ signatures at the earliest stage of event processing, ultimately enabling more efficient data collection for measurements of Higgs couplings, di-Higgs production, and searches for new physics.

\subsection{The Large Hadron Collider}

Located at the European Organization for Nuclear Research (CERN) on the border of Switzerland and France, the Large Hadron Collider (LHC) is a 27-kilometer circular particle collider that probes the fundamental nature of matter through high-energy proton and heavy-ion collisions \cite{history}. It was constructed in the underground tunnels that previously housed the Large Electron-Positron (LEP) collider, which was decommissioned in 2000. Since 2008, the LHC has been operational, currently accelerating proton bunches at a center-of-mass energy ($\sqrt{s}$) of $13.6$ TeV ($6.8$ TeV per beam) \cite{run3}. The LHC also facilitates CERN's heavy-ion research programs by colliding nucleons with an energy of $5.36$ TeV per nucleon pair \cite{ions}. At four interaction points, superconducting magnets direct counter-rotating beams into collision within detectors such as CMS (Compact Muon Solenoid), ATLAS (A Toroidal LHC Apparatus), LHCb (Large Hadron Collider beauty), and ALICE (A Large Ion Collider Experiment), where, under optimal conditions, data are continuously recorded over extended periods. 

The primary source of the proton bunches is CERN's Linear Accelerator 4 (Linac4) \cite{linac}, which accelerates negative Hydrogen ions, H$^{-}$, up to a kinetic energy of $160$ MeV. As these ions traverse a series of radiofrequency (RF) cavities, they undergo a process that strips their electrons, producing protons for injection into the Proton Synchrotron Booster (PSB) for further acceleration ($\text{H}^-\rightarrow$~p(uud)~$+$~$2$e$^-$).

The PSB \cite{psb1,psb2}, consists of four superimposed synchrotron rings that operate in parallel. Within the PSB, the protons are accelerated to $2$ GeV using combined-function magnets and RF cavities. These cavities apply oscillating electromagnetic fields to increase the protons' energy, while dipole and quadrupole magnets ensure their confinement within the accelerator's circular trajectory. The PSB also serves to improve beam quality by increasing brightness and reducing transverse emittance before transferring the beam to the Proton Synchrotron (PS).

At the PS \cite{ps}, a large 628-meter synchrotron, the protons undergo further acceleration to an energy of $26$ GeV. The PS employs conventional electromagnets to bend the proton bunches along a circular path, while RF cavities provide energy boosts at each turn. The PS plays a crucial role in beam manipulation, performing splitting, bunch rotation, and other RF gymnastics to tailor the beam structure for subsequent stages. It also acts as a crucial distribution hub, feeding various experiments and accelerator systems, including the Antiproton Decelerator \cite{antiproton} and the ISOLDE \cite{isolde} facility.

Following the PS, the protons enter the Super Proton Synchrotron (SPS) \cite{sps}, an accelerator with a circumference of $6.9$ km, making it the second-largest machine in the CERN accelerator complex. Within the SPS, the protons are accelerated from $26$ GeV to $450$ GeV. This acceleration is achieved using a combination of powerful dipole magnets, which guide the beam through the synchrotron, and RF cavities that provide energy gain. The SPS serves multiple purposes, acting as an injector for the Large Hadron Collider (LHC) and supplying beams to fixed-target experiments such as NA61/SHINE \cite{shine} and the North Area physics program \cite{na}.

\begin{figure}[h]
    \centering
    \includegraphics[width=\linewidth]{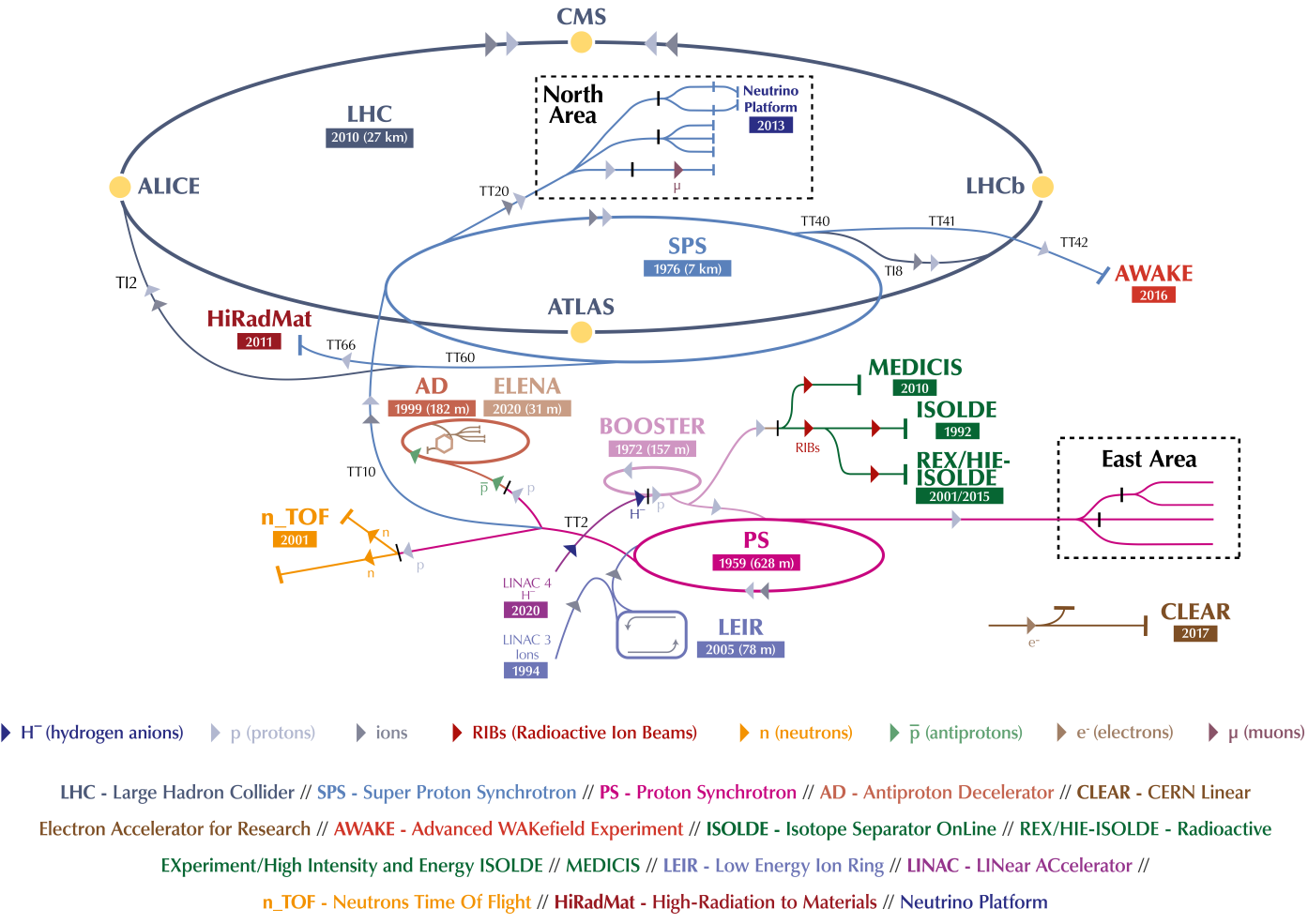}
    \caption{Schematic View of The CERN Accelerator Complex and Particle Acceleration Paths \cite{complex}}
    \label{fig:complex}
\end{figure}

Once the protons reach $450$ GeV in the SPS, they are extracted and transferred via the TI2 and TI8 beamlines to the LHC (see Figure \ref{fig:complex}). These transfer lines use precise magnetic steering to guide the beams into the LHC ring, where they are then captured and further accelerated to their final energy of $6.8$~TeV per beam, leading to the total center-of-mass collision energy of $13.6$~TeV.

From the experiments around the LHC ring, ATLAS and CMS are general-purpose detectors, designed to explore a broad range of high-energy physics phenomena, including the properties of the Higgs boson, searches for potential new particles such as supersymmetric states or dark matter candidates, and precision measurements of Standard Model processes, including electroweak interactions and quantum chromodynamics \cite{cmsdesign,atlasdesign}. Their complementary designs allow cross-verification of results, enhancing the robustness of discoveries.

In contrast, LHCb is optimized for studying b-hadrons, particles containing bottom (beauty) quarks, to investigate charge-parity (CP) violation, which plays a role in understanding the observed dominance of matter over antimatter in the universe \cite{lhcb}. By analyzing rare decays and mixing phenomena in heavy-flavor physics, LHCb provides indirect tests of the Standard Model and potential hints of new physics.

Meanwhile, ALICE specializes in ultra-relativistic heavy-ion collisions, primarily using lead nuclei, to recreate and study the quark-gluon plasma (QGP) --- a deconfined state of matter that existed microseconds after the Big Bang \cite{alice}. By examining QGP properties, ALICE provides insights into the strong interaction and the early universe's thermal evolution.

\subsection{LHC Luminosity and Pileup}

At each bunch crossing, multiple proton-proton (pp) collisions occur. The number of collisions per bunch crossing is proportional to the instantaneous luminosity, $\mathcal{L}$, and can be calculated through the following expression:
\begin{equation}
    n = \frac{\mathcal{L} \cdot \sigma}{f},
\end{equation}
where $n$ is the number of collisions, $\mathcal{L}$ is the instantaneous luminosity measured in units of cm$^{-2}$ s$^{-1}$, $\sigma$ denotes the cross section of the event in units of cm$^{2}$, and $f$ is the frequency of bunch crossings. The instantaneous luminosity, defined as the number of potential collisions per unit area per second, can be expressed as:
\begin{equation}
    \mathcal{L} = \gamma \frac{N^2\;f_{\text{rev}}\;n_{\text{bunch}}}{4\pi\;\beta^*\;\epsilon_n},
\end{equation}
where $N$ is the bunch population (particles per bunch), $f_{\text{rev}}$ is the frequency of revolution, $\gamma$ is the relativistic gamma factor, $n_{\text{bunch}}$ is the number of proton bunches per beam, $\beta^*$ is the evaluated beta function at collision point, and $\epsilon_n$ is the normalized transverse beam emittance.

\begin{figure}[ht]
    \centering
    \includegraphics[scale=0.6]{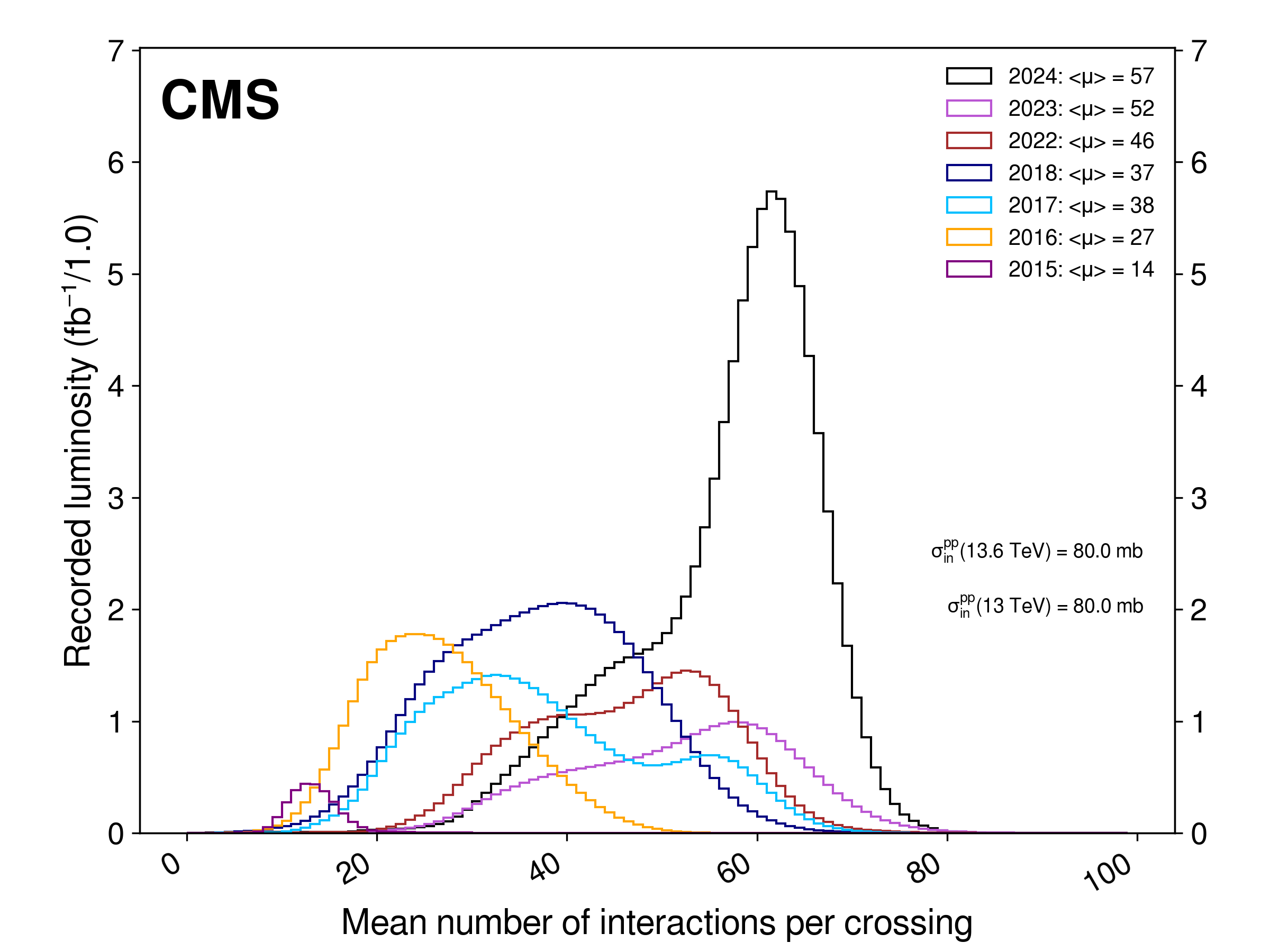}
    \caption{Luminosity vs. Pileup as Recorded by the CMS Detector During the 2015-2024 Data-Taking Period With Cross Section Estimates for Inelastic PP Collisions (Runs 2 and 3) \cite{pilecms}}
    \label{fig:pileup}
\end{figure}

On average, there are:
\begin{equation}
    n_{\text{average}} = \frac{\mathcal{L}\cdot \sigma_{\text{pp}}}{f\cdot n_{\text{bunch}}},
\end{equation}
where $\sigma_{\text{pp}}$ is the cross section for inelastic pp collisions (estimated to be $78.1\pm2.9$ mb for collisions at center-of-mass energy of $13$ TeV, with an approximation of $80$ mb being a sufficient estimate for $\sqrt{s}=13.6$ TeV collisions \cite{pileup}), and $n_{\text{average}}$ is the average number of pp collisions per bunch crossing, also known as pileup. Figure \ref{fig:pileup} demonstrates the pileup recorded by the CMS experiment, which has risen throughout the LHC's operational years. In 2024, the pileup reached a value of $n_{\text{average}}\approx 62$, which is expected to further increase to $140-200$ after the High-Luminosity LHC (HL-LHC) upgrade scheduled for 2030 \cite{hlu}. 

As the average number of interactions per proton bunch crossing increases, the occupancy of the detector's readout channels grows accordingly, leading to significant challenges in event reconstruction and data processing. The high-luminosity environment of future collider upgrades, such as the HL-LHC, will push detectors to operate under extreme conditions, with hundreds of simultaneous proton-proton interactions occurring in each bunch crossing. This high pileup environment introduces substantial background noise, making it increasingly difficult to distinguish the signal of interest from unwanted contributions arising from QCD processes. To cope with these challenges, it is essential to develop advanced trigger systems capable of rapidly selecting relevant events in real-time, preventing data overload and ensuring that the most physics-rich collisions are retained for further analysis. Additionally, sophisticated algorithms must be implemented to accurately reconstruct particle trajectories and efficiently associate them with the correct primary interaction vertex, mitigating the effects of pileup and enhancing the precision of physics measurements. The development of these intelligent data processing techniques is crucial to maximizing the scientific potential of next-generation colliders, enabling discoveries in Higgs boson physics, precision Standard Model tests, and potential new physics beyond the current theoretical framework.

\subsection{The CMS Detector}

The Compact Muon Solenoid (CMS) detector is a multi-purpose apparatus designed to study proton-proton and heavy-ion collisions at $\sqrt{s} = 14$ TeV ($7$ TeV per beam, and $2.75$ TeV per nucleon in heavy-ion collisions) \cite{cms}. It measures the properties of particle jets, leptons, photons, and missing transverse energy (MET), and is capable of tracking and identifying muons, electrons, and hadrons. The design luminosity of the experiment is $10^{34}$ cm$^{-2}$ s$^{-1}$ for pp collisions and $10^{27}$ cm$^{-2}$ s$^{-1}$ for heavy-ion collisions.

\begin{figure}[h]
    \centering
    \includegraphics[width=0.8\linewidth]{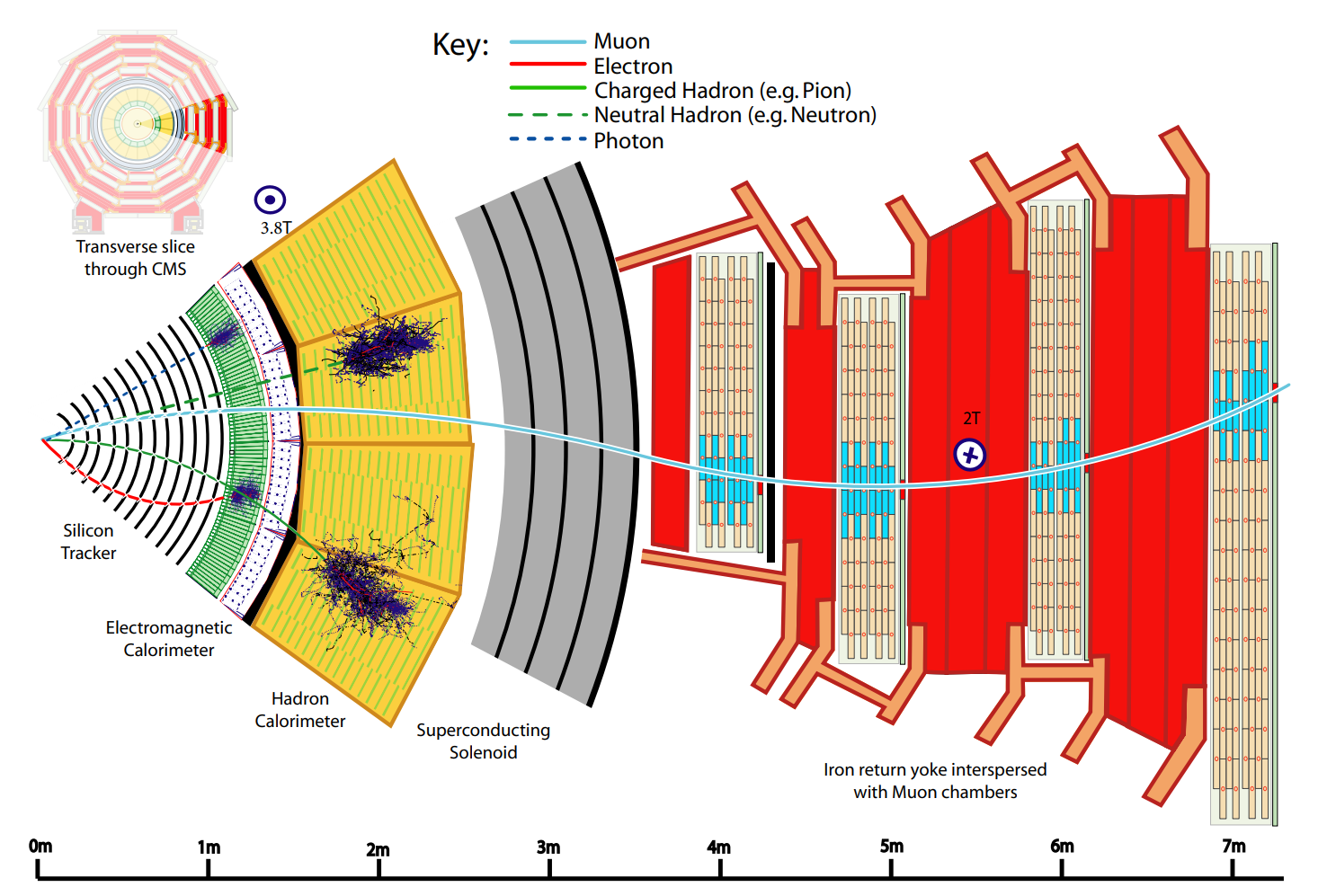}
    \caption{Particle Interactions in a Transverse Slice of the CMS Detector~\cite{cms2}}
    \label{fig:cms}
    \small
    Lines depict trajectories of muons (solid blue), electrons (red), charged and neutral hadrons (solid and dashed green, respectively), and photons (blue dashed). Blue-highlighted muon chambers indicate particle detection, while dark blue splashes in the ECAL and HCAL represent energy deposits.
\end{figure}

The CMS detector features many cylindrical detection layers that are arranged concentrically around the beam axis \cite{cms2}. Figure \ref{fig:cms} illustrates a schematic representation of particle trajectories for different species as they traverse the detector's subsystems.

At the interaction point, where proton-proton collisions occur, charged particles first pass through the Silicon Tracker, a finely segmented system of silicon pixel and strip detectors. The tracker provides precise spatial measurements of charged particle trajectories, allowing for momentum reconstruction based on their curvature in the presence of a $3.8$ T magnetic field.

Beyond the tracker, particles enter the Electromagnetic Calorimeter (ECAL), which is designed to measure the energy of electrons and photons with high precision. The ECAL consists of lead tungstate (PbWO$_{4}$) crystals that produce scintillation light when traversed by high-energy particles. Due to the strong electromagnetic interaction, electrons and photons initiate electromagnetic showers within the ECAL and deposit most, if not all, of their energy before coming to a stop.

Following the ECAL is the Hadron Calorimeter (HCAL), responsible for measuring the energy of strongly interacting particles such as protons, pions, and kaons. The HCAL consists of alternating layers of dense absorber material (brass or steel) and plastic scintillators, enabling the detection of hadronic showers through energy deposition.

Unlike electrons and hadrons, muons interact minimally with both the ECAL and HCAL, allowing them to penetrate these layers with minimal energy loss. This is primarily because muons, being much heavier than electrons, lose significantly less energy through bremsstrahlung radiation. Instead, they predominantly lose energy through ionization, which results in a more gradual energy loss as they travel through matter. Therefore, muons are subsequently detected in the Muon Chambers, which are embedded within the iron yoke that surrounds the solenoid magnet. The yoke serves as a return path for the magnetic field and provides additional shielding. The muon system consists of gaseous detectors, including Drift Tubes (DTs), Cathode Strip Chambers (CSCs), and Resistive Plate Chambers (RPCs), which enable precise muon momentum measurement and trigger capabilities.

By combining data from all these subsystems, the CMS detector can accurately reconstruct particle trajectories, identify different particle species, and measure their properties with high precision. Additionally, it can infer the presence of non-interacting particles, such as neutrinos, by calculating MET, which plays a crucial role in many physics analyses, including searches for new particles.


\subsubsection{Jet Tagging and Reconstruction}

Accurate reconstruction of particle trajectories in the CMS detector is essential for measuring momentum and inferring particle types based on signatures in various detector components. Charged particles, such as electrons, muons, or charged hadrons, experience a Lorentz force in the $3.8$ T uniform magnetic field generated by the Superconducting Solenoid \cite{cms2}. This deflection can be used to determine the charge and momentum of a particle based on its trajectory recorded within the Silicon Tracker. As an example, Figure \ref{fig:cms} shows the bending paths of a pion ($\pi^+$) and a muon ($\mu^+$), both appearing as concave-down arcs in the image due to their like charge. In contrast, the negatively charged electron ($e^-$) follows a concave-up trajectory. Note that the concavity is relative to the orientation of the figure and not an absolute physical descriptor.

At the CMS detector, offline particle tagging begins with the Particle Flow (PF) algorithm \cite{pf}, which plays a central role in event reconstruction. Introduced in 2009 and deployed in CMS physics analyses starting in 2010, the PF algorithm was initially validated using simulated Monte Carlo (MC) events and quickly became a standard reconstruction technique. Designed to fully exploit the combined granularity and resolution of the tracking detectors, calorimeters, and muon systems, PF reconstructs collision events by utilizing information from all subdetectors to generate a comprehensive list of final-state particles, including photons, electrons, muons, and hadrons. Once individual particles are identified using PF, hadronically decaying tau ($\tau$) leptons and composite objects such as jets are reconstructed from the resulting particle collection.

Isolated electrons and photons are primarily identified through the ECAL, where they deposit their energy in distinct electromagnetic showers. These showers exhibit characteristic spatial and energy profiles enabling precision measurements of both the energy and position of incident particles. Electrons are further identified by matching ECAL clusters with charged-particle tracks reconstructed in the inner tracking detector, while photons, being neutral, are identified based solely on their energy deposits and the absence of associated tracks.

Jets originating from b-quark hadronization pose a distinct identification challenge due to the presence of b-hadrons, which decay a few millimeters from the primary interaction point, resulting in displaced secondary vertices. The identification of such b-jets, or b-tagging, employs algorithms such as DeepCSV and DeepJet \cite{deepjet}, which use high-resolution tracking information, processed through deep neural networks. This approach is especially critical in analyses targeting final states involving b-quarks, such as Higgs boson decays to bottom quark pairs, including boosted topologies where collimated b-jets may be reconstructed as a single large-radius jet and identified using substructure-based b-tagging techniques. Accurate identification of b-jets is relevant due to the prevalence of bottom quarks in final states of Higgs boson decays and various BSM scenarios, where enhanced couplings to third-generation quarks are often predicted.

Muon identification relies on a dedicated system of muon chambers placed at the outermost layers of the detector, beyond the calorimeters. Muons are highly penetrating particles and interact minimally with both electromagnetic and hadronic calorimeters. The muon chambers provide complementary tracking information by recording the trajectories of these particles, particularly aiding in momentum measurement through the curvature of their paths in the magnetic field. By combining data from the inner tracking system and the muon chambers, the detector achieves improved resolution and reliability in muon identification, efficiently distinguishing them from other particles and backgrounds.

Accurate particle tagging enables the identification of rare and complex physics signatures within the vast number of collisions occurring at the CMS detector. However, with a nominal bunch crossing frequency of $40$ MHz, corresponding to 40 million proton-proton interactions per second and a $25$ ns time separation between events \cite{trigger}, the sheer volume of data generated across the tracking, calorimetry, and muon detection systems is immense. Since only a small fraction of these collisions produce physically significant events, and data storage is inherently limited, the CMS detector employs a sophisticated Trigger System designed to drastically reduce the data acquisition rate. This system ensures that only events of potential scientific interest are retained for further analysis, allowing efficient selection of the most relevant interactions while discarding background and low-energy processes.

\subsection{The CMS Trigger System}


The high frequency of bunch crossings along with the comparatively large amount of data ($5$ MB) per bunch crossing imposes strict constraints on the design and operation of the CMS Trigger System. To efficiently manage event selection, the CMS Trigger is structured as follows \cite{triggerrun3}:

\begin{itemize}
    \item \textbf{The Level-1 Trigger (L1T)}: A low-latency system implemented using custom electronics, such as Application-Specific Integrated Circuits (ASICs) and Field-Programmable Gate Arrays (FPGAs), which operate in real-time (online) to process and filter initial collision data. The L1T receives energy and position measurements, known as trigger primitives (TPs), from the calorimeters and muon detectors. Using firmware, the L1T subsystems reconstruct jets, photons, electrons, hadronically decaying $\tau$ leptons, and muons while also computing their energy sums. Notably, due to limited tracking information at this stage during Phase-1, the L1T cannot fully distinguish between photons and electrons, classifying both as electromagnetic objects. This processed information is then sent to the L1T Global Trigger, which uses a configurable set of selection algorithms, called seeds, collectively known as the L1T Menu, to decide whether an event should be retained for further analysis. If an event is labeled of interest, the data is passed to the High-Level Trigger (HLT) for additional processing. The L1T reduces the input rate from $40$ MHz to $100$ kHz (Run 2) and up to $110$~kHz~(Run 3), outputting a decision within $3.8$ $\mu$s after a collision occurs.\footnote{The LHC operates in multi-year periods known as Runs. Run 1 took place from 2009 to 2013, followed by a two-year shutdown. Run 2 lasted from 2015 to 2018, with another long shutdown (LS2) from 2019 to 2022 for upgrades. Run 3 began in 2022 and is expected to continue until mid-2026. Each Run features improvements in collision energy, luminosity, and detector performance \cite{timeline}.}
    \item \textbf{The High-Level Trigger (HLT)}:  Executes advanced algorithms on a dedicated processor farm to process events accepted by the L1T. It performs full event reconstruction using data from the Tracker, ECAL, HCAL, and Muon Detectors, applying refined selection criteria to further reduce the event rate for offline storage and analysis. The HLT runs on commercial Central Processing Units (CPUs) and Graphics Processing Units (GPUs), employing heterogeneous algorithms that can execute efficiently on both architectures. The algorithms at the HLT are designed to run faster than those used in offline reconstruction, prioritizing speed while maintaining sufficient precision. Rather than always running full event reconstruction, the HLT applies selected fast reconstruction algorithms in multiple steps. Each step includes a filter, and if an event fails to pass a filter, processing is terminated early to save resources. From the input stream of $110$ kHz, the HLT reduces it down to about $1.75$ kHz (Run 3), retaining only $4.55$\% of the events selected by the L1T.\footnote{The value of $1.75$ kHz was reported in 2023 by Ref. \cite{triggerrun3}. Additionally, the parking rate, which includes events stored for later reconstruction when computing resources become available, was higher, around $2.5$-$3$ kHz.}
\end{itemize}

From the terabytes of data generated each second in the CMS experiment, only about $0.01\%$ is stored for further analysis. To handle the immense data volume, the Data Acquisition (DAQ) system regulates data transfer between sub-detectors and the trigger system, provides buffering and temporary storage, and ensures the efficient flow of data \cite{daq}. It plays a crucial role in processing and transferring selected events data to permanent storage. Integrated with the DAQ is the Data Quality Monitoring (DQM) system \cite{dqm}. In online mode, the DQM obtains a small subset of the detector data within seconds to minutes after collisions. This data is partially reconstructed in real time to monitor detector health, identify performance anomalies, and ensure stable data-taking conditions. For offline analysis, a larger subset of the data, referred to as the Express Stream, is reconstructed with an approximate $1$-hour latency, allowing for early feedback on data quality and detector calibrations. The complete dataset then undergoes Prompt Reconstruction, which typically begins $48$ hours after data collection, although actual latency may vary depending on operational conditions. Beyond this, further reprocessing, such as Delayed Reconstruction or Re-Reconstruction, may take place weeks or months later, incorporating improved calibrations, updated algorithms, or revised reconstruction parameters.

\subsubsection{The Level-1 Trigger}

\begin{figure}
    \centering
    \includegraphics[scale=0.3]{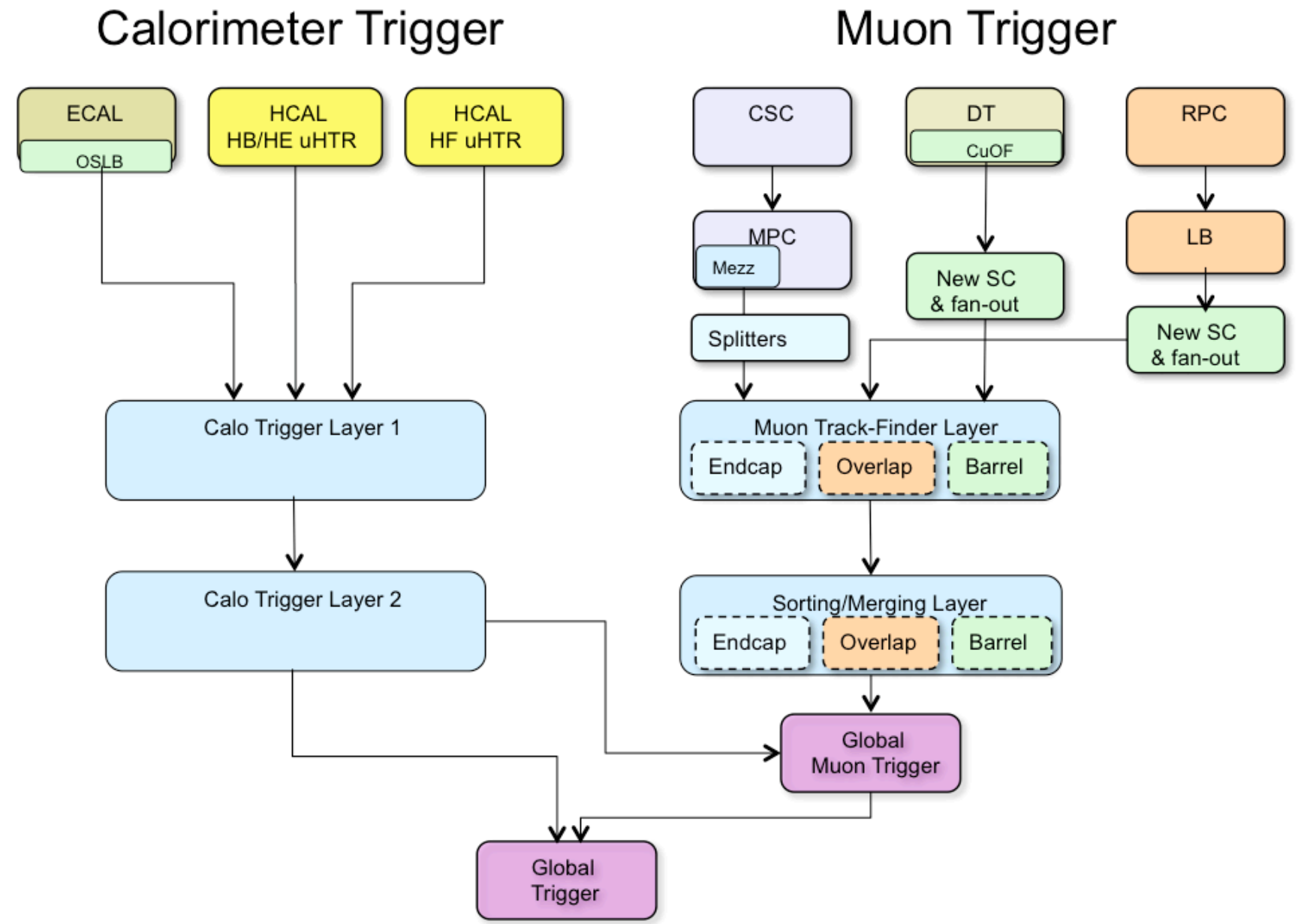}
    \caption{Dataflow of the L1T During Following Phase-1 Upgrades \cite{triggraph}}
    \label{fig:l1t}
\end{figure}

The initial decision of whether an event contains information that can lead to new physical discoveries is carried out by the Level-1 Trigger (L1T). Due to strict latency and resource constraints, highly optimized algorithms are implemented on custom electronics, primarily using FPGA devices in the L1T, with ASICs used in front-end detector electronics for signal digitization and preprocessing. These algorithms range from basic arithmetic operations to more advanced techniques, including pattern recognition and ML methods. The L1T processes calorimetry and muon detector data in real-time, outputting a trigger decision with a latency of approximately $3.8$$\mu$s following a collision. The diagram presented in Figure \ref{fig:l1t} illustrates the processing sequence and decision-making logic of the L1T system during Phase-1 of the CMS experiment, where trigger decisions are made based on reconstructed physics objects derived from calorimeter and muon detector data.

The Level-1 Calorimeter Trigger begins with the Trigger Primitive Generator (TPG) circuits in the ECAL, HCAL, and Forward Hadronic Calorimeter (HF), which process detector signals to compute energy sums. These sums are calculated within each Trigger Tower (TT), the fundamental unit of calorimeter granularity that represents small, discrete regions of the detector. In the trigger's original design, TPs were processed by the Regional Calorimeter Trigger (RCT) which identified electron and photon candidates, determined whether they were isolated or part of a jet, and computed regional energy sums \cite{trig}. The RCT further identified "quiet regions", or areas with minimal calorimetric activity, which helped in distinguishing isolated muons from those produced in dense hadronic environments. This information was transmitted to the Global Muon Trigger (GMT), which applied muon isolation cuts by evaluating the surrounding calorimetric energy and tracking activity. Concurrently, the RCT forwarded processed calorimetric information to the Global Calorimeter Trigger (GCT), where calculations to determine the total ($E_T$) and missing ($E_T^{\text{miss}}$) transverse energy were performed.

Following Run 1 of the LHC, the L1 Calorimeter Trigger underwent major upgrades that increased the granularity of the HCAL and ECAL detectors, enhanced the processing architecture, and improved data throughput. As part of this upgrade, the GCT and RCT were decommissioned and functionally replaced by a time-multiplexed, two-layer processing system: Layer-1, implemented using Calorimeter Trigger Processor cards (CTP7), performs regional data formatting and pre-processing; Layer-2, based on Master Processor cards (MP7), executes full-event calorimetric object reconstruction and computes global energy sums. These changes are summarized as follows \cite{up1,up2,up3}:
\begin{itemize} 
\item Upgrade of data transfer links, allowing for a tenfold increase in speed (to $10$ Gigabits per second) 
\item Upgrade to latest generation FPGAs and Xilinx Virtex 7, which utilize VIVADO as a High-Level Synthesis (HLS) software 
\item Replacement of legacy Versa Module Eurocard (VME)-based electronics with the more compact and scalable MicroTCA architecture 
\end{itemize}

These upgrades facilitated advanced algorithmic capabilities, including improved jet clustering, refined selection criteria, and more precise energy and spatial resolution. Additionally, the hardware architecture significantly reduced the overall trigger latency, allowing quicker trigger decisions and lower dead-time. The adoption of FPGA-based processing allowed for the implementation of higher-complexity selection algorithms, such as ML-based trigger systems.

\begin{figure}[ht]
    \centering
    \includegraphics[scale=0.25]{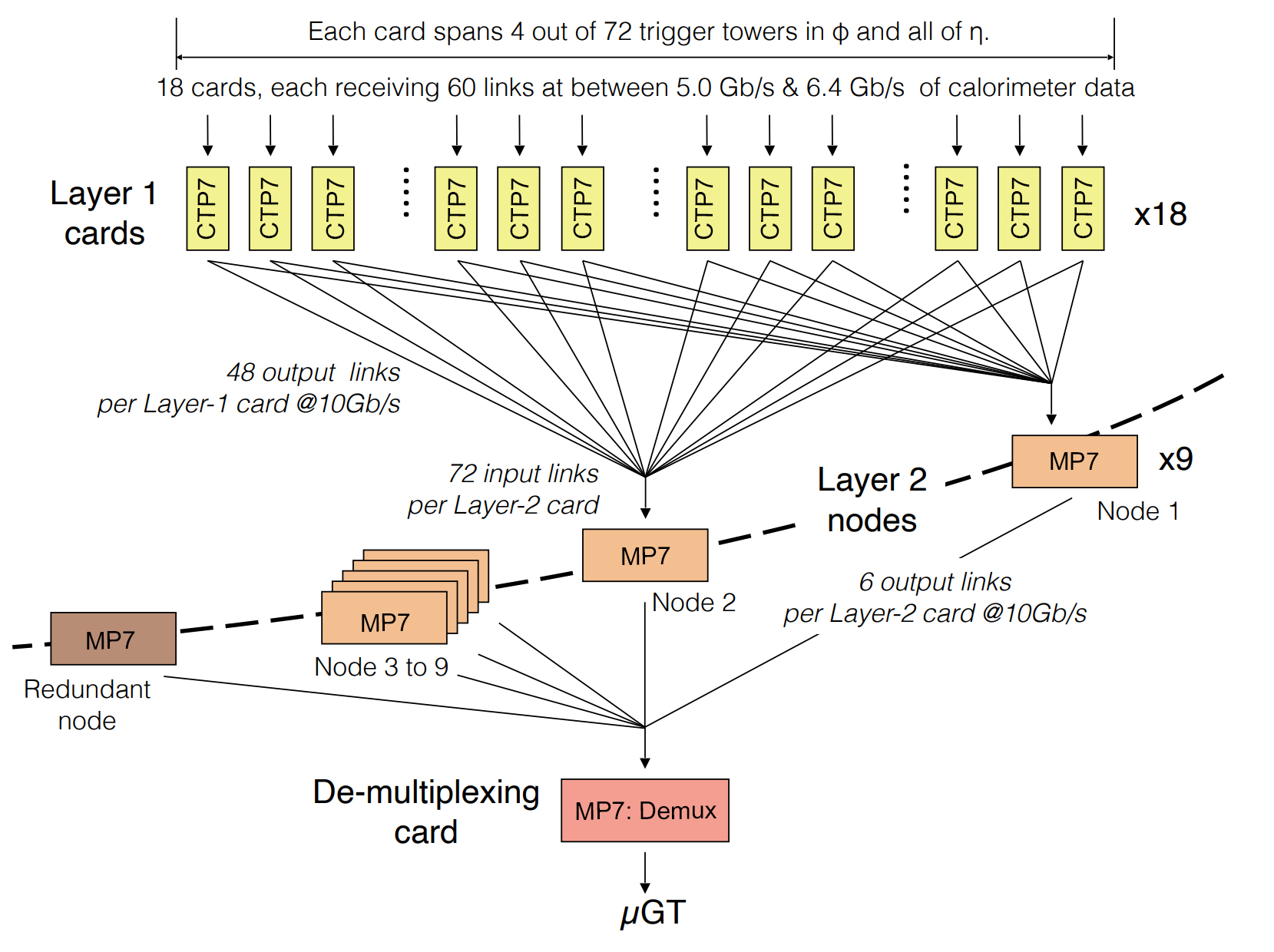}
    \caption{Schematic View of CTP7 and MP7 Cards Constituting the L1 Calorimeter Trigger Following Phase-1 Upgrades \cite{run2muon}}
    \label{fig:l1tmux}
\end{figure}

The calorimeter detector is segmented into 72 tower regions in the azimuthal angle ($\phi$), each covering $5^\circ$. In Layer 1 of the Calorimeter Trigger (see Figure \ref{fig:l1tmux}), these tower regions are grouped and processed by 18 CTP7 cards. Each CTP7 card spans the entire pseudorapidity ($\eta$) space and handles data from a distinct $20^\circ$ segment in $\phi$, collectively covering the full $360^\circ$ range.\footnote{For a schematic illustration of the $\eta$-$\phi$ coordinate system see Appendix C.}

The processed data from the CTP7 cards is then transmitted to the MP7 cards in Calorimeter Trigger Layer 2. The MP7 contain fully pipelined calorimeter algorithms that identify particle candidates and compute global energy sums. Each card takes in $72$ input links and has access to full TT granularity. Selected trigger candidates are then sent to an MP7 demultiplexer board (Demux), which formats the information appropriately for the Global Trigger (GT), also referred to as the microGT ($\mu$GT).

The Level-1 Muon Trigger system has a different operating logic than the Calorimeter Trigger. In the legacy system, a major component of track reconstruction in the RPC subsystem was the Pattern Comparator Trigger (PACT), which performed fast pattern matching by comparing RPC strip hit patterns against predefined templates to identify muon candidates and assign them approximate positions in $\eta$-$\phi$ space \cite{trig}. In this architecture, each of the three muon subdetectors (DT, CSC, and RPC) independently generated trigger primitives and reconstructed standalone muon tracks, which were then forwarded to the GMT for merging and selection. In contrast, the Phase-1 upgrade introduced a unified system in which information from all muon subdetectors is combined early in the trigger chain. Tracks are reconstructed within three distinct pseudorapidity regions using dedicated processors: the Barrel Muon Track Finder (BMTF) for the barrel region, the Overlap Muon Track Finder (OMTF) for the transition region between barrel and endcap, and the Endcap Muon Track Finder (EMTF) for the endcap region \cite{run2muon}.

Each of the three muon track finders receives input from the relevant muon subsystems based on their detector region: the BMTF uses DT and RPC data, the OMTF combines information from all three systems in the transition region, and the EMTF reconstructs muons using CSC and RPC data. Each track finder is segmented in $\phi$ to process data in parallel, with the BMTF divided into twelve $30^\circ$ sectors, and the OMTF and EMTF each divided into twelve $60^\circ$ sectors spanning the two endcaps \cite{run2muon}. Within each sector, trigger primitives are used to reconstruct muon candidates, assign charge, estimate transverse momentum based on track curvature in the magnetic fringe field, and assign a quality score. Up to $36$ muon candidates per processor are transmitted to the upgraded GMT, referred to as microGMT ($\mu$GMT), which replaces the legacy GMT. The $\mu$GMT removes duplicates across regional boundaries, sorts candidates based on a combination of $p_T$ and quality, and sends the top-ranked muons to the $\mu$GT for the final L1T decision.

The data processed and selected by the upgraded calorimeter and muon trigger subsystems are sent to the $\mu$GT, which applies predefined logical algorithms to produce the final L1T decision. Upon issuing an L1 Accept (L1A), the $\mu$GT signals the Trigger Control System (TCS) to synchronize subsystem timing and initiate the DAQ readout process \cite{trig}.

\subsubsection{The High Level Trigger}


The High-Level Trigger (HLT) is the second stage of the CMS trigger system, analyzing event data from all CMS sub-detectors with information content comparable to offline reconstruction, though some algorithmic steps are simplified \cite{daq}. It is built using commercial CPUs and GPUs. The data analysis algorithms can execute on either type of processor, with a preference for GPUs when available. With relaxed latency constraints, the software running on the HLT resembles the offline CMS analysis tools which provide a greater degree of accuracy when performing calculations.

The HLT utilizes the Particle Flow algorithm to perform real-time reconstruction of physics objects with high precision. Introduced into the HLT in $2011$, PF was initially applied to $\tau$ lepton identification using combined tracking and calorimetric information. In $2012$, its role was expanded to include full jet reconstruction and the calculation of missing transverse energy ($E_T^{\text{miss}}$). Within the HLT, PF enables the identification of individual particles such as electrons, muons, photons, and hadrons by integrating information from the tracker, calorimeters, and muon systems. This detailed event interpretation enables the HLT to reconstruct composite physics objects with better resolution than the L1T and to apply selection criteria that are broadly aligned with offline analysis strategies while operating within the real-time constraints of the online DAQ system.

The HLT comprises numerous software modules, each designed to execute well-defined tasks, which are systematically organized into multiple trigger paths \cite{wikihlt}. Each HLT trigger is specifically optimized to process a distinct category of physics objects and event information, ensuring an efficient and targeted event selection process. To initiate the execution of an HLT trigger path, it must be seeded by at least one L1T bit. This requirement enables the initial filter module within the HLT to identify and extract the relevant event data by referencing the L1 objects encoded in the corresponding L1 seed. Consequently, the HLT leverages L1 trigger information to streamline event selection, reducing computational overhead while maintaining high selection efficiency.

\subsection{High Luminosity LHC and CMS Phase-2 Upgrades}

To expand the physics reach of the LHC experiments, in 2030, the collider is scheduled to launch with a substantial upgrade. Currently, the LHC achieves a nominal luminosity of approximately $2\times 10^{34}$ cm$^{-2}$ s$^{-1}$ and an integrated luminosity of $65$ fb$^{-1}$ \cite{hllhc1}. Following the High Luminosity LHC (HL-LHC) upgrade, these values are expected to increase to $7.5\times 10^{34}$ cm$^{-2}$ s$^{-1}$ for the peak and $4000$~fb$^{-1}$ for the integrated luminosity (at least $250$ fb$^{-1}$ per year) \cite{hllhc1}. As a result, the average pileup is expected to increase from approximately $60$ to about $200$. This change impacts the experiments' ability to measure, select, and store data given the higher density of collisions per unit time.

Consequently, the CMS detector is undergoing significant Phase-2 upgrades to increase data granularity and improve triggering systems for efficiency and pileup mitigation. These upgrades target key subsystems, including tracking, calorimetry, muon detection, and triggering systems, ensuring the experiment remains sensitive to rare physics processes, precision measurements, and BSM searches.

The Phase-2 upgrades introduce:
\begin{itemize}
    \item A redesigned tracking system with enhanced granularity and real-time momentum discrimination.
    \item The High Granularity Endcap Calorimeter (HGCAL) to replace the existing endcap calorimeters, providing improved energy resolution and radiation hardness.
    \item Enhancements to the muon detection system, expanding coverage and improving track resolution in the forward regions.
    \item A new L1T architecture, incorporating more computationally expensive ML algorithms and increased latency to accommodate the higher event rates.
\end{itemize}

These advancements will enable CMS to cope with the extreme data rates and complexity of collisions at the HL-LHC while preserving and enhancing its ability to efficiently reconstruct and analyze events.

\subsubsection{Upgrades to the Tracking and Calorimetry Systems}

The upgraded Silicon Tracker will feature a highly granular design with 25 times the output channels of its Phase-1 predecessor, ensuring improved performance in high-pileup conditions \cite{phase2}. Additionally, in this configuration, the Tracker Endcap Pixel (TEPX) and the Tracker Barrel 2 Strip (TB2S) detectors will serve as real-time luminometers.

In the calorimetry systems, the front-end (FE) electronics will be replaced. The upgrade is meant to achieve $30$ ps time resolution for electrons and photons of $30$ GeV at a rate of $40$ MHz \cite{phase2}. Another key implementation is the Very Front-End electronics (VFE), which are meant to resolve and filter out anomalous signals, that result from direct particle impacts on the Avalanche Photodiodes (APDs). Furthermore, to mitigate the aging of the detector's electronics, the operating temperature for the APDs will be lowered from $18$ $^{\circ}$C to $9$ $^{\circ}$C.

To achieve system harmonization and improve efficiency the hadron barrel calorimeter's back end (BE) is set to adopt Advanced Telecommunications Computing Architecture (ATCA) boards \cite{phase2}. These ATCA boards will not only facilitate data readout and trigger primitive generation but also manage clock distribution to FE components. This unified approach to using ATCA boards across different subsystems ensures a streamlined and cohesive data acquisition and processing framework, enhancing both reliability and maintainability.

During Phase-2, the HCAL and ECAL endcaps are undergoing a major upgrade, replacing the existing systems with the High Granularity Calorimeter (HGCAL). The HGCAL is designed to operate in the high pileup environment of the LHC, aiming to enhance precision in particle flow reconstruction, improving sensitivity for vector boson fusion and scattering, allowing more precise jet substructure reconstruction, and extending the reach for long-lived particle searches \cite{phase2}. The system integrates $6$ million silicon sensor channels, covering $620$ m$^{2}$ near the interaction point, and $250,000$ scintillator tiles read out by silicon photomultipliers (SiPMs) across $370$ m$^{2}$ in lower fluence hadronic regions. These components work together to ensure high-resolution detection and robustness against radiation damage, facilitated by a carbon dioxide cooling system maintaining temperatures at $-30$ $^{\circ}$C.

The High Granularity Calorimeter Read-Out Chip (HGCROC) used in the detection modules will feature a dynamic range to read out signals originating from high-energy photons, as well as minimum ionizing particles (MIPs) \cite{phase2}. The lower energy signals will be digitized using a 10-bit Analog-to-Digital Converter (ADC), whereas the higher energy signals will be reconstructed using the time over threshold (ToT) method. Both Online and Offline information processing will use ML-assisted pattern recognition algorithms to achieve jet clustering and particle reconstruction.

\subsubsection{Muon System Upgrade}

The muon system will undergo upgrades on the DT, CSC, and RPC detectors which will be enhanced with more efficient electronics to increase their performance and cope with the 10-fold increase in muon production rates \cite{phase2}. In the high-background, high-rate regions new detectors will be installed intended to extend the geometric range from $2.4$ to $2.8$ in $|\eta|$, enhance tracking, and allow for a bending angle measurement at the trigger level.

\subsubsection{Level-1 Trigger Phase-2 Upgrades}

To meet the demands of the HL-LHC, where up to $200$ simultaneous proton-proton interactions per bunch crossing are expected, the CMS L1T system will undergo a substantial Phase-2 upgrade involving major improvements to both hardware and trigger algorithms \cite{l1phase2}. The upgraded L1 Trigger will feature extended latency and bandwidth, enabling the integration of information from high-granularity sub-detectors, such as the new tracking system and the high-granularity endcap calorimeter, directly into the trigger decision. A key innovation is the introduction of a correlator layer, which combines inputs from multiple subsystems to reconstruct complex physics objects with improved resolution and selectivity. These enhancements are essential not only to maintain efficiency under HL-LHC conditions but also to improve the purity and precision of triggered events, ensuring the system remains sensitive to rare and high-value physics processes. Additionally, the planned integration of tracking information at the L1 during Phase-2 will enable real-time reconstruction of charged particle trajectories, providing precise spatial and momentum information for efficient pileup suppression and improved object identification. This functionality will be implemented through the Track Trigger, a key component of the Phase-2 architecture. The Phase-2 input to the L1 Trigger can be summarized as follows \cite{l1phase2}:
\begin{itemize}
    \item \textbf{Tracker:} Data will be included from the Outer Tracker at a rate of $40$~MHz. This allows for local $p_T$ measurements to be performed using FE electronics. In such a way, the read-out rate of soft (low transverse momentum) interactions can be reduced $10$-fold through selection on the local $p_T$. Studies have demonstrated that $97\%$ of particles created in pp collisions at $14$ TeV have $p_T<2$ GeV, making soft interactions a significant portion of the measured processes \cite{study}. It is expected for approximately $15,000$ stubs per bunch crossing to be sent to the Track Finder (TF) TPG, which will reconstruct the trajectories with minimal latency of $5$ $\mu$s, which includes the transmission time from the detector ($1$ $\mu$s). A subset of $200$ tracks will be sent to the L1 Trigger, which will use $100$ bits per track to encode the parameters with no degradation in performance. This increased precision and efficiency in TP input will enhance the L1 Trigger's accuracy and performance.
    \item \textbf{Electromagnetic Barrel Calorimeter:} For Phase-2, the ECAL barrel trigger primitive generator (EB TPG) will be relocated from the on-detector electronics to the back-end system, where it will receive crystal-level data directly from the detector. The primary goal for the EB TPG upgrade is to enable input data calibration and apply digital filtering to extract precise energy and timing information. The data granularity will increase from one TT to a $5\times5$ array of crystals per tower 
    \item \textbf{Hadron Barrel and Forward Calorimeters:} The Phase-2 upgrade of the CMS Hadron Barrel (HB) and Forward (HF) calorimeters aims to enhance the back-end electronics and partially replace front layer scintillator tiles if required due to radiation damage anticipated during the HL-LHC. The upgrade maintains the current readout channel count, transverse segmentation, and longitudinal readout depths established after the Phase-1 upgrade. The HB TPG will utilize the same hardware as the ECAL Barrel to streamline development and operational resources. Signals from four depth segments per TT will be sampled at $40$ MHz, corrected for pedestal, gain, and response, and then summed, with peak detection algorithms applied. The HF detector will retain its Phase-1 electronics but will be supplemented by reusing Phase-1 HB and Hadron Endcap (HE) back-end cards to meet the increased L1A rate demands. Both HB and HF TPGs will feature advanced encoding and signal suppression algorithms to improve calibration, lepton isolation, MIP identification, and overall energy reconstruction.
    \item \textbf{High Granularity Endcap Calorimeter:} The HGCAL will feature a new high granularity sampling design, utilizing both silicon and scintillator sensors. The calorimeter will have 52 sensitive layers per endcap, with 28 in the electromagnetic section and 24 in the hadronic section, with only half of the electromagnetic section layers contributing data to the L1 Trigger. The calorimetry TP data will be sums of individual channels, referred to as trigger cells, implemented in both the silicon and scintillator regions. These values form tower maps of $E_T$ covering any $\eta-\phi$ grid. The Endcap Calorimeter Trigger (ECT) TPG processes this data in two stages: first, by forming two-dimensional clusters within each layer from trigger cells and summing tower data to form a single $\eta-\phi$ grid, and then by combining all 2D clusters in depth to form 3D clusters. The tower maps and 3D clusters from the ECT TPs will be input to the L1 Trigger during Phase-2.
    \item \textbf{Muon Barrel:} The barrel muon system will replace the existing DT and RPC TPGs to enhance efficiency, spatial resolution, and timing precision. The trigger primitive generation will be managed by $84$ processor boards, similar to those used for barrel muon track-finding, handling data transmission rates of $30.7$ Tb s$^{-1}$ per sector from the DT system and 0.3 Tb s$^{-1}$ from the RPC system via 10 Gb s$^{-1}$ links. Studies on DT stub identification algorithms suggest possible data formats that include precise hit positions to improve track-finding accuracy. Independent paths for DT and RPC primitives will reduce sensitivity to detector issues while combining both sources is expected to optimize performance. 
    \item \textbf{Muon Endcap:} In the Muon Endcap detector, the CSC TPG electronics will be upgraded, maintaining the Phase-1 data format but with improved stub reconstruction algorithms to address high pileup inefficiencies. These improvements include better ghost track cancellation, reduced pre-trigger deadtime, optimized pattern recognition, and enhanced timing, with data transmitted via $588$ optical links, each operating at $3.2$ Gb s$^{-1}$. The RPC detectors will retain their data format with an upgrade to faster link speeds. In contrast, the new iRPC detectors will have no $\eta$ segmentation, but will extrapolate the position in $\eta$ through two precision timing measurements. The Gas Electron Multiplier (GEM) detectors will provide L1 Trigger information through reconstructed hit clusters and integrated GEM-CSC track stubs, enhancing local reconstruction efficiency, particularly in regions prone to CSC aging. GEM TPs will be transmitted via $252$ links at $10$ Gb s$^{-1}$, with integrated stubs boosting efficiency by up to $30\%$ in specific areas. For the GEM ME0 (Muon Endcap station 0), multi-layer stubs will be reconstructed on-detector to minimize link requirements.
\end{itemize}

Most of the aforementioned upgrades aim to increase the data quality received by the L1T, with some pre-selection being done at detection level. The higher granularity, precision, and transfer speed of the TP data is expected to improve the performance of the L1T which is intended to employ complex jet clustering and tagging algorithms that are aided by machine learning programs.

\begin{figure}[ht]
    \centering
    \includegraphics[scale=0.4]{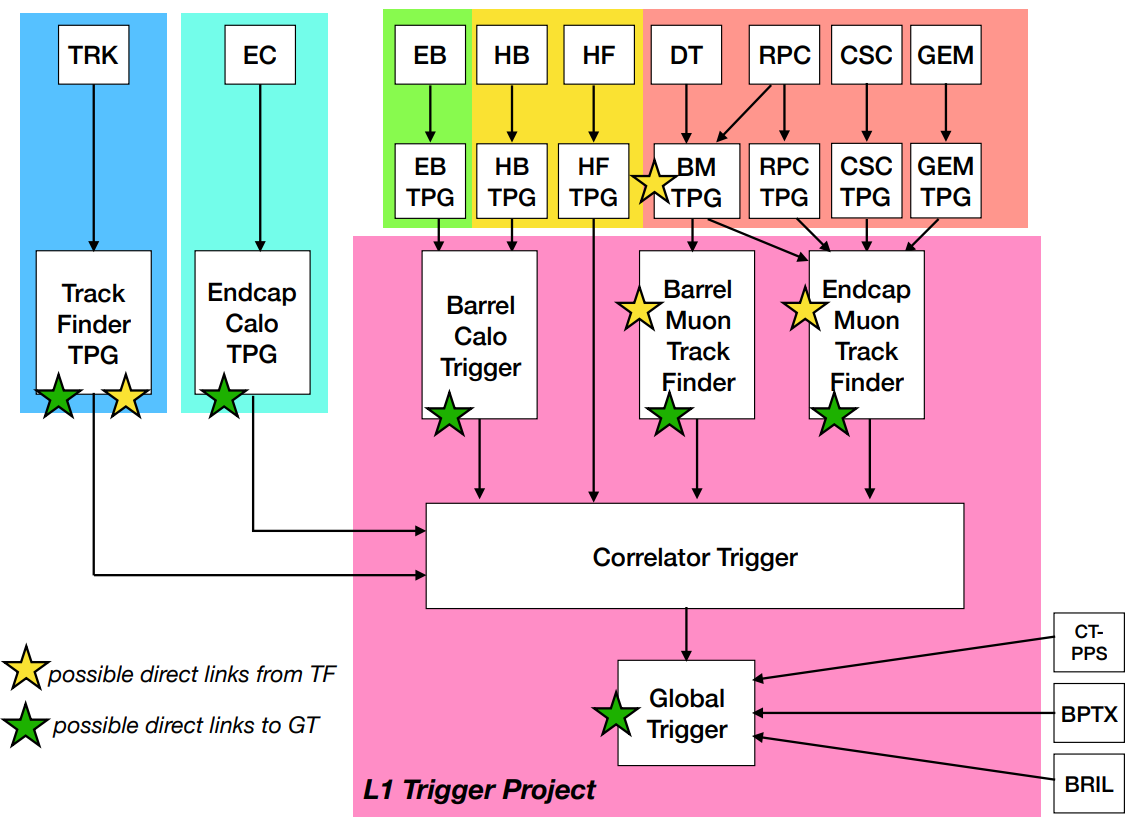}
    \caption{High-Level Diagram of the Phase-2 L1 Trigger Showing Arrows for Established Paths and Direct Links Under Investigation \cite{l1phase2}}
    \label{fig:l1phase2}
\end{figure}

Due to the increased pileup and information availability, the latency of the L1T, which is the time available to produce an L1A signal following a collision, will be increased from $3.8$ $\mu$s in Phase-1 to $12.5$ $\mu$s in Phase-2, with a maximum rate of $750$ kHz \cite{l1phase2}. The high-level view of the planned Phase-2 L1 Trigger setup is depicted in Figure \ref{fig:l1phase2}, which highlights data flow from various subdetectors into the Correlator Trigger, which feeds into the GT for L1A decision processing. Solid lines indicate established data paths, while additional links under investigation (marked with green and yellow stars) include potential direct connections from upstream systems to the Track Finder (TF) and Global Trigger (GT).

A significant change from the Phase-1 setup is the inclusion of the Correlator Trigger (CT), which is necessitated by the introduction of the Track Trigger. The CT performs event reconstruction by combining information from the central tracker, calorimeters, and muon systems \cite{l1phase2}. The online selectivity of this layer is designed to approach the performance benchmarks of offline reconstruction in the HLT. Unlike the setup used during Phase-1 of the experiment, Phase-2 will enable tracking information to be available at the L1T stage. Reconstruction will be performed using four dataflow paths that utilize the upgraded sub-detector components: Tracking Trigger path (initiated from TRK in Figure \ref{fig:l1phase2}), Calorimeter Trigger path (initiated from EC, EB, HB, and HF in Figure \ref{fig:l1phase2}), Muon Trigger path (initiated from DT, RPC, CSC, and GEM in Figure \ref{fig:l1phase2}), and Particle-Flow Trigger path (embedded in two layers in the CT). All paths feed into the CT, which then transmits them to the Global Trigger. With minimal latency, the GT outputs the L1A decision to the Trigger Control and Distribution System (TCDS), which then initiates the DAQ readout chain. While each path contributes to event reconstruction at L1, some subsystems also provide standalone trigger objects with limited correlation to other detectors. These objects may have reduced resolution and higher fake rates but can improve trigger efficiency or aid in commissioning and validation.  

\newpage

\newpage\stepcounter{section}
\section*{Chapter {II}: Boosted Jets, Higgs Boson Decays, and Di-Higgs Production}
\addcontentsline{toc}{section}{Chapter {II}: Boosted Jets, Higgs Boson Decays, and Di-Higgs Production}
\setcounter{figure}{0}
\markboth{Chapter {II}: Boosted Jets, Higgs Boson Decays, and Di-Higgs Production}{}

\subsection{The Standard Model}

The Standard Model (SM) of Particle Physics provides a theoretical framework that describes all known particles and their interactions, with the exclusion of gravity. It is considered a gauge theory because its fundamental interactions arise from requiring local gauge invariance under specific symmetry transformations. The theory is built on the gauge group $\text{SU}(3)_{\text{C}}\times\text{SU}(2)_{\text{L}}\times\text{U}(1)_{\text{Y}}$, which reduces to $\text{SU}(3)_{\text{C}}\times \text{U}(1)_{\text{EM}}$ after spontaneous symmetry breaking via the Higgs mechanism. In this notation, L denotes the left-handed nature of weak isospin, while Y represents the weak hypercharge. Mathematically, this means that the Lagrangian remains invariant under local transformations of these groups, requiring the introduction of gauge bosons (gluons, W/Z bosons, and the photon) as force carriers. The Higgs field, which will be further discussed in Chapter II, Section \ref{hm}, plays a crucial role in electroweak symmetry breaking by acquiring a vacuum expectation value, thereby giving mass to the $\text{W}^\pm$ and Z bosons, while leaving the photon massless.



\begin{figure}[ht]
    \centering
    \includegraphics[width=0.65\linewidth]{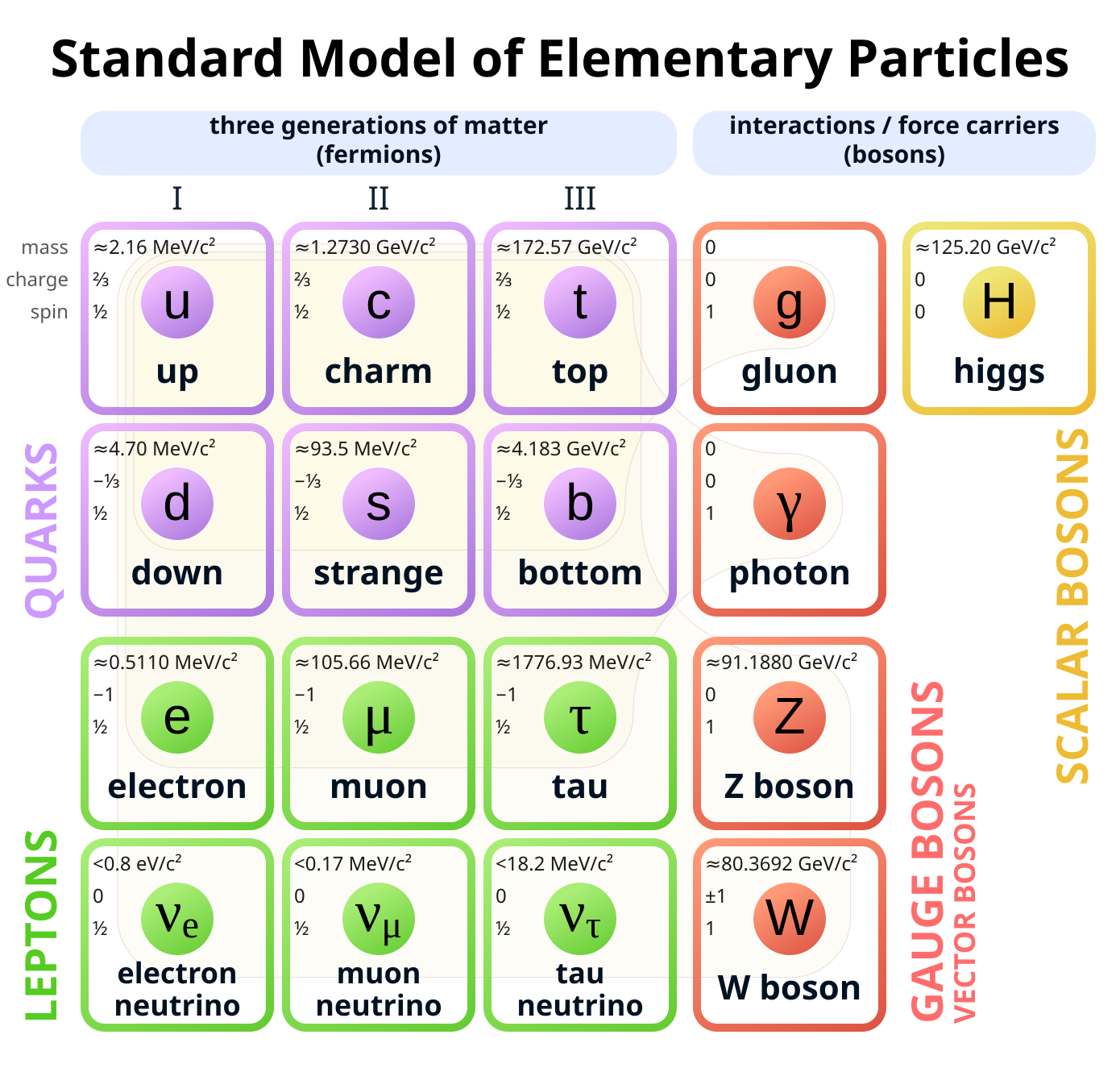}
    \caption{Standard Model of Particle Physics}
    \label{fig:sm}
    \small
    The Standard Model diagram depicting the bosons (force carrying particles) and the three generations of matter fermions. Possible interactions between species are highlighted, with the mass, charge, and spin shown. Each fermion has an antimatter counterpart which is omitted in this diagram.
\end{figure}

As shown in Figure \ref{fig:sm}, there are two (major) classes of particles \cite{book}:
\begin{itemize}
    \item \textbf{Fermions}: Fundamental particles that have half-integer spin. In the SM, all fermions except neutrinos acquire mass through the Higgs mechanism. The dynamics of the 12 fundamental fermions is governed by the Dirac equation ($(i\gamma^\mu\partial_\mu-m)\Psi(x)=0$), though neutrino masses may require an extension such as the Majorana formalism \cite{majoranacern}. Fermions follow Fermi-Dirac statistics and thus obey the Pauli exclusion principle.
    \item \textbf{Bosons}: Force-carrying particles that have integer spin. The vector bo-sons, which have spin-1, mediate three of the four fundamental forces: the Strong Nuclear Force (gluon), the Weak Nuclear Force (W and Z bosons), and the Electromagnetic force (photon). Gravity is not included in the Standard Model but is hypothesized to be mediated by the graviton (spin-2). The Higgs particle is a scalar boson, having spin-0. 
\end{itemize}

In quantum chromodynamics (QCD), the strong nuclear force is given as an interaction between colored quarks. The symmetry group for gauge transformations in the case of QCD is given by $\text{SU}(3)_\text{C}$, where C denotes color. The gauge boson for the strong interaction, the gluon, is massless and does not carry hypercharge. Additionally, electroweak interactions are not affected by quark color changes. This implies that $\text{SU}(3)_C$ transformations commute with $\text{U}(1)_\text{Y}$ and $\text{SU}(2)_\text{L}$, making the Standard Model Lagrangian invariant under $\text{SU}(3)_\text{C} \times \text{SU}(2)_\text{L} \times \text{U}(1)_\text{Y}$ transformations. However, gauge invariance forbids explicit mass terms for gauge bosons. In the SM, spontaneous symmetry breaking via the Higgs mechanism allows electroweak bosons and fermions to acquire mass while preserving gauge invariance.


\subsubsection{The Higgs Mechanism}\label{hm}

For massive particles to exist in the Standard Model, the physical vacuum must break some of the gauge symmetries present in the SM Lagrangian \cite{chris}. Specifically, the electroweak symmetry $\text{SU}(2)\text{L} \times \text{U}(1)\text{Y}$ is spontaneously broken to the electromagnetic subgroup $\text{U}(1)_{\text{EM}}$, which corresponds to the unbroken gauge symmetry of the vacuum \cite{weak}.

The central idea of the Higgs mechanism is the existence of a scalar field permeating all of space. This would entail a non-zero vacuum expectation value (VEV). The Higgs field causes a symmetry breakdown from $ \text{SU}(2)_{\text{L}} \times \text{U}(1)_{\text{Y}}$ to $\text{U}(1)_{\text{EM}}$, which induces mass by modifying the vacuum structure. Additionally, this accurately models the mass ratios of the Z and W$^\pm$ bosons in terms of the Weinberg angle, while adding an additional particle degree of freedom.

Due to the spontaneous symmetry breaking, the Higgs field is required to be charged under both $ \text{SU}(2)_\text{L}$ and $\text{U}(1)_{\text{Y}}$ \cite{chris}. Given that the smallest $ \text{SU}(2)_\text{L}$ multiplet is the doublet, this can be taken to be the minimal choice for a Higgs field description:

\begin{equation}
    \Phi (x) = \begin{bmatrix}
        \phi^+(x)\\
        \phi^0(x)
    \end{bmatrix},
\end{equation}
for:
\begin{equation}
    \phi^+(x) = \frac{1}{\sqrt{2}}\Big( \phi^+_1(x)+i\phi^+_2(x) \Big),
\end{equation}
\begin{equation}
    \phi^0(x) = \frac{1}{\sqrt{2}}\Big( \phi^0_1(x)+i\phi^0_2(x) \Big),
\end{equation}
where $\phi_1^+,\phi_2^+,\phi_1^0,$ and $\phi_2^0$ are real and constitute the four degrees of freedom in the Higgs field. The kinetic energy ($T$) of this field can be expressed as:
\begin{equation}\label{ken}
    T(\Phi^\dagger,\Phi) = (D_\mu\Phi)^\dagger(D^\mu\Phi),
    \end{equation}
where $D_\mu$ is the $ \text{SU}(2)_{\text{L}} \times \text{U}(1)_{\text{Y}}$ gauge-covariant derivative expressed as:
\begin{equation}
D_\mu = \Big( \partial_\mu +ig'YB_\mu-igW_\mu^aT^a  \Big)
\end{equation}
where:
\begin{itemize}
    \item $\partial_\mu$ is the kinetic term,
    \item $ig'YB_\mu$ is contributed by the abelian $\text{U}(1)_{\text{Y}}$ symmetry with $g'$ being the coupling constant for the interactions under the symmetry, $\text{Y}$ is the hypercharge of the field, and $B_\mu$ is the gauge field associated with the $\text{U}(1)_{\text{Y}}$ group,
    \item and $-igW_\mu^aT^a $ is contributed by the $\text{SU}(2)_{\text{L}}$ symmetry, where g is the coupling constant, $W_\mu^a$ ($a=1,2,3$) are the gauge fields associated with the $\text{SU}(2)_{\text{L}}$ group, and $T^a$ are the Pauli matrices multiplied by one half.
\end{itemize}

Similarly, the potential energy ($V$) of the $\Phi$ field can be expressed as:
    \begin{equation}\label{pot}
    V(\Phi^\dagger\Phi) = -\mu^2\Phi^\dagger\Phi+\lambda(\Phi^\dagger\Phi)^2,\quad \mu^2>0,\;\lambda>0,
\end{equation}
where $\lambda$ is the coupling strength of the four-point Higgs interaction and $\mu$ is the mass parameter \cite{chris}.

Based on the equation above, the minimal value for the potential energy does not occur when $\Phi=0$, but rather at a finite value. This would imply a non-zero VEV, evaluated to be $vev=\langle\Psi\rangle=\Psi_{min}$, with $\Psi_{min}$ expressed as:
\begin{equation}
    \Psi_{min} = \frac{1}{\sqrt{2}}\begin{bmatrix}
        0\\\nu
    \end{bmatrix},\quad \text{where }\;\nu = \sqrt{\frac{\mu^2}{\lambda}}.
\end{equation}

The spontaneous symmetry breaking can be easily seen due to the fact that the ground states of the physical vacuum given by $ \text{SU}(2)_{\text{L}} \times \text{U}(1)_{\text{Y}}$ gauge transformations of $\Phi_{min}$ are not equal to it \cite{chris}. This, however, still implies that the Lagrangian symmetries are intact and it is the dynamical selection of the vacuum from the self-interacting potential in Equation \ref{pot} that has reduced the physical vacuum's symmetries.

With respect to $\Phi_{min}$, excitations of the field can be parameterized as:
\begin{equation}\label{killmepls}
    \Phi(x) =\frac{1}{\sqrt{2}} e^{i\mathbf{\xi}(x)\times\mathbf{\tau}}\begin{bmatrix}
        0\\
        \nu+H(x)
    \end{bmatrix},
\end{equation}
where $\mathbf{\xi}(x)$ are excitations of $\Phi_{min}$ along the potential minimum, $\mathbf{\tau}$ are the Pauli matrices, and $H(x)$ is an excitation in the radial direction that corresponds to the prediction of a free particle state \cite{chris}. Based on this, an expansion of $V(\Phi)$ with respect to $\Phi_{min}$ gives:
\begin{equation}\label{expanded}
    V(H) = -\frac{1}{4}\mu^2\nu^2+\mu^2H^2+\lambda \nu^2H^3+\frac{1}{4}\lambda H^4.
\end{equation}
Therefore, the predicted particle, the Higgs boson, should have a mass of $m_H=\sqrt{2\mu^2}=\sqrt{2\lambda}\nu$.

From Equation \ref{ken} the mass of the W boson can be derived:
\begin{equation}
    (D_\mu\Phi)^\dagger(D^\mu\Phi) = \frac{1}{4}g^2W_\mu^iW^{j\mu}\Phi^\dagger \tau_i\tau_j\Phi+...,
\end{equation}
where summation over indices is implied. For $i=j$, $\tau^2=1$, therefore the term becomes:
\begin{equation}
    \frac{g^2\nu^2}{8}\Big( (W_\mu^-)^\dagger W^{-\mu}+(W_\mu^+)^\dagger W^{\pm\mu}  \Big),
\end{equation}
using the fact that charged W boson states are given by $W^\pm_\mu=2^{-\frac{1}{2}}(W_\mu^1\mp iW_\mu^2)$ \cite{chris}. This term in the Lagrangian corresponds to a W boson particle with mass $M_{W^+}=M_{W^-} = \frac{g\nu}{2}$. This is simply a coupling constant multiplying the VEV term, $\nu$, which is indicative of the Higgs mechanism. Similarly, by considering the coupling of neutral gauge fields to the Higgs doublet the mass of the Z boson can be recovered as:
\begin{equation}
    M_Z^2 = \frac{\nu^2}{4}(g^2+g'^{2}) = \frac{M_{W^\pm}^2}{\cos^2\theta_W},
\end{equation}
where $\theta_W$ is the Weinberg angle (also known as the weak mixing angle) \cite{chris}. This is fully derived in Appendix B. The photon, on the other hand, remains massless due to the preservation of $\text{U}(1)_{\text{EM}}$ symmetry. Additionally, fermions also interact with the Higgs field, acquiring mass through Yukawa interactions.

\subsubsection{Higgs to Bottom-Antibottom Quark Decay Mode}\label{hot}

The existence of the Higgs mechanism was confirmed in 2012, when the Higgs boson was discovered by the CMS collaboration \cite{cmshiggs}, which was soon followed by results presented by the ATLAS experiment \cite{atlashiggs}. Since then, many Higgs decay modes have been measured, such as $H\rightarrow ZZ\rightarrow 4l$, $H\rightarrow W^+W^-$, $H\rightarrow b\bar{b}$, $H\rightarrow \gamma\gamma$, $H\rightarrow e^+e^-$, $H\rightarrow \mu^+\mu^-$, and $H\rightarrow \tau^+\tau^-$.

The lifetime of the Higgs boson is short, resulting in a small time-of-travel in the detector before it decays. The lifetime can be calculated using the branching ratio and the reduced Planck constant through the following equation:
\begin{gather}
    \tau_H = \frac{\hbar}{\Gamma_H},
\end{gather}
where $\tau_H$ is the lifetime of the Higgs, $\Gamma_H$ its full decay width, and $\hbar$ is the reduced Planck constant. Using the most recent value for $\Gamma_H$ as reported by the Particle Data Group (2024)\footnote{$\Gamma_H = 3.7^{+1.9}_{-1.4}$ MeV \cite{codata}}, in addition to the CODATA value for the Planck constant\footnote{$\hbar = 6.582 119 569...\times 10^{-16}$ eV s \cite{pdghiggs}}, the following lifetime is computed:
\begin{gather}\label{tau}
    \tau_H = \Big(1.78^{+1.08}_{-0.60}\Big)\times 10^{-22}\;\text{s},
\end{gather}
where the error was estimated using the usual error propagation method \cite{codata, pdghiggs}. The above estimate is within the error margin of the theoretically computed value of $1.6\times10^{-22}$ s \cite{higgscernyas}.

With a branching fraction of $(53\pm 8)\%$, the most common Higgs decay is to a bottom-antibottom quark pair \cite{pdghiggs}. This decay channel was observed at the LHC in 2018 through the VH production mode (Higgs in association with a vector boson) \cite{atlashbb}. A significant obstacle to measuring this decay was the high presence of QCD background, which made it difficult to isolate and reconstruct the signal.

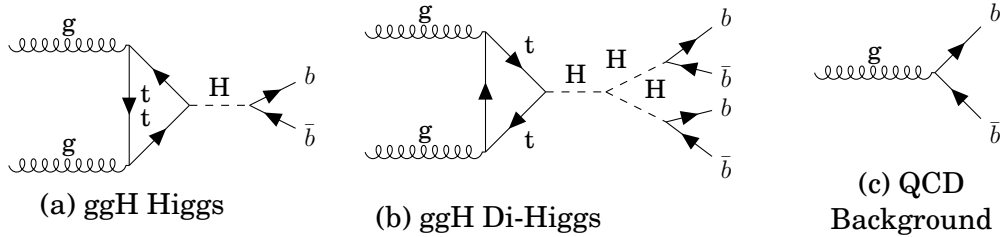
\begin{figure}[ht]
  \centering

  \begin{minipage}{.2\textwidth}
    \centering
    \begin{tikzpicture}[scale=0.8, transform shape]
             \begin{feynman}
    \vertex (gluon1) at (0,0);
    \vertex (g1) at (2,0);
    \vertex (gluon2) at (0,-2);
    \vertex (g2) at (2,-2);
    \vertex (higgs) at (3,-1);
    \vertex (h) at (4,-1);
    \vertex (b) at (5,-0.5) {$b$};
    \vertex (antib) at (5,-1.5) {$\bar{b}$};

    \diagram* {
     (gluon1) -- [gluon, edge label=g] (g1),
     (gluon2) -- [gluon, edge label=g] (g2),
    (g1) -- [fermion] (g2),
    (g2) -- [fermion, edge label=t] (higgs),
    (higgs) -- [fermion, edge label=t] (g1),
    (higgs) -- [scalar, edge label=H] (h),
    (h) -- [fermion] (b),
    (antib) -- [fermion] (h),
    };
  \end{feynman}
    \end{tikzpicture}
    \subcaption{ggH Higgs}\label{ghhH}
  \end{minipage} \hspace{1.2cm}
    \begin{minipage}{.2\textwidth}
    \centering
    \begin{tikzpicture}[scale=0.8, transform shape]
             \begin{feynman}
    \vertex (gluon1) at (0,0);
    \vertex (g1) at (2,0);
    \vertex (gluon2) at (0,-2);
    \vertex (g2) at (2,-2);
    \vertex (higgs) at (3,-1);
    \vertex (h) at (4,-1);
    \vertex (b) at (5,-0.5);
    \vertex (antib) at (5,-1.5);
    \vertex (1) at (6,0.25) {$b$};
    \vertex (2) at (6, -0.75) {$\bar{b}$};
    \vertex (11) at (6,-1.25) {$b$};
    \vertex (22) at (6, -2.25) {$\bar{b}$};

    \diagram* {
     (gluon1) -- [gluon, edge label=g] (g1),
     (gluon2) -- [gluon, edge label=g] (g2),
    (g2) -- [fermion] (g1),
    (higgs) -- [fermion, edge label=t] (g2),
    (g1) -- [fermion, edge label=t] (higgs),
    (higgs) -- [scalar, edge label=H] (h),
    (h) -- [scalar,edge label=H] (b),
    (h) -- [scalar,edge label=H] (antib),
    (b) -- [fermion] (1),
    (2) -- [fermion] (b),
    (antib) -- [fermion] (11),
    (22) -- [fermion] (antib),
    };
  \end{feynman}
    \end{tikzpicture}
    \subcaption{ggH Di-Higgs}\label{aaa}
  \end{minipage}\hspace{2.2cm}
  \begin{minipage}{.2\textwidth}
    \centering
    \begin{tikzpicture}[scale=0.8, transform shape]
     \begin{feynman}
    \vertex (g) at (0,0);
    \vertex (int) at (2,0);
    \vertex (b) at (3,1){$b$};
    \vertex (antib) at (3,-1){$\bar{b}$};

    \diagram* {
     (g) -- [gluon, edge label=g] (int);
     (int) -- [fermion] (b);
     (antib) -- [fermion] (int);
    };
  \end{feynman}
    \end{tikzpicture}
    \subcaption{QCD Background}\label{d}
  \end{minipage}
  \small
  \caption{ggH Production Mechanism of single Higgs and Di-Higgs}{
  \small
  Diagram \ref{ghhH} depicts a single Higgs boson decaying into two b quarks. Shown in Diagram \ref{aaa} is the Higgs self-coupling mechanism resulting in a final state with four b quarks. A similar diagram can be drawn for the process in Figure \ref{b}, with each Higgs boson decaying into a $b\bar{b}$ pair. For comparison, Diagram \ref{d} shows a QCD background process that can mimic these Higgs decay signatures.}
\end{figure}

The choice to develop an ML trigger system trained on simulated $gg\rightarrow H\rightarrow b\bar{b}$ events is driven by a combination of experimental practicality and strong physics motivation. Among all Higgs production modes, gluon fusion has the highest cross section at the LHC, making it the most statistically rich source of Higgs events. By concentrating on a single, well-understood channel with distinct kinematic properties, systematic uncertainties can be reduced and events can be more accurately simulated. This results in more robust training data for ML models, ultimately enhancing QCD background rejection and signal efficiency. Improved trigger-level $H\rightarrow b\bar{b}$ tagging can contribute to more precise measurements of the bottom quark Yukawa coupling in single-Higgs production, and facilitate better sensitivity to di-Higgs final states relevant for probing the Higgs self-coupling discussed in Chapter II, Section \ref{didi}.

The bottom quark Yukawa coupling, $y_b$, is a fundamental parameter in the Standard Model, governing the strength of the interaction between the Higgs field and the bottom quark. Precise measurements of $y_b$ are essential for testing the proportionality between fermion masses and their couplings to the Higgs boson, a core prediction of the Higgs mechanism. Any deviation from the SM expectation could signal new dynamics in the Higgs sector or the presence of additional BSM interactions.

Beyond testing the SM prediction for the bottom Yukawa coupling, measurements of $H \rightarrow b\bar{b}$ events also offer sensitivity to potential BSM effects \cite{main}-\cite{6}. In particular, analyzing high-$p_T$ (boosted) $H \rightarrow b\bar{b}$ decays provides an alternative approach for probing the top quark Yukawa coupling, complementary to the $t\bar{t}H$ production mechanism (see Appendix A, Diagram \ref{htt}). Moreover, at high transverse momentum, the process $gg \rightarrow H \rightarrow b\bar{b}$ becomes sensitive to virtual contributions from heavy BSM particles in the gluon-fusion loop, offering a potential window into new physics through deviations in the Higgs kinematic distributions.

\subsubsection{The Higgs Potential, Self-Coupling, and Di-Higgs Production}\label{didi}

In addition to coupling with SM particles, the Higgs boson is theoretically predicted to exhibit self-interactions, as described by the structure of the Higgs potential. This phenomenon, referred to as Higgs self-coupling, is encoded in the scalar potential of the SM Higgs field, given in Equation \ref{pot}. Upon spontaneous symmetry breaking, the Higgs field acquires a vacuum expectation value and the potential can be perturbatively expanded around the physical Higgs field to yield interaction terms, including a trilinear self-coupling term proportional to $\lambda$ (see Equation \ref{killmepls}).

Experimentally, the trilinear Higgs self-coupling is most directly accessible via processes involving the production of Higgs boson pairs, commonly referred to as "di-Higgs" production. Although the total di-Higgs production cross section is influenced by multiple contributions --- including box diagrams and triangle diagrams involving the trilinear vertex --- deviations in the measured rate or kinematic distributions from the SM predictions can be used to constrain or extract the value of $\lambda$.

It is important to emphasize that di-Higgs production does not offer a model independent measurement of the Higgs potential in its entirety. Instead, it provides empirical access to the cubic term of the potential, and by extension, to the coefficient $\lambda$ when combined with an independent determination of $m_H$. Given the established relation $m_H^2 = 2\lambda \nu^2$, the self-coupling constant $\lambda$, and hence the entire shape of the scalar potential, can be inferred and cross-validated with experimental measurements of both $m_H$ and $\lambda$.

Experimentally, precise determination of $\lambda$ is of critical importance, as any deviation from the SM prediction would signal new physics. Such deviations could arise from extended scalar sectors, modified Higgs dynamics, or non-standard electroweak symmetry breaking mechanisms.

Mathematically, a di-Higgs measurement is aimed to determine $\kappa_\lambda$, a constant defined as:
\begin{gather}
    \kappa_\lambda = \frac{\lambda_{\text{ex}}}{\lambda_{\text{SM}}},
\end{gather}
where $\lambda_{\text{ex}}$ is the experimentally measured value of $\lambda$, and $\lambda_{\text{SM}}$ is the SM predicted value, which can be obtained through a prior measurement of the Higgs boson mass (and knowledge of $v$). If the SM is correct, $\kappa_\lambda$ is expected to be $1$. Any experimentally significant deviation from this value is a clear indicator of BSM physics.

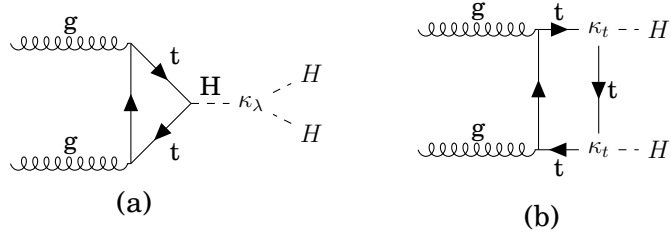
\begin{figure}[h]
  \centering

  \begin{minipage}{.2\textwidth}
    \centering
    \begin{tikzpicture}[scale=0.8, transform shape]
             \begin{feynman}
    \vertex (gluon1) at (0,0);
    \vertex (g1) at (2,0);
    \vertex (gluon2) at (0,-2);
    \vertex (g2) at (2,-2);
    \vertex (higgs) at (3,-1);
    \vertex (h) at (4,-1) {$\kappa_\lambda$};
    \vertex (b) at (5,-0.5) {$H$};
    \vertex (antib) at (5,-1.5) {$H$};

    \diagram* {
     (gluon1) -- [gluon, edge label=g] (g1),
     (gluon2) -- [gluon, edge label=g] (g2),
    (g2) -- [fermion] (g1),
    (higgs) -- [fermion, edge label=t] (g2),
    (g1) -- [fermion, edge label=t] (higgs),
    (higgs) -- [scalar, edge label=H] (h),
    (h) -- [scalar] (b),
    (antib) -- [scalar] (h),
    };
  \end{feynman}
    \end{tikzpicture}
    \subcaption{}\label{a}
  \end{minipage}\hspace{2cm}
  \begin{minipage}{.2\textwidth}
    \centering
   \begin{tikzpicture}[scale=0.8, transform shape]
             \begin{feynman}
    \vertex (gluon1) at (0,0);
    \vertex (g1) at (2,0);
    \vertex (gluon2) at (0,-2);
    \vertex (g2) at (2,-2);
    \vertex (higgs) at (3,-2){$\kappa_t$};
    \vertex (higgs2) at (3,0) {$\kappa_t$};
    \vertex (h2) at (4,-2) ;
    \vertex (b) at (4,0) {$H$};
    \vertex (antib) at (4,-2) {$H$};

    \diagram* {
     (gluon1) -- [gluon, edge label=g] (g1),
     (gluon2) -- [gluon, edge label=g] (g2),
    (g2) -- [fermion] (g1),
    (higgs) -- [fermion, edge label=t] (g2),
    (g1) -- [fermion, edge label=t] (higgs2),
    (higgs2) -- [fermion, edge label=t] (higgs),
    (higgs2) -- [scalar] (b),
    (higgs) -- [scalar] (antib),
    };
  \end{feynman}
    \end{tikzpicture}
    \subcaption{}\label{b}
  \end{minipage}%
  \small
  \caption{Di-Higgs Production Processes Through the Gluon-Gluon Fusion Mechanism}\label{di}
\end{figure}

Figure \ref{di} illustrates two Feynman diagrams contributing to di-Higgs production through the gluon-gluon fusion (ggF) mechanism. Although diagram \ref{b} is independent of the Higgs self-coupling, its amplitude interferes with that of diagram \ref{a}, whose triple-Higgs vertex is parametrized by $\kappa_\lambda$ and is thus sensitive to the self-coupling. As a result, the overall di-Higgs production rate is determined by the coherent sum of both contributions, including interference effects \cite{dihiggs}. By comparing the measured di-Higgs rates and distributions to theoretical predictions, one can extract the value of $\kappa_\lambda$.

Due to a high branching ratio, the $HH\rightarrow b\bar{b}b\bar{b}$ final state offers a particularly promising avenue for di-Higgs measurements. However, this channel is also subject to significant QCD multi-jet backgrounds, necessitating refined strategies for distinguishing the signal. One powerful approach is to exploit boosted topologies, where the Higgs boson has a high transverse momentum and its decay products are collimated. In such scenarios, jet substructure and advanced b-tagging techniques can be used to identify Higgs candidates more efficiently and suppress the large background. The trigger algorithm developed in this thesis, WOMBAT, is specifically designed to locate boosted $H\rightarrow b\bar{b}$ jets at the Level-1 Calorimeter Trigger, 
aiming for high efficiency in event selection and jet tagging. It is important to note that WOMBAT does not target di-Higgs production specifically, but rather enhances sensitivity to individual $H\rightarrow b\bar{b}$ decays. Meanwhile, other decay channels (e.g. $HH\rightarrow b\bar{b}\gamma\gamma$, $HH\rightarrow b\bar{b}l\nu l\nu$, and $HH\rightarrow b\bar{b}\tau\tau$)  can provide complementary measurements of the Higgs self-coupling, each offering different sensitivities and systematic uncertainties.

\subsection{Jet Clustering}

High-energy collisions that involve quarks and gluons in the final state often represent some of the most interesting processes in particle physics. Because these colored partons have extremely short lifetimes after a collision and cannot exist as free particles due to color confinement, they hadronize into jets, which are collimated streams of hadrons. As these hadrons propagate through the detector, they may undergo secondary decay or scattering processes (often called particle showers), further contributing to the observed final-state signature.

In online data analyses, jets are typically reconstructed via the PF algorithm, which uses calorimeter tower information. For more precise offline analyses, clustering algorithms such as the anti-k$_{\text{T}}$ technique are commonly employed \cite{antikt}. This algorithm iterates over all detected particles, identifies those that are nearest neighbors in phase space, and decides whether they should be merged, as summarized by
\begin{gather}
    d_{i,j}=\frac{\Delta_{i,j}^2}{R^2}min(p_{T,i}^{-2},p_{T,j}^{-2}),\;\begin{cases}
        \text{if: }d_{i,j}<p_{T,i}^{-2}\quad \text{then: combine,}\\
        \text{if: }d_{i,j}>p_{T,i}^{-2}\quad \text{then: stop,}\\
    \end{cases}
\end{gather}
where $\Delta_{i,j}^2 = (\eta_i-\eta_j)^2+(\phi_i-\phi_j)^2$, $i,j$ are particle indices, and $R$ is the jet radius parameter, commonly selected to be $0.4$ (known as AK4 jets) or $0.8$ (AK8 jets). This procedure ensures that jets are robustly identified and clustered in a manner that reflects the underlying parton kinematics while accounting for the spatial distribution of the final-state particles.

\subsubsection{Boosted Jets}

\begin{figure}[ht]
    \centering
    \includegraphics[scale=0.4]{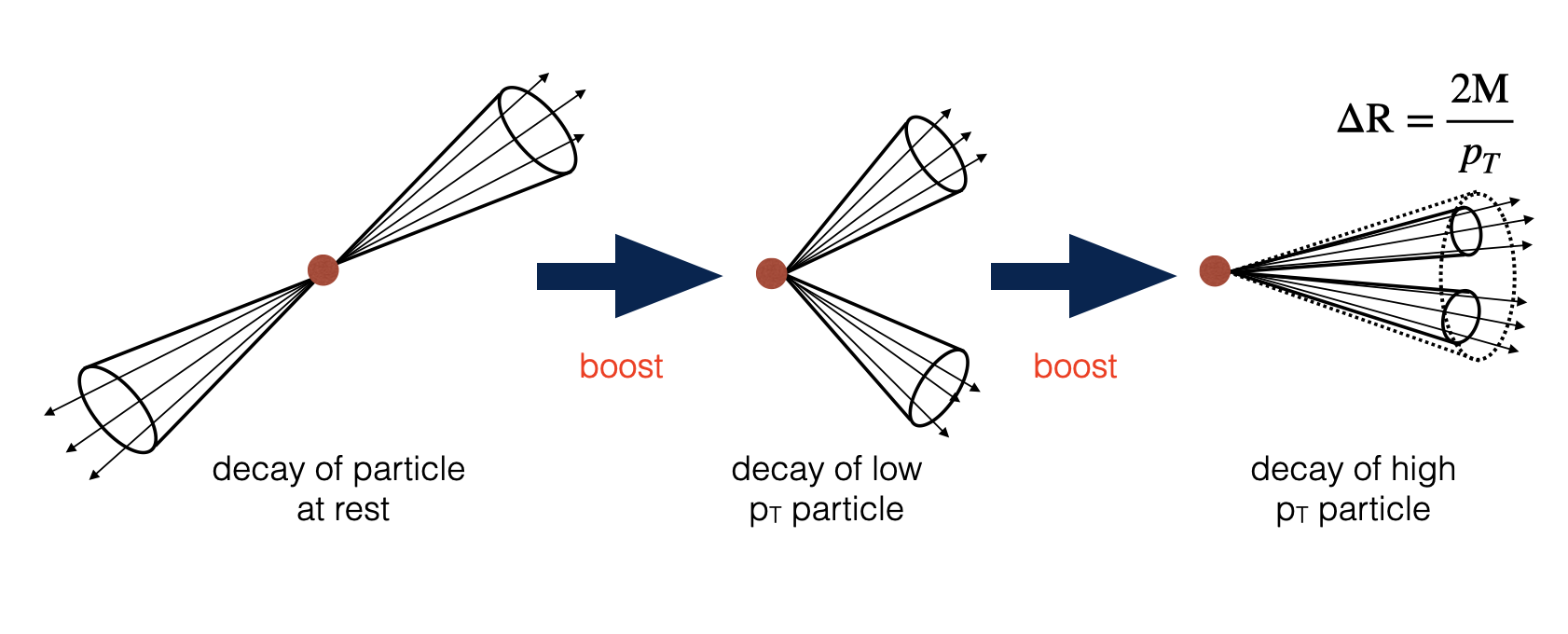}
    \caption{Visualization of Particle Decay Collimation With Increasing $p_T$}
    \label{fig:boost}
\end{figure}

Boosted jets comprise a class of high-transverse-momentum ($p_T$) events observed at the CMS detector. These jets are typically produced by the decay of massive particles (e.g., the Higgs boson or the top quark) that acquire significant Lorentz boosts. As a result, their decay products are highly collimated, often merging into a single jet-like structure, as depicted in Figure \ref{fig:boost}. This high degree of collimation makes it increasingly challenging to resolve individual decay products and accurately measure their kinematic properties.

One of the primary difficulties in detecting boosted jets at the L1T stage is the substantial background from QCD processes. Distinguishing signals of interest (e.g., those originating from Higgs bosons) from this background requires precise energy and momentum measurements, which can be difficult to achieve within the strict real-time constraints of the L1T system. Moreover, the granularity of the L1 readout is often insufficient to fully resolve jet substructure, including the presence of multiple subjets within a single, merged jet.

The upcoming Phase 2 upgrades to the CMS L1T are designed to address these challenges by increasing detector granularity and enhancing real-time processing capabilities. These improvements will facilitate more efficient identification of boosted jets and better discrimination of their internal substructure. Additionally, emerging ML techniques show significant promise for further enhancing the performance of L1-based jet identification \cite{mltrig}. Methods such as deep neural networks (DNNs) and boosted decision trees (BDTs) can be trained on extensive datasets (both simulated and real) to identify complex patterns indicative of boosted jets. By implementing these algorithms on FPGAs, it is possible to achieve low-latency and rapid, high-volume data processing, thereby maintaining sensitivity to rare processes such as Higgs boson decays into $b\bar{b}$ pairs while substantially reducing background contamination. These ML-based strategies have demonstrated enhanced signal purity and lower false-positive rates, thus improving the overall efficiency and physics reach of the trigger system.

\subsection{WOMBAT: Motivation}

The extremely collimated nature of boosted jets from high $p_T$ Higgs decays presents a unique challenge for real-time event selection at the L1T. While the Phase 2 upgrades to the CMS calorimeter and readout electronics will enhance spatial resolution and data processing capabilities, fully leveraging this improved hardware to identify boosted Higgs bosons in their dominant $b\bar{b}$ decay mode still requires specialized algorithms. To address this, ML-based trigger systems are being developed for L1T electronics, extending the physics reach of the current setup while serving as prototypes for Phase 2.

\begin{figure}[ht]
    \centering
    \includegraphics[width=\linewidth]{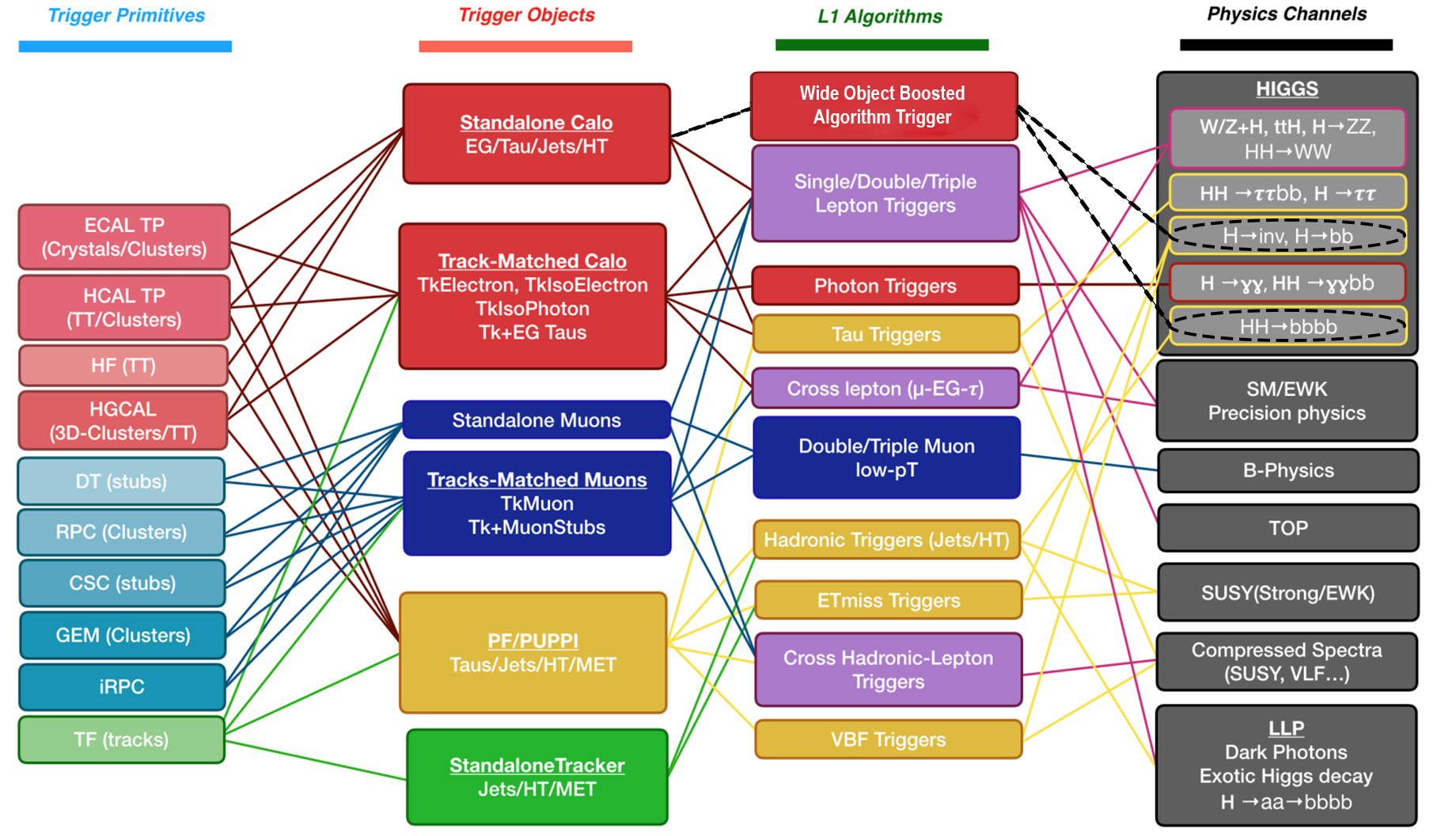}
    \caption{Phase-2 Physics Reach Based on L1T System \cite{phyreach} (modified to include WOMBAT)}
    \label{fig:phy}
   The first column shows links between TPs from different systems and the associated trigger objects (second column), which use L1 Algorithms (third column) to reach a specific physics goal shown (fourth column). Dashed black lines represent new links formed by the WOMBAT standalone calorimeter jet tagging algorithm.
\end{figure}

Efficient event tagging at the L1T is essential for enriching datasets with relevant processes, enabling more precise measurements. However, existing boosted $H\rightarrow b\bar{b}$ algorithms rely on deterministic energy sum calculations, which are computationally expensive for real-time execution.

This thesis presents WOMBAT (Wide Object ML Boosted Algorithm Trigger), an ML-based system designed for implementation in the Calorimeter Layer 1 (CaloLayer1) of the L1T for boosted $gg\rightarrow H\rightarrow b\bar{b}$ jet tagging and clustering. The primary motivation for this trigger is to improve rapid event selection for Yukawa coupling measurements, di-Higgs studies, as well as BSM searches discussed in Chapter II, Sections \ref{hot} and \ref{didi}. While the WOMBAT algorithm is not explicitly designed to isolate di-Higgs production, it enhances sensitivity to boosted $H\rightarrow b\bar{b}$ decays at the L1 Calorimeter Trigger. This improved tagging efficiency increases the likelihood of capturing rare signatures, including those from di-Higgs events and BSM processes that manifest through modified kinematics or excesses in the $b\bar{b}$ final state. In an online implementation, such enriched datasets would be passed to the HLT for further refinement and potential signal isolation.

WOMBAT takes raw data from the CTP7 cards in the calorimeter with minimal pre-processing. It identifies boosted $H\rightarrow b\bar{b}$ jet clusters in the TPs and outputs the center coordinates of the leading-order jets in indexed $\eta-\phi$ space. In this context, WOMBAT is considered a standalone calorimeter trigger, which refers to a system that makes decisions based solely on calorimeter information without relying on inputs from other subdetectors, such as the Silicon Tracker or muon systems.

Although WOMBAT was developed for the Phase-1 L1T system, it serves as a proof-of-concept demonstrating that ML-based jet tagging is feasible within current hardware constraints. While not designed for Phase-2, WOMBAT illustrates the potential of ML-based triggers, which are expected to perform even more effectively under the upgraded architecture, benefiting from increased bandwidth, finer granularity, and enhanced processing capabilities. In this context, WOMBAT-inspired systems could serve as standalone calorimeter triggers for identifying boosted $H\rightarrow b\bar{b}$ decays at the L1T. As shown in Figure \ref{fig:phy}, which presents the projected physics reach for Phase-2 triggers, WOMBAT-like systems form new, critical links between low-level TPs and high-level Higgs physics. Unlike traditional missing $E_T$ or VBF-based selections, WOMBAT targets the dominant ggH production mode using only calorimetric information, enhancing L1A efficiency for $b\bar{b}$ final states.

\newpage
\newpage\stepcounter{section}
\section*{Chapter {III}: Data Structure, Samples Used, and Data Pre-processing}
\addcontentsline{toc}{section}{Chapter {III}: Data Structure, Samples Used, and Data Pre-processing}
\setcounter{figure}{0}
\markboth{Chapter {III}: Data Structure, Samples Used, and Data Pre-processing}{}

\subsection{Datasets and Monte Carlo Samples}\label{montecarlo}

For ML training and evaluation purposes, simulated Monte Carlo (MC) events were generated using the MadGraph5\_aMC@NLO event generator \cite{mcnlo}, which models the hard scattering matrix element at next-to-leading order (NLO) in QCD. The event generation includes up to two additional partons in the matrix element calculation, allowing for the explicit simulation of final states with up to two extra jets originating from the hard process. This matrix element multiplicity is matched to the parton shower using the MLM merging scheme to avoid double-counting of emissions between the hard scattering and the subsequent parton showering \cite{mlm}. The inclusion of multi-parton matrix elements significantly improves the modeling of complex, high-multiplicity final states characteristic of boosted Higgs boson production, particularly in $H\rightarrow b\bar{b}$ decays where the decay products may be reconstructed as a single large-radius jet. To select events within the boosted regime, a transverse momentum threshold of $p_T > 250$ GeV is applied at the generator level to the Higgs boson. This requirement enhances the signal-to-background ratio in the dataset used to train and evaluate WOMBAT which triggers on boosted topologies.

The generated events are interfaced with Pythia8 \cite{pythia} for parton showering and hadronization, which simulate the evolution of colored partons into colorless hadrons, including soft and collinear QCD radiation, underlying event activity, and hadron decays. The matching between the matrix element and parton shower is carefully handled to preserve the accuracy of high-$p_T$ observables while maintaining infrared safety.\footnote{Infrared safety refers to the requirement that physical observables remain insensitive to the emission of soft gluons or collinear splitting of partons, ensuring theoretical predictions are well-defined and stable. Proper matrix element and parton shower matching preserves this property by avoiding divergences and double-counting in soft/collinear regions of phase space.} Final-state hadrons are processed through a dedicated CMS trigger simulation framework, based on Geant4 \cite{geant}, which emulates the detector response relevant for L1 TPs, including calorimeter digitization, trigger tower granularity, and electronic response effects.

To evaluate the trigger rate, Zero Bias (ZB) data was utilized. This dataset consists of events recorded solely based on the occurrence of a bunch crossing, without any additional physics-based trigger conditions. As its name suggests, ZB data is inherently unbiased, making it a representative snapshot of the full range of detector activity following a collision, including background noise, low-energy interactions, and pileup effects.

This makes ZB data especially valuable for assessing the performance and expected rates of trigger algorithms, such as WOMBAT, under realistic LHC running conditions. Since only a small fraction of all collisions produce events of physical interest, passing ZB events through the WOMBAT trigger provides insight into how the algorithm behaves in the presence of high event rates, noise, and pileup. This provides a metric for how frequently the trigger issues an L1A decision under realistic conditions. Since the data acquisition system cannot record every event due to bandwidth and storage limitations, the goal of any trigger system is to maintain a low acceptance rate while maximizing efficiency for selecting physics-rich events.

The ZB data used was taken during Run 3 of the LHC, in a period of stable beam and detector conditions known as Era C of 2023. In particular, the sample is \texttt{ZeroBias/} \texttt{Run2023C-PromptReco-v1/MINIAOD}, which has an integrated luminosity of $0.64$ fb$^{-1}$, as calculated through the Brilcalc framework \cite{bricalc}.

Passing ZB data through the trigger algorithm was carried out using the CMS Remote Analysis Builder (CRAB) framework. CRAB provides a streamlined interface for submitting and managing large-scale distributed computing jobs across the CMS grid infrastructure. It enables efficient processing of extensive datasets like ZB by handling job distribution, resource allocation, and output collection, all while ensuring consistency and scalability across the analysis workflow.

CRAB jobs process ZB and MC samples into n-tuples, flat ROOT-based data structures typically stored in \texttt{TTree} format, which encode per-event physics objects (e.g., jets, muons, trigger primitives) as branches of C++-type arrays or scalar variables. For the ZB dataset \texttt{ZeroBias/Run2023C-PromptReco-v1/MINIAOD}, custom CMSSW analyzers traverse the MINIAOD event content to extract quantities relevant for L1T emulation and ML inference. These include calorimeter TPs, generator-level information (such as $\eta$, $\phi$, and $p_T$), jet substructure variables, and full collections of physics objects such as AK8 jets, subjets, and tau seeds stored as \texttt{TLorentzVector} arrays. These branches are serialized using ROOT's high-throughput Input/Output (I/O) backend and compressed to optimize disk usage and access speed. The resulting n-tuples contain the subset of events that pass the WOMBAT trigger emulation, as well as those that pass the Single Jet 180 algorithm, detailed in Chapter IV, Section~7.



\begin{figure}[ht]
    \centering
    \includegraphics[width=0.7\linewidth]{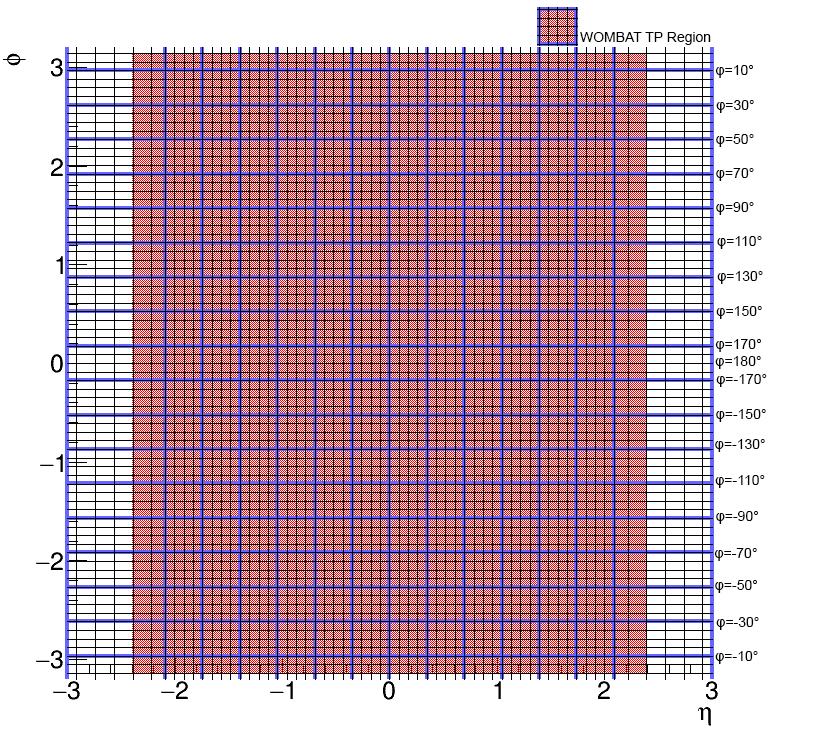}
    \caption{Phase-1 CMS Calorimeter Trigger Tower Segmentation}
    \label{fig:ecal}
    {\small
    Black grids represent TTs which cover approximately $(0.087)\times (0.087\text{ radians})$ in $\eta\times \phi$ space. Blue grids comprise of $4\times 4$ TTs, which are the fundamental units of each CaloLayer1 TP region. All trigger regions marked in red are used as input to WOMBAT. The $|\eta|>2.4$ region is excluded. The TP regions amount to $14\times 18$ in $\eta\times \phi$.
    }
\end{figure}

\subsection{Trigger Primitives Input}\label{tpinput}

As a standalone calorimeter trigger, WOMBAT fully relies on TP information from the ECAL and HCAL barrel and endcap detectors. These calorimeters provide coverage within a pseudorapidity range of $|\eta|<3$ and encompass the full azimuthal angle, $0\leq \phi <2\pi$. Due to the geometry of the detector, the barrel (associated with $\phi$) and endcap (associated with $\eta$) calorimeter TPs require different analysis approaches. For a geometric view of the detector refer to Appendix C.

The CMS calorimeter segmentation is illustrated in Figure \ref{fig:ecal}, where the red-shaded region denotes the $14\times18$ input grid in $\eta\times\phi$ used by WOMBAT. Each blue-outlined CaloLayer1 TP region comprises a $4\times4$ array of TTs, shown in black. Due to L1T computational constraints, WOMBAT's ML models operate at the coarser CaloLayer1 TP granularity rather than full TT resolution. Consequently, model predictions span a $14\times18$ index space in $\eta\times\phi$.

To recover TT-level precision, WOMBAT manually identifies the maximum $E_T$ TT within each selected CaloLayer 1 TP region, assigning $H\rightarrow b\bar{b}$ jet locations accordingly. This dimensionality reduction enables a more tractable ML architecture with $252$ input features, significantly fewer than the full TT set.\footnote{To illustrate the scaling challenge, WOMBAT Master Model (W-MM) selects $3$ jet centers from $252$ regions, yielding $252^3 = 16,003,008$ possible outputs under independent sampling with replacement ($63,504$ for the WOMBAT Apprentice Model, W-AM). At TT granularity, assuming each CaloLayer1 TP region contains $4\times 4$ TTs, the input space expands to $252\times 16 = 4032$ points. This results in $4032^3=65,548,320,768$ possible outputs for W-MM and $4032^2 = 16,257,024$ for W-AM-a dramatic increase in model complexity. Such high-resolution modeling exceeds the resource capacity of current L1T FPGAs, making it impractical due to prohibitive computational and latency constraints.} Each input feature corresponds to the summed $E_T$ within a TP region, retaining key kinematic information while supporting low-latency inference.

\begin{figure}[ht]
  \centering
  \begin{minipage}[b]{0.49\textwidth}
    \centering
    \includegraphics[width=\linewidth]{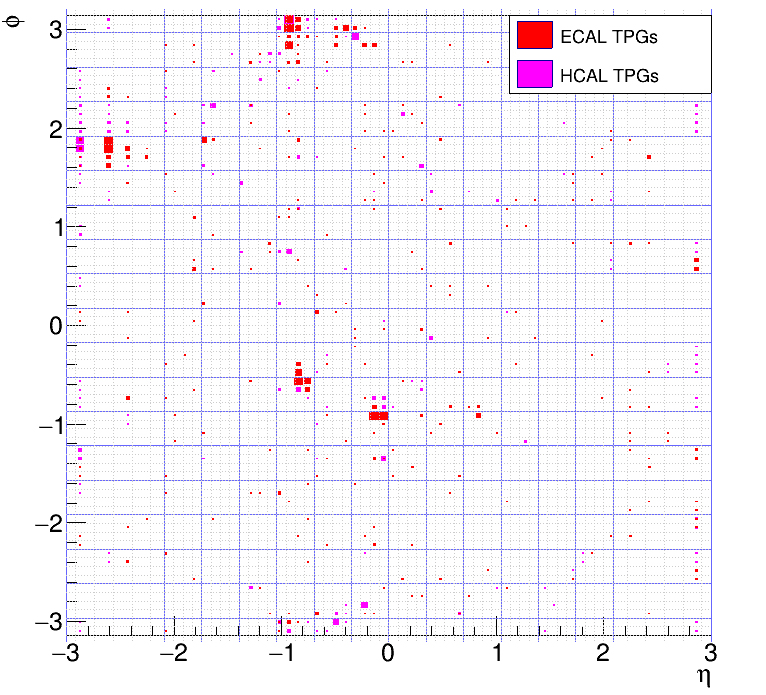} 
    \subcaption{Raw TP Display}
    \label{fig:image11}
  \end{minipage}
  \hfill
  \begin{minipage}[b]{0.49\textwidth}
    \centering
    \includegraphics[width=\linewidth]{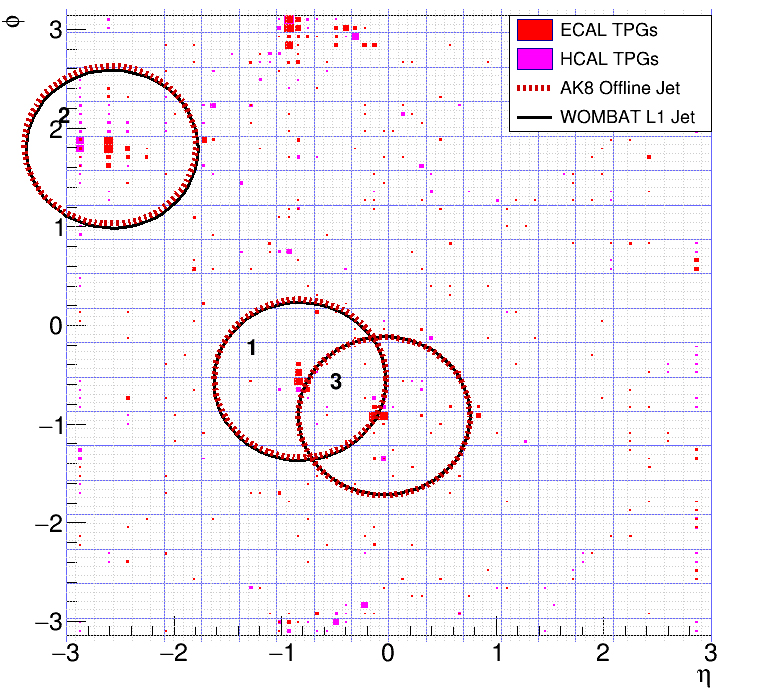} 
    \subcaption{Processed TP Display}
    \label{fig:image22}
  \end{minipage}

    \caption{Raw and Processed Calorimeter TP Display (Event~3468)}
  \label{fig:joint12}
  {\small
  Figure \ref{fig:image11} is a display of raw calorimeter values with associated TTs. In Figure \ref{fig:image22}, this event was processed through offline reconstruction (AK8 Jets) and the WOMBAT trigger system to locate $H\rightarrow b\bar{b}$ jet centers.  
  }
\end{figure}

An example TP input can be seen in Figure \ref{fig:joint12}, which contains boosted $H\rightarrow b\bar{b}$ jets with $p_T$ in the range of $150.8$ GeV (jet $3$) to $220.8$ GeV (jet $1$). This event is extracted from the MC dataset used for an efficiency evaluation of the WOMBAT trigger system. While the legend uses labels such as HCAL and ECAL TPGs instead of TPs, this refers to the same underlying data. The term Trigger Primitive Generator (TPG) denotes the hardware or firmware responsible for producing TPs from raw calorimeter signals. As a result, “TPGs” is often used interchangeably with “TPs” to indicate the output of this processing step. In this context, the labels represent the four-vector quantities produced by the TPGs. 

Visually, due to the high level of activity, this TP grid contains physics signatures of potential interest, with multiple $H\rightarrow b\bar{b}$ jet candidates. Ideally, a trigger system should be able to declare this event an L1A by resolving jet substructure and locating relevant $H\rightarrow b\bar{b}$ decay products. 

Figure \ref{fig:image22} illustrates WOMBAT's jet-tagging performance, benchmarked against an offline AK8 reconstruction algorithm, which utilizes high-granularity inputs from multiple detector subsystems. In this event, both algorithms identified the same $H\rightarrow b\bar{b}$ candidates. While a detailed discussion of the algorithms and analysis is provided in Chapters IV and V, it is worth noting that this event was deliberately selected to highlight the role of jet substructure and TP patterns in $b$-tagging. Notably, both algorithms rejected a mid-energy cluster near $(\eta,\phi)\approx (-1,3)$ as a boosted Higgs candidate, likely due to latent features in the TP data.

As shown in Figure \ref{fig:ecal}, WOMBAT's ML models restrict input TPs to $|\eta| < 2.4$ due to non-uniform sampling in $\eta$. However, predictions can still map to edge TTs, enabling jet tagging across the full TP grid. The outputs of WOMBAT are indexed by CaloLayer1 trigger regions rather than individual TTs, and precise jet locations are resolved during the conversion from index space to real coordinates. Consequently, predictions at $\eta$ indices $0$ or $13$ correspond to CaloLayer1 regions that encompass $|\eta|\geq 2.4$, allowing tagging in those outer regions despite input limitations.

\subsection{WOMBAT Data Processing and Label Generation}

WOMBAT accepts lower-granularity input from the HCAL and ECAL TPs in a fixed-precision integer format. Each calorimeter region encodes the transverse energy as a $10$-bit unsigned integer, quantized uniformly over the interval $[0,1023]$. Each increment value corresponds to one least significant bit (LSB), representing the smallest resolvable energy increment in hardware. This ensures that the algorithm meets strict latency constraints while maintaining a high degree of accuracy. When deployed online, WOMBAT is designed to require minimal input pre-processing, mainly related to data formatting (see Chapter IV, Section 5). However, for model training, labels were manually computed, and input processing was required to ensure compatibility with the model's discretized output.

Although the MC samples contain extensive event information, only a small subset is used for model training and evaluation. The pre-processing pipeline begins with a filtering algorithm that selects only boosted $H\rightarrow b\bar{b}$ events from a ROOT file, which are subsequently converted to an HDF5 format. This filtering is performed by referencing the generator-level particle identifier (genID), ensuring that only Higgs boson events are retained for training. Once the events of interest have been isolated, the calorimeter region information (c-region) is extracted and reshaped into an $14\times 18$ grid, corresponding to the segmentation of the CMS calorimeter. These TPs encompass a large segment of pseudorapidity space ($|\eta|\leq 2.4$) and the entire span of $\phi$.

In addition to the c-regions data, other kinematic features-such as the $p_T$ of the Higgs boson, as well as its generator-level pseudorapidity (genEta) and azimuthal angle (genPhi) are also extracted. To ensure compatibility with the c-regions grid structure, genEta and genPhi undergo a transformation from real-space coordinates to an indexed space representation. This transformation is directly tied to the CaloLayer1 TP regions, as each covers a specific portion of the calorimeter, and the mapping of genEta and genPhi onto the indexed space aligns with this segmentation. The transformed coordinates, referred to as indexed Eta (iEta) and indexed Phi (iPhi), span fixed integer ranges from 0 to 13 for iEta and from 0 to 17 for iPhi. 

Although the extracted iEta and iPhi values were not used in WOMBAT's architecture, they were essential for developing the label-generating algorithm. Its purpose is to identify the highest energy leading order (LO) clusters (corresponding to high-$p_T$ jets) in each c-region. The algorithm uses a maximum filter operation \cite{scripy}, which applies a $3\times 3$ sliding window to detect local maxima by comparing each element to its neighbors. A connected-component labeling step groups contiguous maxima, defining distinct energy clusters. For each extremum, the $E_T$ and corresponding indexed coordinates (iPhi, iEta) are extracted. A thresholding step filters out low-energy noise, and the remaining extrema are ranked by $E_T$. Depending on the model, the top three (or two, for the WOMBAT Apprentice model) peaks are selected and their coordinates are used as training and evaluation labels for the model.

After WOMBAT generates predictions in indexed space, these are converted to real $\eta$-$\phi$ coordinates. Each selected CaloLayer1 region is scanned to identify the TT with the highest energy deposit, which is then designated as the predicted jet center and converted into physical coordinates. Although WOMBAT operates on TP region-level inputs without TT-level granularity, a lightweight post-processing step resolves jet positions at TT-level precision.

\newpage
\newpage

\stepcounter{section}
\section*{Chapter {IV}: WOMBAT Architecture, Performance, and FPGA Implementation}
\addcontentsline{toc}{section}{Chapter {IV}: WOMBAT Architecture, Performance, and FPGA Implementation}
\setcounter{figure}{0}
\markboth{Chapter {IV}: WOMBAT Architecture, Performance, and FPGA Implementation}{}

\subsection{Deep Neural Networks: Background}

Ever since the initial proposal in 1943, Deep Neural Networks (DNNs) have been recognized for their ability to learn representative features from complex high-dimensional data, making them well-suited for tasks such as real-time triggering and classification in proton-proton collision events \cite{dnn}. In particular, Convolutional Neural Networks (CNNs) have become a cornerstone for processing grid-structured data. The standard mathematical definition of convolution is \cite{cnn}:
\begin{equation}
    s(t) = (x\cdot w)(t)=\int x(a)w(t-a)\;da,
\end{equation}
which is an operation describing how the signal input function, $x(a)$, is weighted with the signal $w(t)$, which can be thought of as a filter applied to $x(t)$. For two-dimensional data, such as TPs, the convolution can be represented as:
\begin{equation}
    S(i,j)=(I\cdot K)(i,j)=\sum_m\sum_nI(i\cdot s +m, j\cdot s+n)K(m,n),
\end{equation}
where $I$ denotes the input image, $K$ represents a filter, $(i,j)$ are indices, and $s$ is the stride parameter.

While CNNs excel at spatial feature extraction, this thesis introduces an innovative hybrid approach that integrates an autoencoder (AE) within the CNN architecture. AEs are unsupervised neural networks designed to learn compressed representations of input data by encoding it into a lower-dimensional latent space and then reconstructing it. This two-step process, involving an encoder and a decoder, enables AEs to remove noise, extract meaningful latent features, and facilitate efficient data compression. Latent features, also known as latent variables or hidden representations, are the underlying factors inferred by the model during training.

The WOMBAT architecture incorporates the proposed Embedded Deterministic Autoencoder (EDA) to compress the $\phi$ dimension while preserving the granularity of $\eta$.\footnote{During early development, the CNN achieved a maximum R$^2$ of 0.89, while the proposed EDA model reached 0.98 under identical training conditions. Since higher R$^2$ indicates improved accuracy, performance was further validated on unseen data to rule out overfitting.} This design choice is motivated by the consistent resolution and cyclic nature of $\phi$,  which enables the EDA to extract intricate features that conventional CNNs might overlook.  In contrast, maintaining higher-resolution $\eta$ information ensures effective local feature extraction across network layers, which is critical for a trigger system. Although downsampling both dimensions would improve computational efficiency, it significantly reduces the model's ability to resolve jet substructure effectively.

Due to the high complexity of this algorithm, WOMBAT was developed as a knowledge diffusion framework in which a large EDA-based model, referred to as the WOMBAT Master Model (W-MM), serves as a teacher model. The W-MM generates labels and transfers structured knowledge to a simplified CNN model, referred to as the WOMBAT Apprentice Model (W-AM), enabling it to learn essential patterns and generalize effectively while maintaining computational efficiency. Both models are evaluated using the same criteria and software, however, only the W-AM was deployed in firmware due to latency and resource constraints.

To evaluate WOMBAT's performance and establish a baseline for comparison with ML-based approaches, a fully deterministic, rule-based algorithm was implemented on FPGA hardware. The Jet Event Deterministic Identifier (JEDI), originally referred to as "Bit Pattern", is a manually engineered pipeline that mirrors the trigger-level reasoning including fixed thresholding, lookup table corrections, and spatial pattern matching. The input is identical to WOMBAT, a $14\times 18$ grid of CaloLayer1 TP regions, each quantized to $10$ bits. By computing $3\times 3$ energy sums, JEDI identifies localized high-energy deposits indicative of jet activity. These sums are filtered through a spatial pattern matching logic that is pre-defined to capture signatures of boosted $H\rightarrow b\bar{b}$ decays. Unlike the WOMBAT trigger system, which learns complex spatial and energetic correlations directly from the TP data with minimal manual input, JEDI relies entirely on predefined logic for jet tagging. This makes the comparison between these two algorithms especially compelling, as it highlights the fundamental contrast between data-driven learning and rule-based L1T classification.

\subsection{WOMBAT Models Architecture}

The high-level structure of WOMBAT can be summarized as follows:

\begin{itemize}
\item \textbf{WOMBAT Master Model} (W-MM): A large CNN model incorporating an EDA architecture, designed to maximize performance without significant constraints on resource usage or latency. It outputs the location of either two or three jets.
\item \textbf{WOMBAT Apprentice Model} (W-AM): An 8-bit quantized CNN model built using the quantized Keras (QKeras) library \cite{qkeras}. It features a custom threshold layer and is designed to output the location of exactly two jets. Optimized for FPGA implementation, it adheres to strict latency and resource usage constraints.
\item \textbf{WOMBAT Apprentice Skeleton Model} (W-ASM): A streamlined variant of W-AM, lacking custom layers and featuring a single output in the form of a dense layer. Used solely for HLS4ML \cite{hls4ml} code generation, whereas the custom layers and weights of W-AM are manually implemented in firmware.
\end{itemize}

A schematic overview of the models can be seen in Appendix D.

\subsubsection{WOMBAT Master Model Architecture}\label{structure}

W-MM is implemented using TensorFlow's Keras API \cite{tens} and incorporates a combination of convolutional layers, batch normalization, and activation functions to extract and encode spatial features. A key innovation in the architecture is the EDA, which compresses the cyclic $\phi$  dimension while maintaining high-resolution information in $\eta$. While the W-MM is computationally intensive, it serves as a teacher model in a knowledge distillation framework, training the more efficient W-AM for real-time deployment in firmware-constrained environments.

\subsubsection{Embedded Deterministic Autoencoder}\label{eda}

WOMBAT's EDA architecture can be represented by:
\begin{gather}
    z = f_{\Omega}(x),\quad
    \hat{x} = g_{\Phi}(z),
\end{gather}
where $f_\Omega$ is the encoder function parametrized by the set $\Omega$, $x$ is the input, $g_\Phi$ is the decoder function parametrized by the set $\Phi$, and $\hat{x}$ is the output.

\subsubsubsection{Encoder Function and Custom Layers}

More explicitly, given the input $x\in \mathbb{R}^{18\times14\times1}$, the encoder function $f_\Omega$ is composed of three main stages:

\begin{itemize}
    \item $1.$ Pre-processing:
\end{itemize}

During pre-processing, a value of $30$ GeV is subtracted from each TP region. This is performed through a modified ReLU operation which can be written as:
\begin{gather}
    x_{pre} = \max \{x(i,j)-30,0\}.
\end{gather}

It is relevant to note that this operation is encoded in the WOMBAT's structure and does not need to be performed externally.

\begin{itemize}
    \item $2.$ Convolutional Feature Extraction:
\end{itemize}

The preprocessed input is then passed through a series of custom encoder blocks defined as:
\begin{gather}
    E(y;f,k,s)= \text{ReLU}\Bigg(\text{BN}\Big(\text{Conv2D}(C(y);f,k,s)\Big)\Bigg),
\end{gather}
where $C(y)$ is the custom circular padding function with input $y$, $f$ is the number of filters, $k$ is the kernel size (set to $(3\times3)$), and $s$ is the stride (set to $(1,1)$). The layers, BatchNormalization (BN), 2D Convolution (Conv2D), and ReLU are also represented.

Formally, $C(y)$ circularly pads the $\phi$ dimension, while adding constant (zero) padding to $\eta$. Given the input $y\in \mathbb{R}^{\phi\times\eta\times 1}$, where $1$ is the number of channels used by WOMBAT, $C(y)$ for a single sample can be represented as:
\begin{gather}
    C_\phi (y(i,j,1) )= \begin{cases}
        y(\phi-p+i,j,1),\quad 0\leq i <p,\\
        y(i-p,j,1),\quad p\leq i <\phi+p,\\
        y(i-\phi-p,j,1),\quad \phi+p\leq i <\phi+2p,
    \end{cases}
\end{gather}
\begin{gather}
     C_\eta (y(i,j,1)) = \begin{cases}
        0,\quad 0\leq j <q,\\
        y(i,j-q,1),\quad q \leq j <\eta+q,\\
        0,\quad \eta+q\leq j <\eta+2q,
    \end{cases}
\end{gather}
where $p$ is the number of rows that are circularly padded along $\phi$, $(i,j)$ are the row and column indices in the padded output, and $q$ is the number of columns added as zeroes to the $\eta$ dimension. By default, WOMBAT uses only $1$ channel with $\phi=18$ and $\eta=14$. To minimize resource usage, $p$ and $q$ are set to $1$, however, the dynamic implementation of this layer allows for any choice of parameters.

The model features three encoder blocks, where the first two are followed by a MaxPooling layer with a pooling window of $(2,1)$. This operation performs an anisotropic downsampling of the feature map and thus reduces the spatial dimension of the input. With each pooling operation, the effective receptive field of the network increases. This allows deeper layers to capture the broader context and complex jet patterns, making it easier to identify features in the TPs originating from boosted $H\rightarrow b\bar{b}$ events.

Defining $p_n$ such that:
\begin{gather}
   p_n(i,j) =\max\{y_n(2i,j,1),y_n(2i+1,j)\} \;\Rightarrow\;p_n = \text{MaxPool}_{(2,1)}(y_n)
\end{gather}
for $n$ being the index of the layer.

Given these definitions, WOMBAT's EDA encodes the input as:
\begin{gather}
    y_1 = E\Big(x_{pre};32,(3,3),(1,1)\Big),\\
    p_1 = \text{MaxPool}_{(2,1)}\Big(y_1\Big),\\
    y_2 = E\Big(p_1;64,(3,3),(1,1)\Big),\\
    p_2 = \text{MaxPool}_{(2,1)}\Big(y_2\Big),\\
    y_3 = E\Big(p_2;128,(3,3),(1,1)\Big).
\end{gather}

\begin{itemize}
    \item $3.$ Latent Representation:
\end{itemize}

Following the pooling and convolution operations, the output $y_3$ has dimensions of $\mathbb{R}^{4\times 14\times 128}$. This is then flattened and mapped to a latent vector $z\in \mathbb{R}^{128}$ using a dense layer with a ReLU activation function:
\begin{gather}
   f_\Omega = z =  \text{ReLU}\Big(W_f\cdot \text{Flatten}(y_3)+ b_f\Big),
\end{gather}
for a weight matrix $W_f$ and an associated bias term $b_f$.

\subsubsubsection{Decoder Function}

The decoder function $g_\Phi$ maps the latent vector $z$ back to the reconstruction $\hat{x}$ in the original space. The pipeline can be outlined as follows:

\begin{itemize}
    \item $1.$ Dense Layer Projection and Reshaping
\end{itemize}

Initially, $z$ is reshaped into a tensor of dimensions $\mathbb{R}^{4\times 14\times 128}$ through the function:
\begin{gather}
    h = \text{Reshape}\Big(\text{ReLU}(W_g\cdot z+b_g)\Big),
\end{gather}
for a weight matrix $W_g$ and bias term $b_g$.

\begin{itemize}
    \item $2.$ Up-sampling and Reconstruction
\end{itemize}

Following the reshaping operation, reconstruction is performed using a decoder block that mirrors the encoder. This can be defined as:
\begin{gather}
    D(y;f,k,s) = \text{ReLU}\Bigg(\text{BN}\Big(\text{Conv2D}(C(y);f,k,s)\Big)\Bigg).
\end{gather}

The reconstruction process uses up-sampling, which is an operation that increases the spatial dimension of the input, reversing the MaxPooling$_{(2,1)}$ performed by the encoder. Letting $X$ be an input feature map with dimensions $\mathbb{R}^{H\times W}$ and $U$ be the up-sampled output with dimensions $\mathbb{R}^{(2H)\times W}$, for each output pixel $U(i,j)$ the up-sampling can be written as:
\begin{gather}
    U(i,j) = \text{UpSampling}_{(2,1)}(X) = X\Big(\Big[\frac{i}{2}\Big],j\Big),
\end{gather}
where $\Big[\frac{i}{2} \Big]$ implies floor division.

Using this definition, the reconstruction pipeline is:
\begin{gather}
    u_1 = \text{UpSampling}_{(2,1)}\Big(h\Big),\\
    d_1 = D\Big(u_1;128,(3,3),(1,1)\Big),\\
    u_2 = \text{UpSampling}_{(2,1)}\Big(d_1 \Big),\\
    d_2 = D\Big(u_2; 64,(3,3),(1,1)\Big).
\end{gather}

\begin{itemize}
    \item $3.$ Padding and Convolution
\end{itemize}

By this stage, the indexed $\phi$ and $\eta$ jet center predictions are already extracted. To finalize the reconstruction, zero padding is added to the $\phi$ dimension in order to match the expected output size. Given that this does not impact WOMBAT's predictions, it has only a structural purpose. Mathematically, $d_3$ can be defined as $d_3 = \text{Pad}(d_2)$, which gives a compact expression for $g_\Phi$:
\begin{gather}
    g_\Phi = \hat{x} = \sigma\Bigg(\text{Conv2D} \Big(d_3; 1,(3,3), (1,1)\Big) \Bigg),
\end{gather}
where $\sigma$ stands for the sigmoid activation function, defined as:
\begin{gather}
    \sigma(x) = \frac{1}{1+e^{-x}}.
\end{gather}

\subsubsection{Global CNN Structure}\label{globalcnn}

The global CNN structure integrates the EDA into a multi-task framework that simultaneously reconstructs the input and predicts the jet coordinates. In this design, the latent vector extracted by the encoder, $z$, serves as the common feature representation for the two distinct branches: one dedicated to TP reconstruction and another to coordinate regression. Although the reconstructed output is not currently used, it serves as an auxiliary task that guides the learning of robust latent features. 

In the case of W-MM, the latent representation $z\in\mathbb{R}^{128}$ is used to compute a 7-dimensional output, $c\in\mathbb{R}^7$ through the function:
\begin{gather}
    c = \sigma\Big( W_3 \text{ReLU}(W_2z+b_2)+b_3 \Big),
\end{gather}
where $W_2\in\mathbb{R}^{64\times 128}$, $b_2\in\mathbb{R}^{64}$, $W_3\in R^{7\times 64}$, and $b_3\in \mathbb{R}^7$ are trainable parameters. The sigmoid function provides normalization, as it ensures that each element of $c$ lies within the interval $[0,1]$. 

To extract the final outputs, each entry of $c=[c_0,...,c_6]^T$ is mapped to a physical quantity via a custom Lambda layer \cite{lambda} as follows:
\begin{itemize}
    \item Jet 1: $\Big(\phi_1=c_0\times 17, \eta_1 = c_1\times 13\Big)$,
    \item Jet 2: $\Big(\phi_2=c_2\times 17, \eta_2 = c_3\times 13\Big)$,
    \item \texttt{is\_there\_third} - A variable that is $1$ if the TP contains a third jet whose $3\times 3$ region is has $p_T>100$, and $0$ otherwise: $c_4$,
    \item Jet 3: $\Big(\phi_3=c_5\times 17, \eta_3 = c_6\times 13\Big)$.
\end{itemize}

In parallel, the decoder branch reconstructs the input from the same latent vector $z$. By jointly training the coordinate regression and reconstruction tasks, W-MM uses a composite loss function. Although the model is pre-configured to prioritize minimizing the loss in $\phi$ and $\eta$, it is still able to learn the latent features extracted through the EDA.

\subsubsection{WOMBAT Apprentice Model Architecture}\label{apprentc}

The W-AM is built using the QKeras library and incorporates a custom threshold layer. To minimize resource usage, all weights and biases are quantized to $8$ bits. The input, $x\in\mathbb{R}^{18\times14\times 1}$, is equivalent to that of W-MM, and passes through the following ML pipeline:

\begin{gather}
    y_1 = \text{QConv2D}\Big(x;4,(5,5),(1,1)\Big),\\
    y_{pre} = \max\{y_1(i,j)-30,0\},\\
    y_2 = \text{BN}\Big( y_{pre} \Big),\\
    y_3 = \text{QConv2D}\Big(y_2;4,(3,3),(1,1)\Big),\\
    y_4 = \text{BN}\Big( y_3 \Big),\\
    y_5 = \text{QActivation(ReLU)}\Big( y_4 \Big),\\
    y_6 = \text{AvgPool}_{(3,3)}\Big(y_5\Big),\\
    y_7 = \text{BN}\Big( y_6\Big),\\
    z = \text{Flatten}\Big( W_z\cdot y_7+b_z \Big),\\
    z_1 = \text{QActivation(ReLU)}\Big(  z\Big).
\end{gather}
In the above notation, the prefix Q indicates that the layer is quantized and part of the QKeras library. As previously, the associated weight matrix and bias vectors are labeled as $W$ and $b$.

Following this pipeline, the network produces a latent vector $z_2\in\mathbb{R}^33$. This 33-dimensional latent representation is then transformed into the physical quantities $\phi_1$, $\eta_1$, $\phi_2$, and $\eta_2$ via four separate dense (fully connected) layers. Each dense layer performs an affine transformation with its own trainable weight matrix and bias vector, such that:
\begin{gather}
    \phi_1 = \Big(  W_{\phi_1}\cdot z_2+b_{\phi_1} \Big),\\
    \eta_1 = \Big( W_{\eta_1}\cdot z_2+b_{\eta_1} \Big),\\
    \phi_2 = \Big( W_{\phi_2}\cdot z_2+b_{\phi_2} \Big),\\
    \eta_2 = \Big(W_{\eta_2}\cdot z_2+b_{\eta_2} \Big),
\end{gather}
where $W_{\phi_1,\eta_1,\phi_2,\eta_2}\in \mathbb{R}^{1\times 33}$, and  $b_{\phi_1,\eta_1,\phi_2,\eta_2}\in\mathbb{R}^1$.

As shown above, the output of W-AM is fixed at two jet centers, whereas W-MM predicts up to three. In the FPGA implementation, discussed in Chapter IV, Section 5, the expected output is a single dense layer. Since this does not reduce the number of trainable parameters, it does not amount to a significant performance difference. Mathematically, the W-ASM output can be shown as:
\begin{gather}
    \begin{bmatrix}
        \phi_1\\\eta_1\\\phi_2\\\eta_2 \end{bmatrix} = \Big( W_{z_2}\cdot z_2 +  b_{z_2}\Big) = \begin{bmatrix}
            W_{\phi_1}\\
            W_{\eta_1}\\
            W_{\phi_2}\\
            W_{\eta_2}
        \end{bmatrix}\cdot z_2 + \begin{bmatrix}
            b_{\phi_1}\\
            b_{\eta_1}\\
            b_{\phi_2}\\
            b_{\eta_2}
        \end{bmatrix},
\end{gather}
where $W_{z_2}\in \mathbb{R}^{4\times 33}$ and $b_{z_2}\in \mathbb{R}^4$.

Since matrix multiplication is linear, partitioning the transformation into four parts or combining them into a single operation does not change the underlying function that maps $z_2$ to the outputs. In this sense the W-AM and W-ASM models are equivalent, however, the split output offers more compatibility with the analysis software used. This is a choice of representation which does not affect the learning capabilities of either model.

The main difference between W-AM and W-ASM is in the definition of the threshold layer, $y_{pre}$:
\begin{gather}
   \text{W-AM:}\quad y_{pre} = \max\{y_1(i,j)-30,0\},\\
   \text{W-ASM}\quad y_{pre} = \max\{y_1(i,j),0\} = \text{QActivation(ReLU)}\Big( y_1\Big).
\end{gather}

Evaluations demonstrate that W-AM is more effective at noise filtering and capturing latent features in the data. While the W-ASM is initially implemented in FPGAs, the weights, biases, and custom activation layers from a pre-trained W-AM are manually added later. Consequently, the FPGA implementation is fully that of W-AM, with W-ASM serving as an intermediate stage of development.

During the design process of W-AM, three options were considered for the placement of the $p_T$ threshold layer:
\begin{itemize}
    \item Before $y_1$, as the first layer of the model.
    \item Following $y_1$, as the first activation function following convolution.
    \item Not at all due to latency considerations.
\end{itemize}

While the first option had the potential to achieve the highest performance, the increased model complexity led to a significant rise in execution latency.\footnote{There needs to be an activation function following convolution, so adding the threshold layer before $y_1$ increases the model size.} Although this placement logically aligns with the role of an activation layer in preprocessing internal inputs, it introduced an additional $12.5$ ns delay (which translates to 2 additional clock cycles) in FPGA execution, making it less favorable for real-time applications.

\begin{figure}[ht]
    \centering
    \includegraphics[width=\linewidth]{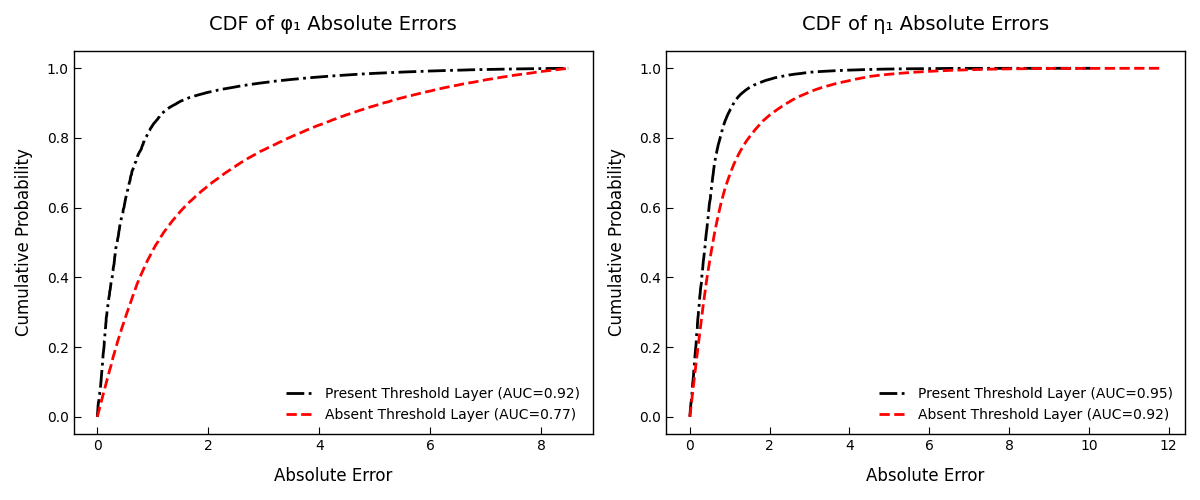}
    \caption{Cumulative Distribution Function Comparison for W-AM With and Without the $p_T$ Threshold Layer}
    \label{evalm}
\end{figure}

However, rather than completely discarding the layer, an alternative approach was to replace the ReLU activation following $y_1$. This substitution resulted in a notable performance improvement compared to the third option, which omitted the threshold layer entirely and retained a standard ReLU activation after $y_1$. This is demonstrated in Figure \ref{evalm}, where a Cumulative Distribution Function (CDF) evaluation is performed on two versions of W-AM. In the absolute error computation for the CDF function, the geometry of the detector is accounted for by treating the $\phi$ dimension as circular. In both instances, the models were trained for $250$ epochs, which roughly corresponded to a global minimum in the loss function, and a set batch size of $32$. Throughout this work, accepted WOMBAT models are graphically depicted in black, while external algorithms are shown in red.

In this analysis, the CDF represents the empirical probability that the absolute error is less than or equal to a given threshold. Denoting the set of absolute errors by $\{q_n\}_{n=1}^N$, for a sample size of N, the normalized CDF is given by:
\begin{gather}
    \text{CDF}(q) = \frac{1}{N}\sum_{n=1}^N \Theta(q-q_n),
\end{gather}
where $\Theta(x)$ indicates the Heaviside function. For sorted $\{q_n\}_{n=1}^N$ this simply becomes:
\begin{gather}
    \text{CDF}(q_n)=\frac{n}{N},
\end{gather}
where $q_n$ is the $\text{n}^{\text{th}}$ smallest absolute error.

Following, the normalized area-under-the-curve (AUC) value is computed using a trapezoidal approximation as follows:
\begin{gather}
    \text{AUC} \approx \frac{1}{\max_n q_n}\sum^{N-1}_{k=1}\frac{q_{k+1}-q_{k}}{2}\Big(\frac{k}{N}+\frac{k+1}{N}  \Big).
\end{gather}

By definition, a larger AUC indicates a steeper CDF increase, signifying fewer errors. Evidently, including the $p_T$ threshold enables the model to better resolve jet substructure, with a significant increase in accuracy in $\phi$. By filtering low-energy signals, the jet's center becomes more well-defined, improving prediction accuracy. The uniform sampling and finer granularity in $\phi$ make it more responsive to the $p_T$ threshold. Given that there is no increase in latency or computational overhead when a QActivation(ReLU) layer is replaced by the custom $p_T$ threshold function, including it in W-AM leads to a significant improvement in predictive power.

Unlike the $p_T$ threshold layer, no optimal solution was found for the $\phi$ circular wrapping function. Implementing it in the model extends the $\phi$ dimension, increasing the number of convolutions per filter. While the stride can be adjusted to compensate, this approach leads to reduced accuracy. In FPGAs, minimizing arithmetic operations is crucial for reducing latency, making padded inputs, regardless of the method used, unfavorable. An alternative approach was attempted by implementing a custom circular Mean Squared Error (MSE) in W-AM for the $\phi$ outputs:
\begin{gather}
    \text{Circular Loss} = \frac{1}{N}\sum_{i=1}^N\Big( \min(|y_{\text{true},i}-y_{\text{pred},i}|, 17 - |y_{\text{true},i}-y_{\text{pred},i}|) \Big)^2,
\end{gather}
where N is the total number of samples, $y_{\text{true}}$ are the $\phi$ labels, and $y_{\text{pred}}$ are the $\phi$ predictions.

\begin{figure}[ht]
    \centering
    \includegraphics[width=\linewidth]{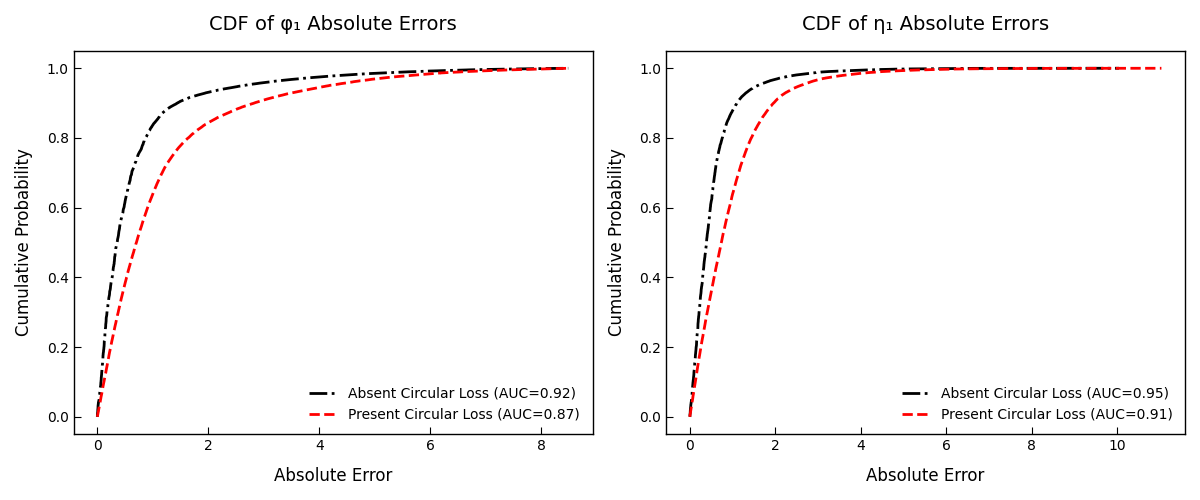}
    \caption{Cumulative Distribution Function Comparison for W-AM With and Without Circular Loss}
    \label{circular}
\end{figure}

This strategy modifies the model's trainable parameters to account for $\phi$ wrapping without any padding. However, as shown in Figure \ref{circular}, this results in lower performance. Partly, the greater complexity of the loss makes it difficult to minimize, but also, due to the simplicity of the model, there is no strict constraint on the range of predictions. This leads to unexpected behavior, such as a large loss, if the model predicts a value close to, but slightly above $17$. It is possible to impose an output value limit through a sigmoid function, however, this adds to the model's complexity and is not optimal for FPGA implementation. Although a circular $\phi$ loss function aligns with the detector's geometry, a regular MSE was used to maximize performance.

\subsection{Performance Overview of the WOMBAT Master and Apprentice Models}

Due to the complexity of the models, W-MM generally outperforms W-AM. This section details a comparison overview through numerous tests conducted on the validation data set.

\begin{figure}[ht]
    \centering
    \includegraphics[width=\linewidth]{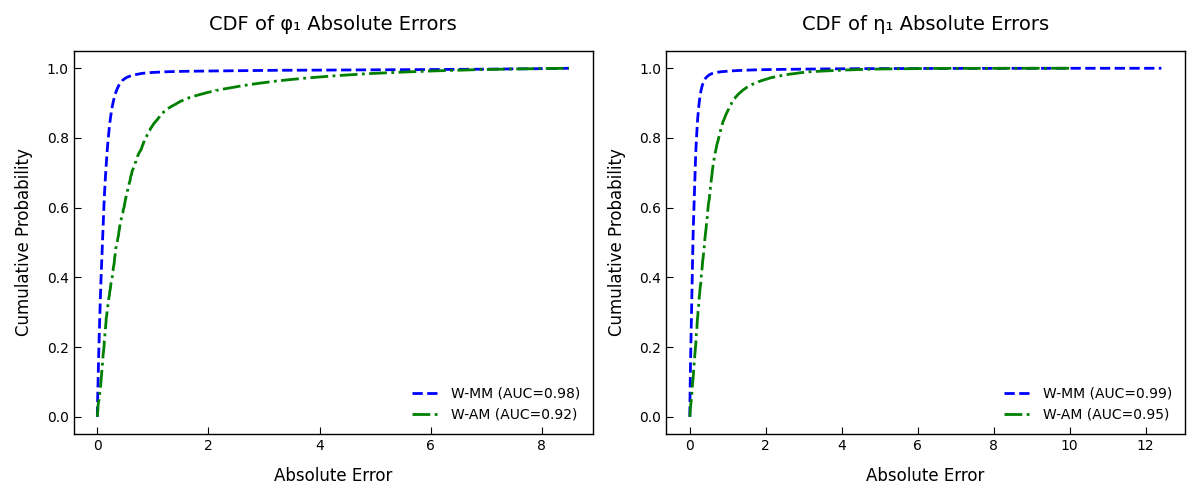}
    \caption{Cumulative Distribution Function Comparison of W-MM and W-AM}
    \label{cdfwombat}
\end{figure}

Figure \ref{cdfwombat} shows that W-MM achieves a higher normalized AUC for both $\phi_1$ and $\eta_1$. Across all CDF analyses (Figures \ref{evalm}, \ref{circular}, and \ref{cdfwombat}), models consistently exhibit lower average AUCs for $\phi$ than for $\eta$, regardless of the architecture. This trend is expected due to the greater granularity in $\phi$, which results in a larger phase space for predictions. Stronger models reduce this discrepancy. For instance, W-MM attains normalized AUCs of $0.98$ for $\phi$ and $0.99$ for $\eta$, outperforming W-AM, which scores $0.92$ and $0.93$, respectively.

\begin{figure}[ht]
    \centering
        \begin{subfigure}[b]{0.49\textwidth}
        \centering
        \includegraphics[width=\textwidth]{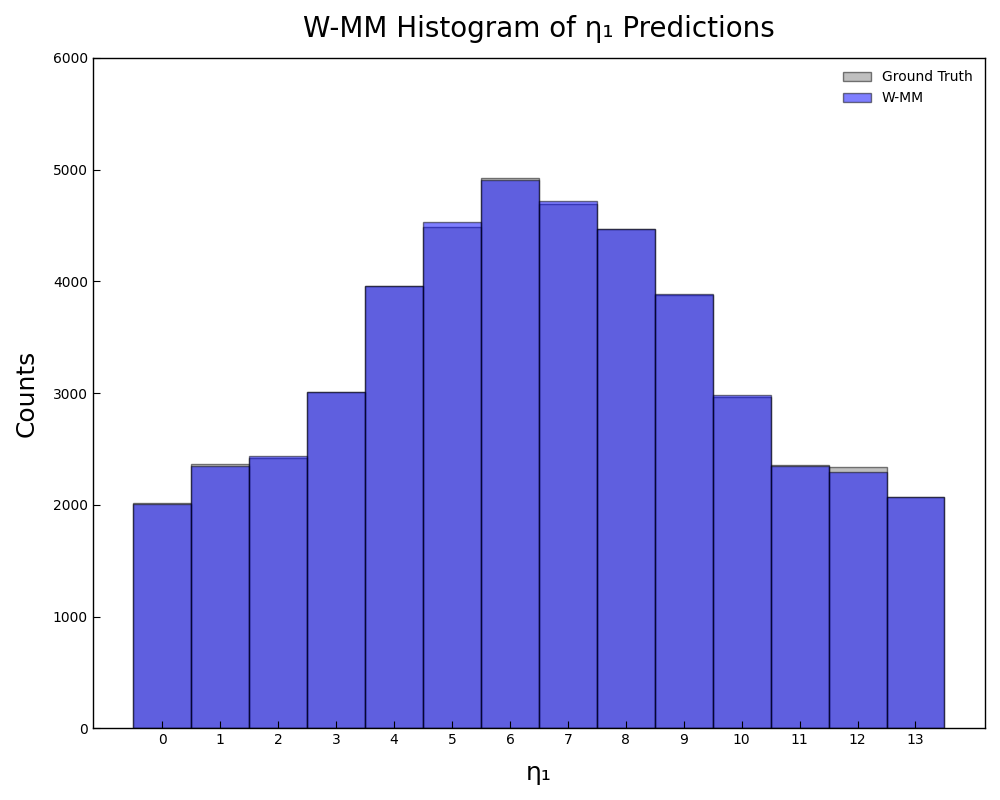}
        \label{fig:subfig2}
    \end{subfigure}
    \hfill
    \begin{subfigure}[b]{0.49\textwidth}
        \centering
        \includegraphics[width=\textwidth]{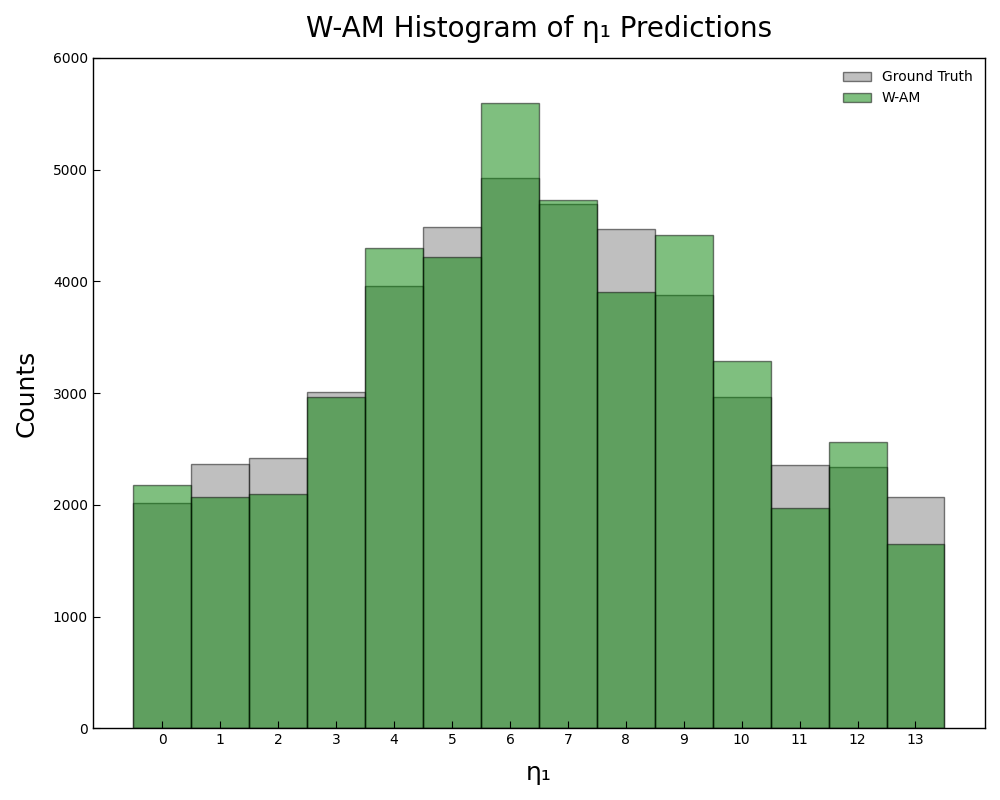}
        \label{fig:subfig1}
    \end{subfigure}
        \begin{subfigure}[b]{0.49\textwidth}
        \centering
        \includegraphics[width=\textwidth]{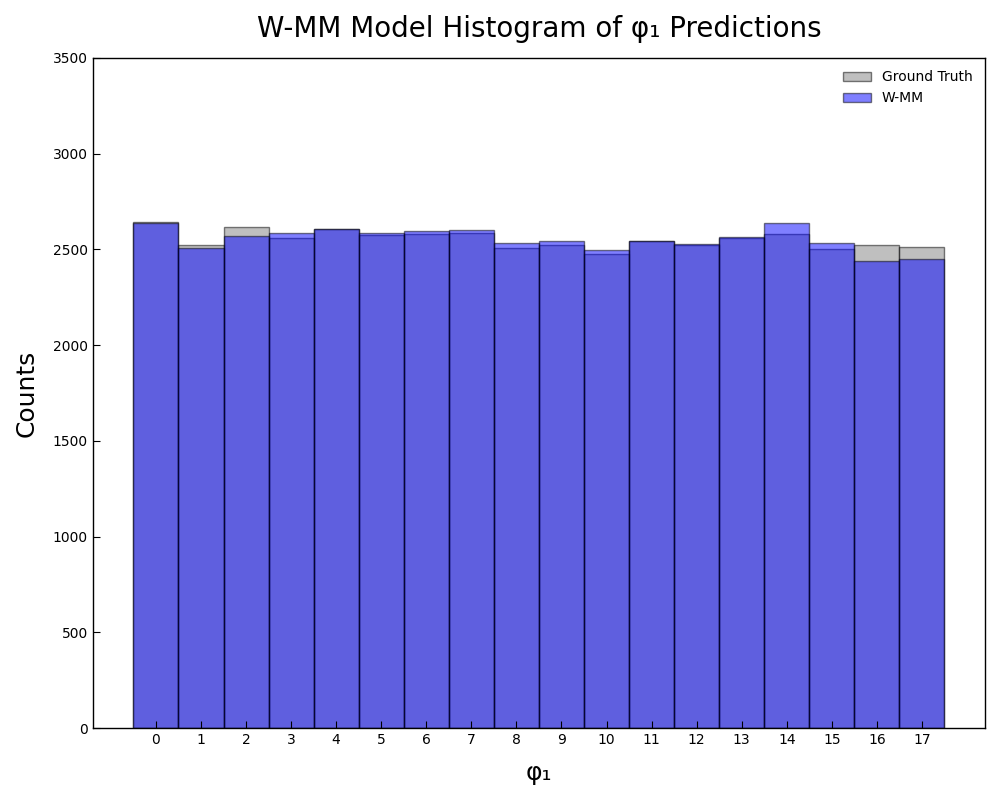}
        \label{fig:subfig4}
    \end{subfigure}
    \hfill
    \begin{subfigure}[b]{0.49\textwidth}
        \centering
        \includegraphics[width=\textwidth]{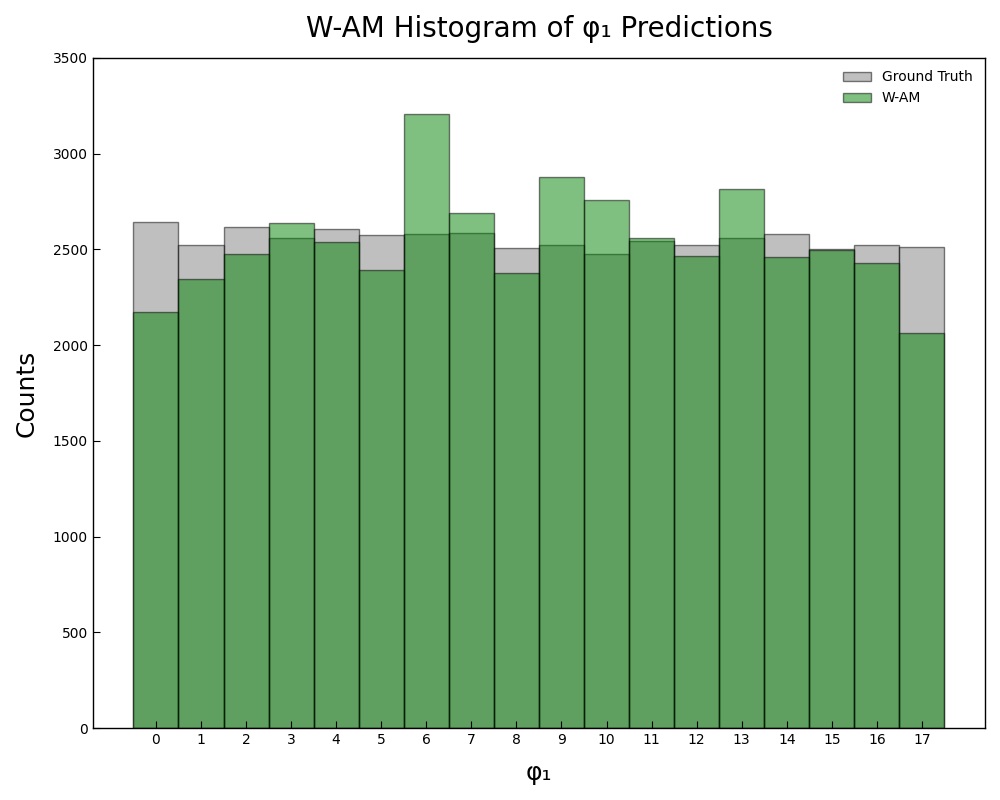}
        \label{fig:subfig3}
    \end{subfigure}
    
    \caption{$\eta$ and $\phi$ Prediction Distributions Compared To Ground Truth for W-MM and W-AM}
    \label{fig:4panel}
\end{figure}

In addition to the CDF score analysis, Figure \ref{fig:4panel} presents the distribution of predicted class counts relative to the ground truth. Optimal performance entails alignment between prediction and ground truth frequencies for each class; deviations indicate prediction inaccuracies. For both $\eta$ and $\phi$, the W-MM model exhibits distributions closely matching the ground truth, with only minor deviations in the high-$\phi$ region. These discrepancies stem from the cyclic nature of $\phi$, which complicates classification near the grid boundaries. Nonetheless, W-MM substantially mitigates these effects compared to W-AM, which displays pronounced discrepancies at both low and high $\phi$ values.

Moreover, W-AM frequently predicts the class value $6$ more often than observed in the ground truth for both $\eta$ and $\phi$. This reflects the model's architectural limitations which constrain its capacity to learn jet substructure, leading to bias toward mid-range predictions that minimize loss. By independently processing $\eta$ and $\phi$ through the EDA architecture, W-MM effectively neutralizes the impact of class imbalance in $\eta$ on $\phi$ predictions. In contrast, W-AM lacks this decoupling mechanism, allowing the central clustering of events in $\eta$ to distort $\phi$ predictions. This is visually evident in the similarity between the $\phi$ and $\eta$ distributions produced by W-AM, particularly in the edge behavior and the artificial peak at $\phi = 6$.

\begin{figure}[ht]
    \centering
    \begin{subfigure}[b]{0.49\textwidth}
        \centering
        \includegraphics[width=\textwidth]{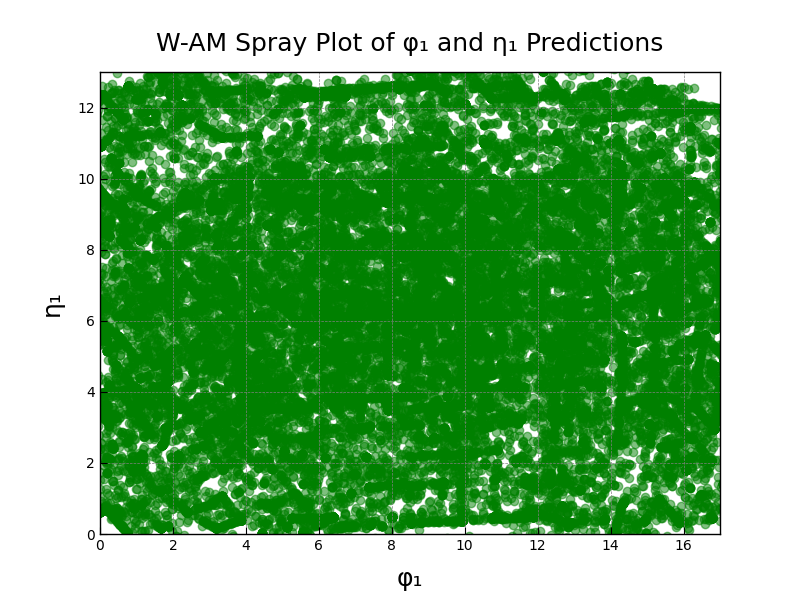}
        \label{fig:subfig1}
    \end{subfigure}
    \hfill
    \begin{subfigure}[b]{0.49\textwidth}
        \centering
        \includegraphics[width=\textwidth]{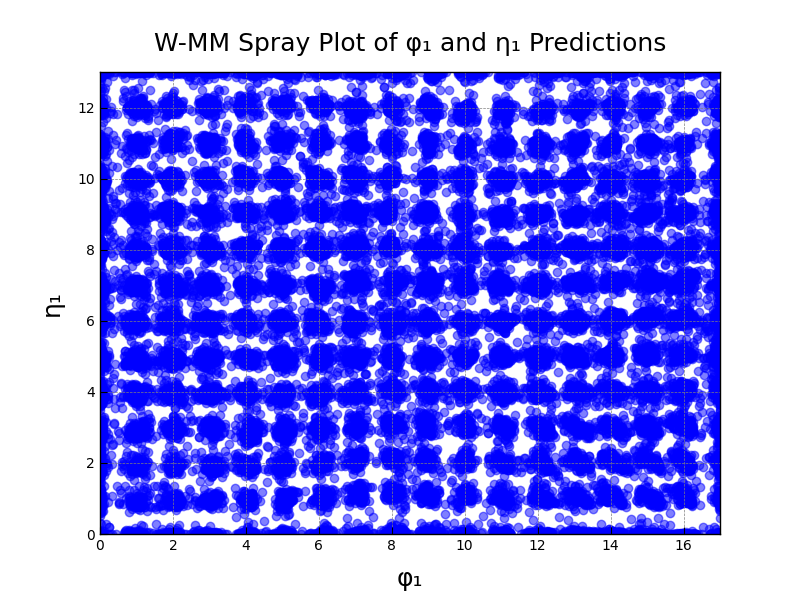}
        \label{fig:subfig2}
    \end{subfigure}
    \caption{Raw Prediction Spray on $\eta-\phi$ Grid for W-MM and W-AM}
    \label{fig:2panel}
\end{figure}

Figure \ref{fig:2panel} presents spray plots of the predicted $(\phi_1, \eta_1)$ coordinates, revealing key differences between W-MM and W-AM. Notably, model outputs are continuous, non-integer values and only rounded post hoc; thus, given the large validation set, a uniform coverage within quantization limits across the $\eta$-$\phi$ grid is expected if predictions are unbiased.

W-AM exhibits a central bias, with a concentration of predictions near the grid center and sparse coverage at the boundaries. This indicates a failure to adequately learn edge-region classes, consistent with the ground truth prediction count discrepancies shown in Figure \ref{fig:4panel}. The lack of edge predictions reflects suboptimal generalization.

In contrast, W-MM not only achieves broader coverage but also exhibits distinct structural formations aligned with integer-valued grid points. This indicates that W-MM has internalized a latent discretization structure inherent to the labels, despite receiving no explicit constraint to output integer values. A key factor enabling this behavior is the use of sigmoid activation in the output layer, which normalizes the continuous predictions which are then mapped to bounded physical coordinates.

Collectively, these findings indicate that W-MM more effectively captures both the global class distribution and the underlying discrete structure of the labels compared to W-AM. However, the architectural complexity and resource demands of W-MM exceed the constraints of the target FPGA, rendering it unsuitable for deployment. In contrast, W-AM represents the highest-performing model that meets the hardware limitations, making it the most viable option for FPGA implementation despite its reduced predictive accuracy.

\subsection{JEDI Architecture}\label{jediarc}

The JEDI algorithm operates on the same TP input as WOMBAT, structured as a $14\times 18$ grid in $\eta\times\phi$ space. Each CaloLayer1 TP region encodes the transverse energy as a $10$-bit fixed-point unsigned integer.

Before cluster formation, JEDI estimates the pileup multiplicity. The number of active regions, P, is obtained through the equation:
\begin{gather}
    \text{P} = \sum_{i=0}^{N_{\text{CR}}-1}\Theta \Big(E^{\text{raw}}_i-E_{\text{thr}}\Big),
\end{gather}
where $E^{\text{raw}}_i$ is the raw (input) $E_T$ of the TP region $i$, $E_{\text{thr}}$ is the $E_T$ threshold, set to $30$ GeV, similarly to WOMBAT, $N_{\text{CR}}$ is the number of regions, $252$ for the $14\times 18$ grid, and $\Theta$ is the Heaviside step function.

The result for P is then quantized to a pileup bin, $b_p$:
\begin{gather}
    b_p = \Big( \frac{\text{P}}{14} \Big),
\end{gather}
which is used to compute $E_i$ (which is equivalent to $E(\eta_i,\phi_i)$) through:
\begin{gather}
    E_i = \max\Big( 0,E_i^{\text{max}}-\Delta_i \Big),
\end{gather}
where $\Delta_i$ is a pileup offset retrieved from a 2D look-up table indexed by $(b_p,E_i^{\text{raw}})$. This table encodes pre-calibrated subtraction values optimized for each pileup bin and energy level. Unlike JEDI, WOMBAT implements this correction using a fixed $\Delta_i=30$ GeV, for all $i$. The $30$ GeV threshold was determined within JEDI to be the most effective on average for distinguishing signal from pileup across a wide range of conditions.

After pileup subtraction, JEDI computes local energy sums over a sliding $3\times 3$ window centered on each non-edge region. This mimics the ML convolution performed by WOMBAT, which has a stride of $1$ and a window size ranging from $3\times 3$ to $5\times 5$. For each region $i$, the clustered energy, $\text{S}_i$ is given by:
\begin{gather}
    \text{S}_i = \sum_{\Delta\eta,\Delta\phi\in\{-1,0,1\}}E (\eta_i+\Delta\eta,\phi_i+\Delta\phi),
\end{gather}
where $(\eta_i,\phi_i)$ are the integer grid coordinates, and $E(\eta,\phi)$ denotes the pileup corrected $E_T$ of the TP coordinates $(\eta,\phi)$. The resulting sum, $\text{S}_i$, is truncated to a $10$-bit unsigned integer, consistent with the input representation. Values exceeding the $10$-bit dynamic range are deterministically saturated, preserving stability in high-occupancy conditions.

A veto condition is imposed on each jet candidate as:
\begin{gather}
    \text{V}_i = (E_C<E_{\text{seed}}) \lor (E_C< \max_k E_k),
\end{gather}
where $E_C$ denotes the transverse energy of the central region from the convolving $3\times 3$ window, $E_k$ represents the $8$ neighboring regions, and $E_{\text{seed}}$ is a fixed parameter, set to $10$ GeV.

Following the veto condition, the algorithm characterizes the local topology of the $3\times 3$ energy deposits. For each cell $m$ in the window, an “active” flag is generated if two conditions are simultaneously met:
\begin{gather}
    A_m = \begin{cases}
        1,\quad \text{if}\; E_m>30\text{ GeV}\;\text{and}\; E_m>\frac{\text{S}_i}{16},\\
        0,\quad \text{otherwise.}
    \end{cases}
\end{gather}
Notably, the division by $16$ is realized by a bit-wise shift ($>>4$), optimizing the latency and resource usage of the operation.

Once 9 boolean flags $A_m$ are computed, they are reorganized into two $3$-bit masks, $r_\eta$ and $r_\phi$, that encode the spatial distribution of the active cells along two geometrical axes:
\begin{gather*}
    r_\eta = \sum_{\eta=0}^2\Big( \bigvee_{\phi=0}^2 A_{\eta,\phi} \Big)2^\eta,\\
    r_\eta = \sum_{\phi=0}^2\Big( \bigvee_{\eta=0}^2 A_{\eta,\phi} \Big)2^\phi.
\end{gather*}

The topologies encoded by $(r_\eta,r_\phi)$ are then compared to a set of allowed bit patterns shown in Table \ref{bitp}. If the $(r_\eta,r_\phi)$ masks fall within one of these categories each, then the $3\times 3$ region passes the veto condition.

\begin{table}[h!]
\centering
\begin{tabular}{c c c}
\textbf{Decimal} & \textbf{Binary} & \textbf{Meaning (bit positions)} \\
\hline
$b_1 = 2$ & 010 & Only the middle row/column is active \\
$b_2 = 3$ & 011 & Top + middle rows/columns active \\
$b_3 = 6$ & 110 & Middle + bottom rows/columns active \\
$b_4 = 5$ & 101 & Top + bottom, but not middle \\
$b_5 = 7$ & 111 & All three rows/columns active \\
\end{tabular}
\caption{Allowed Shape Masks for $r_\eta$ and $r_\phi$ in the JEDI Algorithm}
\label{bitp}
\end{table}

JEDI then maps all calorimeter regions to a larger structure called a super-region through a static and surjective mapping denoted by $s = g(i)$. The $252$ TP regions are partitioned into $24$ super-regions, each spanning $14$ rows in $\eta$ and $3$ columns in $\phi$. Within each super-region, only the highest $E_T$ non-vetoed candidate is kept:
\begin{gather}
    \text{J}_s = \max_{i\in g^{-1}(s)}\{ S_i | V_i=0, (r_\eta,r_\phi) \text{allowed} \},
\end{gather}
where $\text{J}_s$ denotes the highest $E_T$ allowed jet found in super-region $s$. Because $g$ is surjective and disjoint, exactly one jet slot per super-region is filled.

Following this, each of the $24$ potential jet candidates is encoded as a fixed-width jet word and stored in a $64$-element array. The remaining $40$ entries are filled with zero-valued placeholders to conform to the input size requirements of a bitonic sorting network, which operates optimally on arrays of length $2^n$. This padded array is subsequently processed by a bitonic sorting network with a depth proportional to $\log_2(64)$, consisting of $6$ hierarchical stages. Each stage comprises parallel compare-swap units that recursively transform partially ordered bitonic sequences into a fully sorted array, ordered by jet transverse energy. From the original $24$ possible jets, only the top 6 are selected for output.

In the firmware implementation, similarly to WOMBAT, JEDI encodes the selected jet's transverse energy, $J_s$, in the first 10 bits of the output word. The $\eta$ position occupies bits 11 through 18, while the $\phi$ position is stored in bits 19 through 26. Bit 27 is reserved for a potential flag, and the remaining bits, 28 to 31, are currently unused.


\subsection{ML Implementation in FPGA Devices}\label{fpga}

To evaluate firmware compatibility, resource usage, and latency for online deployment, WOMBAT was implemented on Xilinx Virtex-7 FPGA devices. In particular, the model in question is XC7VX690T-2FFG1927I \cite{fpgamodel} where:
\begin{itemize}
    \item \textbf{XC}: Indicates that it is a Xilinx device.
    \item \textbf{7V}: Signifies that it belongs to the Virtex-7 family.
    \item \textbf{X690}: Denotes the presence of approximately $690,000$ logic cells.
    \item \textbf{T}: Classifies the device as having high-speed serial transceivers.
    \item \textbf{2}: Is the speed grade of the device, where a lower number is given to slower operating speeds. The speed grade ranges from $1$ (slowest) to $3$ (fastest).
    \item \textbf{FFG}: It stands for Flip-Chip Fine-Pitch Ball Grid Array, which is a specific type of Ball Grid Array (BGA) package used for integrated circuits.
    \item \textbf{1927}: Represents the total number of pins (electrical connections) on the BGA package.
    \item \textbf{I}: Stands for ''Industrial" which is the temperature grade associated with the device ($-40^\circ$C to $100^\circ$C).
\end{itemize}

The Virtex-7 FPGAs are manufactured using $28$ nm process technology, which allows for high transistor density, reduced power consumption, and enhanced computational efficiency. This advanced fabrication process enables the XC7VX690T-2FFG1927I to integrate approximately 693,120 logic cells, 3,600 Digital Signal Processing (DSP) slices\footnote{A DSP slice is a dedicated computational block inside an FPGA optimized for high-speed arithmetic operations, which are crucial for executing complex mathematical functions with minimal latency.}, and a robust interconnect architecture.

For WOMBAT and JEDI    , FPGA algorithm design and implementation were carried out using High-Level Synthesis (HLS) \cite{hlsbook}, which enables complex algorithms to be developed in high-level languages such as C, C++, or SystemC, and then synthesized into firmware for FPGA deployment. Essentially, HLS streamlines the FPGA design process by automatically converting high-level algorithms into Register Transfer Level (RTL) representations. RTL is a low-level hardware representation that defines the flow of data between registers and the logic operations performed in each clock cycle (CC). Unlike traditional Hardware Description Languages (HDLs) like VHDL \cite{vhdl} or Verilog, which require manual specification of registers and logic gates, HLS abstracts this process, automatically optimizing for area, power, and performance. Key optimizations include loop unrolling, which replicates hardware resources to increase parallelism, and pipelining, which allows overlapping execution of multiple computations to improve latency. 

Once HLS generates RTL, Vivado performs logic synthesis, placement, and routing, mapping the design to the FPGA's configurable logic blocks (CLBs), digital signal processing (DSP) slices, and block random access memories (BRAMs). Performance is validated through timing analysis and hardware-in-the-loop (HIL) testing, ensuring compliance with real-time constraints. To optimize WOMBAT's data pipeline, fixed-point arithmetic replaces floating-point operations, reducing DSP usage and improving computational efficiency. Additionally, memory partitioning distributes data across multiple memory banks to prevent bottlenecks. These optimizations enable the FPGA to meet low-latency requirements critical for online deployment in the CMS L1T.

In the case of WOMBAT, two FPGA designs were explored. When discussing the implementation, there are three main interlinked algorithms:
\begin{itemize}
    \item \textbf{Main Algorithm}: The core function, originally \texttt{algo\_unpacked}, processes fully unpacked TPs received from detector readout links. It subsequently forwards the data through the WOMBAT ML trigger and processes the resulting output.
    \item \textbf{Main WOMBAT Function}: This function contains the WOMBAT ML algorithm, where data is processed through a set of pre-defined and pre-trained layers with their associated weights and biases.
    \item \textbf{Parameters}: This module, included within the WOMBAT function, defines the configuration parameters necessary for HLS. Optimizing these parameters is crucial for minimizing latency while ensuring stable CCs below $6.25$ ns.
\end{itemize}

The first design approach pipelines the main algorithm while inlining the WOMBAT function, ensuring efficient execution at the top level with minimal overhead from function calls. This method ensures a streamlined control structure, reducing scheduling complexity by minimizing function call overhead. At a lower level, fine-grained parallelism optimizes ML computations by processing multiple neurons concurrently. Inlining the ML model significantly simplifies control logic.

The second approach applies the \texttt{DATAFLOW} pragma at the main algorithm level. A pragma is a compiler directive that provides optimization hints that influence how the code is translated into hardware, without altering its functional behavior. This required fully restructuring the main algorithm, as \texttt{DATAFLOW} requires a high-level function that contains nothing but function calls. As a result, all logic implemented in \texttt{algo\_unpacked} was made into separate functions, which includes the call to WOMBAT, with each of these functions being pipelined separately. The dataflow pragma ensures that these functions execute concurrently, without unnecessary stalls. This restructuring maximizes parallelism, reduces latency, and allows WOMBAT to operate in a pipelined manner.

\subsubsection{WOMBAT Firmware Implementation and Optimization Procedure}

After training W-AM and W-ASM, an HLS4ML script was developed to convert the pre-trained QKeras models into HLS implementations. HLS4ML is an open-source Python library designed to translate ML models into FPGA-friendly HLS code, enabling efficient deployment of deep learning models on hardware \cite{duartehls}. In general, the output includes a main model function optimized through the \texttt{DATAFLOW} pragma, associated definitions and parameters, a utility folder containing data buffer and ML layer implementations, and a separate folder storing the extracted weights from the trained model. Although most of these files were used in the WOMBAT implementation, many required modifications. Additionally, the main algorithm had to be developed from scratch to handle data processing and ensure the model was executed correctly, with inputs properly passed and outputs efficiently processed.

To convert W-AM from a TensorFlow model to a hardware trigger system, the following procedure was followed:

\begin{itemize}
    \item \textbf{Stage 1:} Initial Conversion from Python to HLS
\end{itemize}

In the HLS4ML configuration, W-ASM and W-AM are loaded separately. Since W-AM contains the $p_T$ threshold layer, the model needs to be read by the script using a custom object scope. To extract the weights from W-AM, during the conversion, the threshold layer is extracted and replaced with a quantized ReLU. This shortcut allows for the HLS4ML program to extract the model weights in an appropriate format. Even though the underlying structure is modified, the weights corresponding to $y_2$ in Chapter IV, Section \ref{apprentc} are associated with the $p_T$ threshold layer, not the substituted ReLU activation. As a result, the W-AM and W-ASM models are converted in parallel: weights from W-AM are combined with the HLS code output from W-ASM for firmware implementation. This approach requires a manual HLS implementation of the $p_T$ threshold layer, which will be discussed in Stage 2 of the FPGA development process.

The HLS4ML configuration was generated using the latency strategy that aims to minimize inference delay. Each input, weight, bias, and output has a designated precision assignment. The input layer is represented as a 10-bit unsigned integer, while convolutional layers use 8-bit fixed-point precision for weights and biases, with 16-bit fixed-point outputs optimized for resource efficiency using a line-buffered implementation. The dense and activation layers also produce 16-bit fixed-point outputs, ensuring consistency across the model. A reuse factor of 2 is applied globally to balance parallel execution and FPGA resource utilization. For some layers, such as the first convolution, the reuse factor was later manually set to 1 to reduce latency. Essentially, a reuse factor defines how many times a hardware multiplier is reused during computation, balancing resource usage and parallelism.

The model is configured for a $6.25$ ns clock period, which corresponds to one-fourth of the 25 ns bunch crossing interval of proton collisions at the LHC, allowing the design to perform up to four processing steps within each collision cycle. To achieve this, the parallel I/O configuration is used, enabling multiple data inputs and outputs to be processed simultaneously within each CC. This approach optimizes data flow for real-time decision-making, ensuring minimal latency while meeting the stringent processing demands of the CMS Level-1 Trigger. The configuration variable \texttt{part} is set to XC7VX690T-2FFG1927I, specifying the FPGA model used for implementation.

\begin{itemize}
    \item \textbf{Stage 2:} HLS Custom Design and Optimization
\end{itemize}

Following the HLS4ML conversion, a lot of effort was taken to ensure efficient processing of the TPs and output handling. Instead of relying on the base code produced by HLS4ML, a high-level function (\texttt{algo\_unpacked}) was developed to process fully unpacked TP data by converting raw input links into a structured format suitable for FPGA implementation. The function extracts region-specific calorimetry information, reshapes the data, and passes it to the WOMBAT algorithm for analysis. The outputs are then mapped into temporary arrays with careful preservation of reserved control bits, using HLS directives like pipelining, dataflow, array partitioning, and unrolling to ensure efficient resource usage.

Post-processing begins immediately after WOMBAT outputs the jet centers by converting the raw fixed-point outputs into a structured, $32$-bit data word tailored for downstream processing within the FPGA. Specifically, each fixed-point result is first cast into an unsigned $16$-bit value, from which the coordinates, indexed $\eta$ and $\phi$, are extracted via the designated bit ranges discussed in Chapter IV, Section 4. This precise bit allocation guarantees that the jet center information is aligned with downstream data protocols.

Two high-level algorithmic approaches were evaluated:

\begin{itemize}
    \item \textbf{Approach 1}: The algorithm was initially implemented as a monolithic function, \texttt{algo\_unpacked}, integrating data preparation, WOMBAT execution, and output post-processing. Key synthesis pragmas included \texttt{PIPELINE}, \texttt{UNROLL}, and \texttt{LATENCY MAX}/\texttt{MIN}, enabling loop unrolling and parallelism. WOMBAT was inlined to minimize function call overhead, promoting aggressive optimization.
    
    \item \textbf{Approach 2}: This variant modularizes the pipeline by isolating computational stages into discrete functions, coordinated through a top-level controller annotated with the \texttt{DATAFLOW} directive. Unlike Approach 1, WOMBAT is not inlined but treated as a pipelined function block. Despite the increased overhead from non-inlined execution, this strategy yielded superior performance, characterized by reduced latency and no change in resource utilization.
\end{itemize}

The primary distinction lies in the pipelining granularity of WOMBAT. In both designs, individual neural network layers are pipelined; however, Approach 2 initiates pipelining at the top-level WOMBAT function, leading to more efficient resource scheduling and improved timing closure. Although the high-level algorithm in the second approach utilized the \texttt{DATAFLOW} pragma to enable concurrent execution, all auxiliary data processing is structured with explicit data dependencies that enforce a fixed execution order. This ensures deterministic behavior without introducing unintended parallelism.

To optimize resource usage and execution timing, the reuse factor of each layer was manually set, as well as the buffer size and partitions. A higher reuse factor reduces resource usage but increases latency, while a lower reuse factor consumes more resources to achieve faster execution. 

For some computationally expensive layers that rely on convolution and dense operations, the reuse factor was set to $2$, and $1$ for the remaining layers. Since W-AM features two convolutions, the FPGA available resources allowed for one to have a reuse factor of $1$.

The original HLS4ML-generated code included large statically defined buffers with suboptimal partition schemes. In particular, some partitions contained mostly zeros, and substantial portions of the allocated memory went unused. This inefficient memory layout not only led to unnecessary resource consumption but also introduced excessive memory access latency. In some cases, the inflated buffer size and poor utilization even caused synthesis failures due to routing congestion or resource overuse. By manually compacting these buffers and enabling their reuse across multiple operations, both memory footprint and access latency were significantly reduced. This restructuring led to a measurable performance improvement, cutting overall latency by approximately $20$ CCs and enabling successful timing closure under the $6.25$ ns constraint.

Both approaches were optimized to achieve the lowest possible clock cycle period and latency. This design requirement arises from the $25$ ns interval at which proton bunch crossings occur at the CMS. To accommodate data processing within a $25$ ns period, WOMBAT's FPGA target period is set to $6.26$ ns, translating to a rate of about $160$ MHz. This choice allows exactly four CCs to fit within each collision window, ensuring that data can propagate through the pipeline without the risk of missing the next collision's input. To force the pipeline to complete its combinational processing and register updates within $4$ cycles, the top-level function, \texttt{algo\_unpacked}, contains the pragma \texttt{LATENCY MIN}=$4$ and, depending on the approach, the top-level pragma \texttt{PIPELINE} is set to 4. The final implementation achieved a nominal path delay of $5.79$ ns, which translates to the time it takes for the signal to travel through the longest path in the circuit under typical conditions. The additional $1.69$ ns is a safety margin added to account for uncertainties such as process variations, temperature changes, or other real-world factors that might cause the actual delay to be longer than expected. This conservative margin ensures that even if the delay increases slightly under less-than-ideal conditions, the design will still meet the target $6.25$ ns clock period reliably.

Formally, latency refers to the delay between the arrival of a data packet at the beginning of the processing pipeline and the time its corresponding output is produced. The \texttt{DATAFLOW} design associated with Approach 2, exhibited a fixed latency of $22$ clock cycles, translating to $137.5$ ns at $6.25$ ns per clock period. In comparison, the algorithm in Approach 1 achieved a latency of $24$ ns. For both designs, the pipeline's initiation interval is set to 4 cycles, therefore, new data sets can be injected every $25$ ns. This means while any given data packet takes $22$-$24$ cycles to traverse the WOMBAT algorithm, the pipeline is capable of overlapping computation such that it can accept fresh inputs at each $25$ ns boundary.

Once all functions are pipelined and loops unrolled, HLS offers numerous ways to constrain the latency, such as the pragma \texttt{LATENCY MAX}. At the expense of resource usage, it is possible to manually set a maximal latency of execution at any level in the system. However, setting too low of a constraint forces the combinational logic between pipeline registers becomes more complex and longer, thus increasing the critical path delay. As a result, the achievable clock period can rise above the $6.25$ ns target meaning that the processing can't be completed within the $25$ ns window. For example, constraining the latency to $20$ in Approach 2 leads to a clock period of $9.804$ ns, which is not acceptable.

\subsection{FPGA Implementation of JEDI}\label{fpgajedi}

Unlike WOMBAT, which leverages the HLS4ML framework, the JEDI algorithm was manually implemented in HLS and synthesized for the same FPGA model, XC7VX690T-2FFG1927I. The architectural details outlined in Chapter IV, Section 4, are derived directly from the FPGA implementation and are further elaborated in this section from a technical and hardware-centric perspective.

In JEDI, every processing stage is optimized using HLS directives to exploit the FPGA's inherent parallelism and ensure deterministic latency. Key loops-such as those responsible for jet candidate preparation, bit-level data packing, and sorting-are fully unrolled via the \texttt{UNROLL} pragma, thereby enabling simultaneous execution across all candidate channels. In addition, arrays are reshaped and partitioned to provide concurrent access to data elements, minimizing latency and avoiding memory access bottlenecks.

The design employs a $64$-element bitonic sorting network, achieved by zero-padding 24 valid jet words to a full array of $64$ elements. The sorter is created with a depth proportional to $6$ (or $\log_2(64)$), which partitions the network into six hierarchical stages. Each stage consists of parallel compare-swap units, implemented using efficient $10$-bit subtract-and-multiplexer (MUX) circuits, that recursively merge bitonic sequences into a fully ordered output. This custom sorting engine is deeply pipelined so that, after the pipeline has filled (approximately $21-24$ cycles), the network can process one new set of inputs every CC.

Furthermore, auxiliary arithmetic operations such as the 3×3 regional energy summing are handled by a fully unrolled adder tree that guarantees a saturating output within the $10$-bit range. The function that counts active bits in a regional threshold mask employs an HLS pipeline directive (with an initiation interval of $4$ cycles, similar to WOMBAT). These operations, along with extensive bit-slicing for output word formation, culminate in the assembly of compact $32$-bit output words. Each word encodes jet transverse energy, spatial position, and reserved fields, and is eventually concatenated into $128$-bit GT links for transmission to subsequent trigger logic.

\subsection{Analysis Through the CMS Software}\label{singlejet}

The CMS Software (CMSSW) is the official software framework used by the CMS experiment for event reconstruction, simulation, and data analysis. It provides a modular, C++-based environment integrated with Python configuration, enabling scalable processing of detector data within a consistent and reproducible infrastructure.

For trigger systems, performance is primarily evaluated using two key metrics: trigger efficiency and trigger rate. Mathematically, the efficiency per transverse momentum, $\epsilon (p_T)$, can be represented as follows:
\begin{gather}\label{effformula}
    \epsilon(p_T) = \frac{\text{N}_{\text{W-OFFLINE}}(p_T)}{\text{N}_{\text{OFFLINE}}(p_T)},
\end{gather}
where N$_{\text{W-OFFLINE}}(p_T)$ denotes the number of events per $p_T$ bin that pass both the WOMBAT algorithm and the offline selection, and N$_{\text{OFFLINE}}(p_T)$ is the total number of events per $p_T$ bin that pass the offline selection. To evaluate the trigger efficiency, $H\rightarrow b\bar{b}$ MC samples were used, which were generated through the algorithm discussed in Chapter III, Section 1.

For an event to pass both offline selection and the WOMBAT trigger, the following set of requirements must be met:
\begin{itemize}
    \item Must pass a minimum $p_T$ threshold which is proportional to the calculated $E_T$ multiplied by a pre-defined scale factor.
    \item Must have sufficient activity in neighboring regions to the jet's center with $p_T>30$ GeV and $>6.25\%$ of the jet's total $E_T$.
    \item Must not be vetoed by electromagnetic or tau-specific region flags.
    \item Each jet must be geometrically matched to an offline AK8 jet passing within $\Delta$R $< 0.4$. The AK8 algorithm provides a high-resolution reference for jet structure, using full detector information to reconstruct large-radius jets with detailed substructure. Its accuracy makes it ideal for validating and matching trigger-level jets in boosted topologies like $H \rightarrow b\bar{b}$.
    \item The matched offline jet must satisfy the analysis-level selection: $p_T$ above a configurable threshold, presence of exactly two SoftDrop subjets, and at least one associated b-hadron per subjet.
\end{itemize}

In this analysis, WOMBAT is directly compared to an existing L1T boosted $H\rightarrow b\bar{b}$ tagger, known as Single Jet 180. This algorithm is implemented in the Calorimeter Layer 1 and serves as a seed to the HLT. If an event passes the selection criteria of Single Jet 180, a trigger bit is set and the corresponding L1 jet object is passed to the HLT. This initiates a specific HLT path, where more detailed reconstruction and selection are performed based on the L1-provided information. As the name Single Jet 180 suggests, the algorithm applies predefined selection criteria to identify a single boosted $H\rightarrow b\bar{b}$ jet in an event, achieving an efficiency greater than 0.8 for jets with $p_T>180$ GeV. The clustering algorithm forms jets by aggregating energy deposits from a $9\times 9$ grid of trigger towers in $\eta-\phi$ space, with pileup subtraction performed using energy estimates from a surrounding band adjacent to the jet area. For this analysis, the Single Jet 180 trigger uses the same TP granularity as WOMBAT and needs to satisfy the same conditions for an event to be considered a boosted $H\rightarrow b\bar{b}$ jet.

In addition to a trigger efficiency study, the WOMBAT algorithm was evaluated for trigger rates and compared to the Single Jet 180 performance. Mathematically, the total rate calculation can be written as:
\begin{gather}\label{rateseq}
    \text{R}(p_T) = \Bigg( \frac{\text{N}_{\geq p_T}(i)}{\text{N}_{\text{total}}} \Bigg)\times \Bigg( \frac{\text{N}_{h_0}}{\text{N}_{h_1}} \Bigg)\times \Bigg(\frac{40\times 10^6}{10^3}\Bigg)\;[\text{kHz}].
\end{gather}
In Equation \ref{rateseq}, the first term, $\frac{\text{N}_{\geq p_T}(i)}{\text{N}_{\text{total}}}$ represents the cumulative count of events above a threshold $p_T(i)$ normalized by the total number of events, $\text{N}_{\text{total}}$. $\text{N}_{\geq p_T}$ is obtained by:
\begin{gather}
    \text{N}_{\geq p_T}(i) = \sum_{j=i}^{\text{N}_{\text{bins}}} \text{N}_j,
\end{gather}
where $\text{N}_j$ represents the number of events falling into the $j^{\text{th}}$ bin of the distribution histogram corresponding to a specific transverse momentum interval $[p_T^j,p_T^j+\Delta p_T]$. 

The second term in Equation \ref{rateseq}, $\frac{\text{N}_{h_0}}{\text{N}_{h_1}}$ serves as a correction factor to account for differences in the overall normalization between WOMBAT and Single Jet 180. In particular, $\text{N}_{h_0}$ and $\text{N}_{h_1}$ denote the total event counts in the histograms corresponding to the WOMBAT algorithm and the Single Jet 180 trigger, respectively. Finally, the coefficient $\frac{40\times 10^6}{10^3}$ is a conversion factor that translates the normalized event fraction into an absolute trigger rate in kHz. The numerator reflects the nominal bunch crossing frequency of $40$ MHz at the LHC, which is divided by $10^3$ for conversion into kHz.

For the trigger rates analysis, ZB data was used, as it provides an unbiased sample of events selected solely on the presence of a beam crossing, independent of physics activity. This dataset is ideal for rate studies since it reflects the true input conditions to the L1T. The ZB data was accessed using the CMSSW framework and processed through the \texttt{l1Ntuple producer}. To retrieve and analyze this dataset, the CRAB system was used to submit grid jobs for data skimming and ntuple production. The resulting ROOT files were merged and used as input to construct rate histograms corresponding to different trigger configurations.

\newpage

\newpage

\stepcounter{section}
\section*{Chapter {V}: Trigger Rate, Efficiency, and FPGA Implementation Results}
\addcontentsline{toc}{section}{Chapter {V}: Trigger Rate, Efficiency, and FPGA Implementation Results}
\setcounter{figure}{0}
\markboth{Chapter {V}: Trigger Rate, Efficiency, and FPGA Implementation Results}{}

This chapter presents a detailed performance evaluation of WOMBAT's Master and Apprentice models in the context of L1 trigger suitability. To qualify for online deployment, a trigger must satisfy the following key criteria:
\begin{itemize}
    \item \textbf{High Efficiency in Relevant Regime}: The trigger must efficiently select targeted physics signatures, such as boosted jet topologies, by resolving substructure and tagging relevant processes within the desired kinematic regime.
    \item \textbf{Low Rate and Pileup Resistance}: In addition to high efficiency, the trigger must suppress background and low-relevance events to maintain a low L1A rate, ensuring resilience to high pileup conditions, especially during the upcoming HL-LHC era.
    \item \textbf{Acceptable Firmware Resource Usage and Processing Latency}: The design must conform to strict FPGA resource limits and latency requirements imposed by the L1T system, delivering decisions within the allowed time budget for real-time operation.
\end{itemize}

\subsection{Trigger Primitives Displays}

As outlined in Chapter III, Section 2, the event displays visualize TPs originating from the HCAL and ECAL, which constitute the primary inputs to the WOMBAT trigger. TPs represent localized energy deposits derived from calorimeter readout signals and are produced in real-time by dedicated hardware or firmware systems known as trigger primitive generators (TPGs). The event displays further include jet centers predicted by WOMBAT, alongside the offline AK8 jet clustering and tagging results that serve as a reference for evaluating performance.

For W-MM, the predictions, based on visual TP displays, can be broadly categorized into the following groups:
\begin{itemize} 
    \item \textbf{Good Matches}: In these events, WOMBAT predicts jet centers that align with the same trigger regions as those identified by the offline AK8 algorithm across all clusters.
    \item \textbf{Semi-Good Matches}: This category includes events where one or more WOMBAT-predicted jet centers exhibit a slight spatial offset from the corresponding AK8 jets. Despite these deviations, the majority of these jets still satisfy the $\Delta$R $< 0.4$ matching criterion, and the mismatch is primarily visual in nature.
    
    \item \textbf{Poor Matches}: Events in this group are characterized by significant discrepancies between WOMBAT and AK8 predictions. Only some, or in some cases, none, of the WOMBAT-predicted jet centers satisfy the $\Delta$R $< 0.4$ condition relative to the AK8 jets.
    
    \item \textbf{Jet Multiplicity Mismatch}: This group encompasses all events in which the number of jets predicted by WOMBAT differs from the number identified by the AK8 algorithm, irrespective of spatial agreement ($\Delta$R). Representative scenarios include:
    \begin{itemize}
        \item AK8 identifies $4$ jets, WOMBAT predicts $3$ or $2$;
        \item AK8 identifies $3$ jets, WOMBAT predicts $2$;
        \item AK8 identifies $2$ jets, WOMBAT predicts $3$;
        \item AK8 identifies $1$ jet, WOMBAT predicts $2$ or $3$.
    \end{itemize}
\end{itemize}

In the case of W-AM, all categories above apply with the constraint that W-AM always predicts only 2 jets. For an organized collection of TP displays, see Appendix F. Moreover, control plots for all trigger systems discussed, along with accompanying commentary, are provided in Appendix E.

\subsubsection{WOMBAT Master Model TP Displays}

\begin{figure}[ht]
  \centering
  \begin{minipage}[b]{0.49\textwidth}
    \centering
    \includegraphics[width=\linewidth]{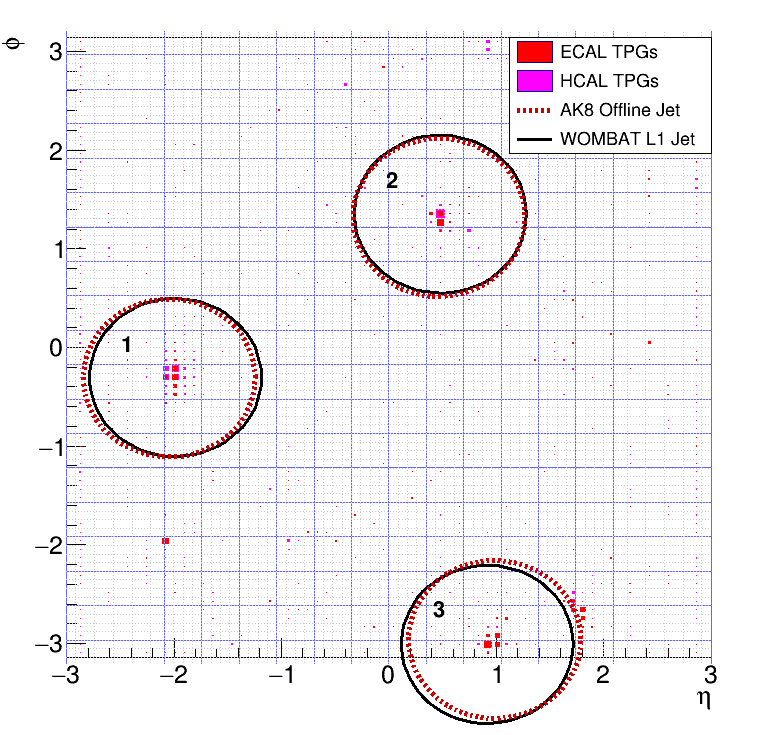} 
    \caption{W-MM Good Match TP Display - Event 2687}
    \label{fig:image112}
  \end{minipage}
  \hfill
  \begin{minipage}[b]{0.49\textwidth}
    \centering
    \includegraphics[width=\linewidth]{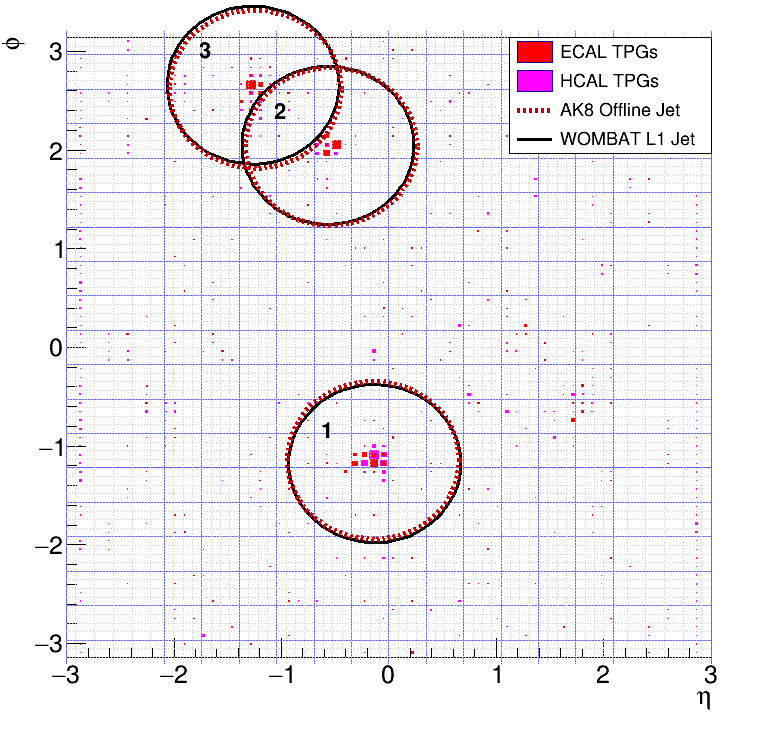} 
    \caption{W-MM Good Match TP Display - Event 2995}
    \label{fig:image222}
  \end{minipage}
\end{figure}

Figures \ref{fig:image112} and \ref{fig:image222} present events classified as “Good Matches,” where the WOMBAT trigger successfully identifies three jets originating from a $H\rightarrow b\bar{b}$ decay. The predicted jet centers exhibit close spatial agreement with those reconstructed by the offline AK8 algorithm. TP cluster numbering corresponds to WOMBAT's output ordering, which is significant for interpreting prediction behavior. Notably, the model consistently resolves the leading and subleading jets but exhibits reduced accuracy in localizing the third cluster. In Figure \ref{fig:image112}, the third jet is slightly displaced toward the leading jet, suggesting reduced confidence in its localization, which can lead to larger inaccuracies with variations in TP structure.

Additionally, Figure \ref{fig:image112} highlights the model's handling of $\phi$ wrapping near the grid boundaries. WOMBAT demonstrates the ability to resolve substructure and accurately predict jet locations even in the $\phi\approx 0$ (or $\phi\approx 2\pi$) region.

\begin{figure}[H]
  \centering
  \begin{minipage}[b]{0.49\textwidth}
    \centering
    \includegraphics[width=\linewidth]{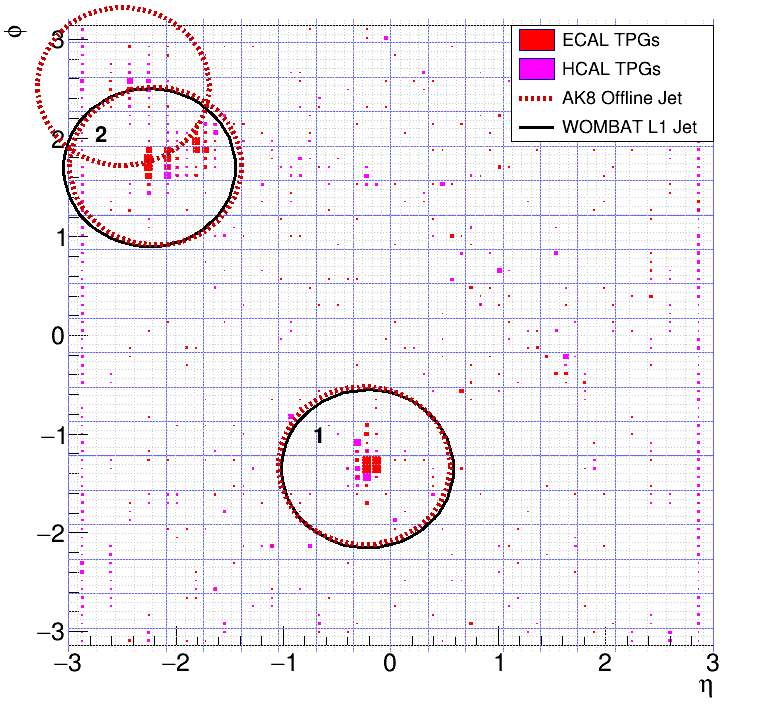} 
    \caption{W-MM Jet Multiplicity Mismatch TP Display - Event 689}
    \label{fig:i}
  \end{minipage}
  \hfill
  \begin{minipage}[b]{0.49\textwidth}
    \centering
    \includegraphics[width=\linewidth]{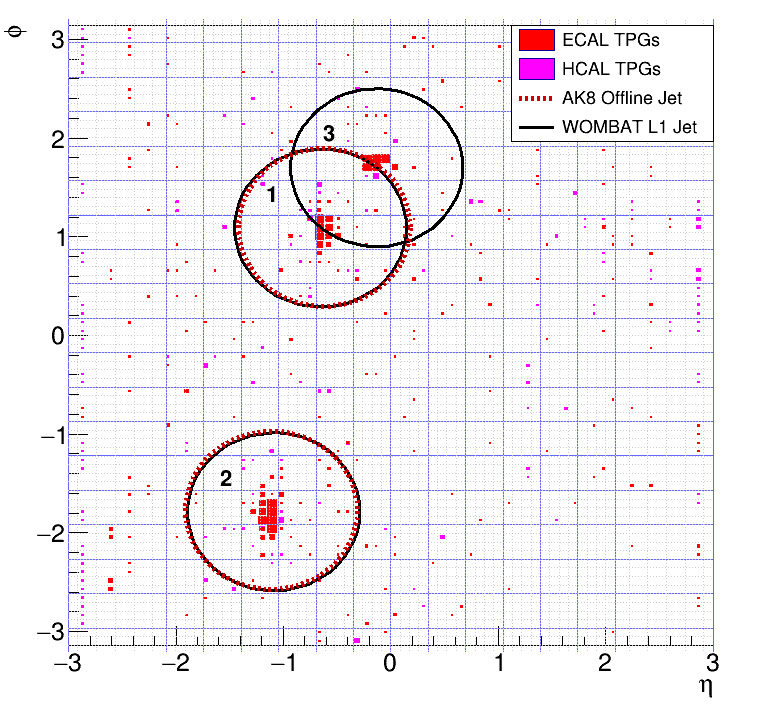} 
    \caption{W-MM Jet Multiplicity Mismatch TP Display - Event 4716}
    \label{fig:i1}
  \end{minipage}
\end{figure}

Although W-MM accurately predicts the jet multiplicity in most events, discrepancies remain. In Figure \ref{fig:i}, WOMBAT predicts two jets, while AK8 reconstruction identifies three, potentially reducing the trigger rate but also lowering efficiency. Conversely, Figure \ref{fig:i1} illustrates an event where WOMBAT predicts three jets despite only two being reconstructed offline, likely increasing trigger rate. Despite efforts to mitigate such mismatches through architectural and hyperparameter optimization, these inconsistencies persist in the trigger algorithm.

Overestimation of jet multiplicity often results from TP readouts that randomly mimic the calorimetric signature of a boosted $H\rightarrow b\bar{b}$ decay. Lacking full trigger tower (TT) granularity, WOMBAT struggles to resolve jet substructure as effectively as offline algorithms. Similarly, underestimation of jet multiplicity typically occurs in high-noise environments, where widespread calorimeter activity leads WOMBAT to misidentify relevant jets as noise, especially if the jet's center falls within the $|\eta|\geq 2.4$ region.

\subsubsection{WOMBAT Apprentice Model TP Displays}

Unlike W-MM, W-AM consistently predicts up to the second-leading $H\rightarrow b\bar{b}$ jets. While this limits trigger efficiency and rate, it ensures model simplicity compatible with FPGA deployment. Although a CNN architecture supports additional jet outputs, W-AM's limited trainable parameters proved insufficient to learn the latent features necessary for reliable three-jet predictions.

\begin{figure}[ht]
  \centering
  \begin{minipage}[b]{0.49\textwidth}
    \centering
    \includegraphics[width=\linewidth]{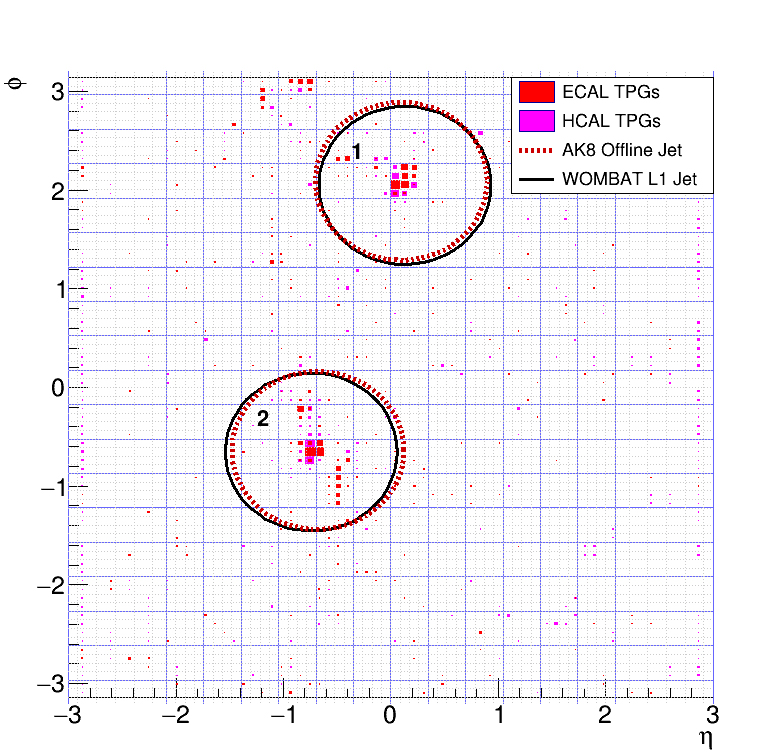} 
    \caption{W-AM Good Match TP Display - Event 3360}
    \label{fig:wam1}
  \end{minipage}
  \hfill
  \begin{minipage}[b]{0.49\textwidth}
    \centering
    \includegraphics[width=\linewidth]{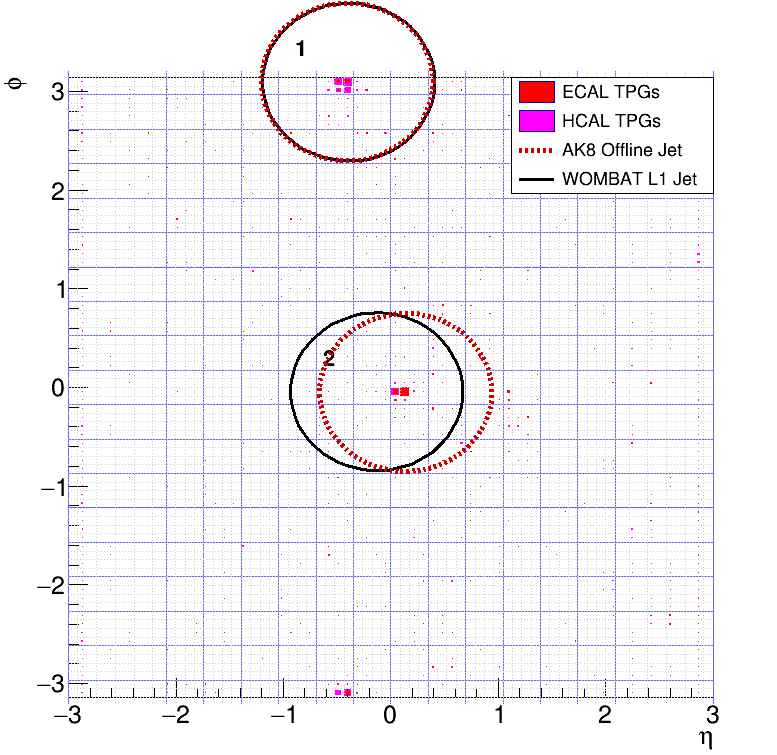} 
    \caption{W-AM Semi-Good Match TP Display - Event 1186}
    \label{fig:wam2}
  \end{minipage}
\end{figure}

Due to knowledge distillation from the larger WOMBAT Master model, W-AM learns the $\phi$-wrapping behavior despite lacking the custom $\phi$-wrapping layer. Figure \ref{fig:wam2} illustrates this behavior, with the first jet output correctly identifying an $H\rightarrow b\bar{b}$ event near $\phi\approx 2\pi$ (or $\phi\approx 0$). This suggests that W-AM captures angular periodicity implicitly. The learned representation generalizes well, even under architectural constraints.


Unlike W-MM, W-AM's predictions are more sensitive to variations in the underlying structure of the calorimeter TPs. As shown in Figure \ref{fig:wam2}, the predicted position of jet $2$ is noticeably displaced toward jet $1$ in $\eta$, which has a higher transverse momentum ($p_T = 558.6$ GeV) relative to jet $2$ ($p_T = 305.4$ GeV). Despite satisfying the $\Delta$R $< 0.4$ matching criterion, W-AM's second prediction is biased toward the more energetic jet due to this imbalance.


For comparison, Figure \ref{fig:wam1} presents a lower-$p_T$ event with jets $1$ and $2$ having $p_T$ of $244.9$ GeV and $183.4$ GeV, respectively. In this range ($\sim150-300$ GeV), increased substructure facilitates more accurate jet identification. However, as partons become increasingly collimated at higher $p_T$ (see Figure \ref{fig:wam2}), W-AM becomes more prone to misidentifying $H\rightarrow b\bar{b}$ decays, especially when there are large $p_T$ imbalances among the jets in an event. Furthermore, since W-AM operates on CaloLayer1 TP regions, which lack the granularity of TTs, its spatial resolution could be insufficient to resolve jet substructure in high-density environments. These phenomena, among others, contribute to reduced efficiency in the high-$p_T$ regime.

\begin{figure}[ht]
  \centering
    \begin{minipage}[b]{0.49\textwidth}
    \centering
    \includegraphics[width=\linewidth]{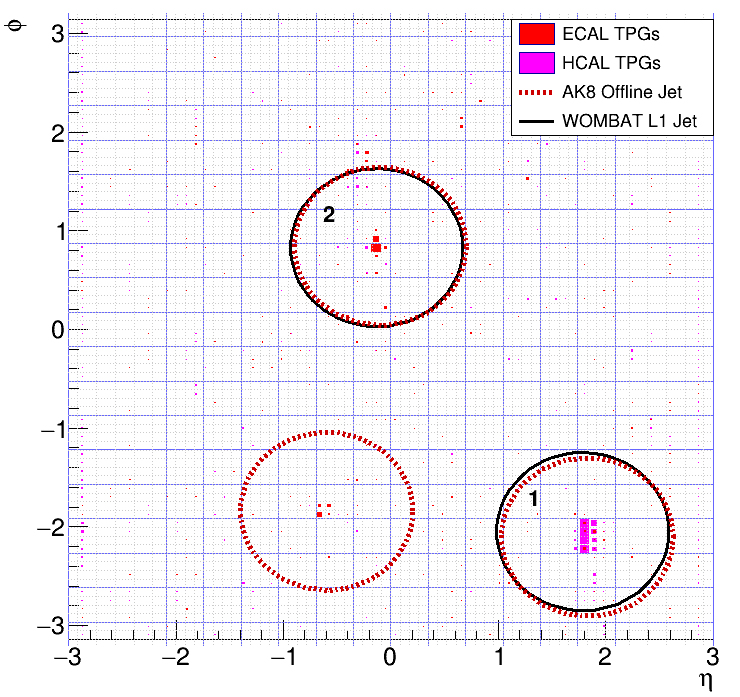} 
    \caption{W-AM Jet Multiplicity Mismatch TP Display - Event 830}
    \label{fig:mult2}
  \end{minipage}
  \hfill
  \begin{minipage}[b]{0.49\textwidth}
    \centering
    \includegraphics[width=\linewidth]{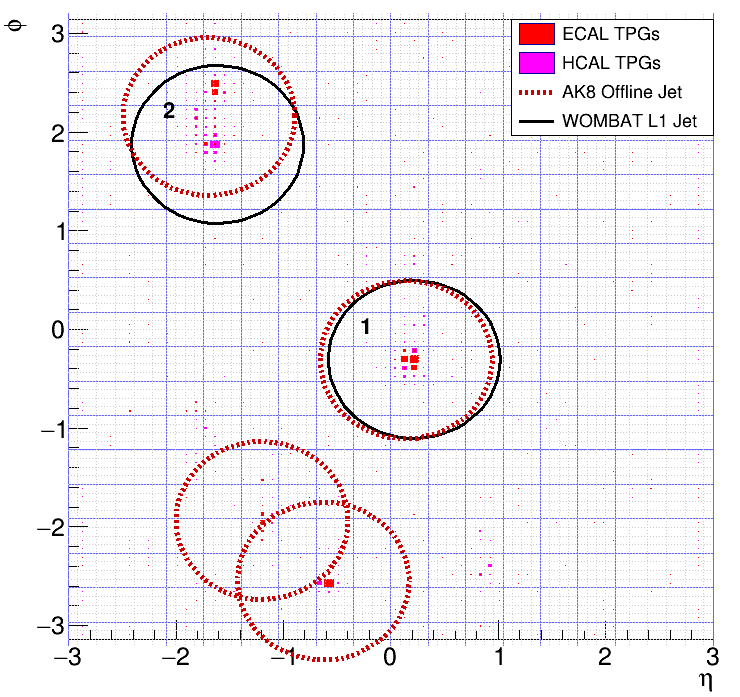} 
    \caption{W-AM Jet Multiplicity Mismatch TP Display - Event 2994}
    \label{fig:mult1}
  \end{minipage}
  \hfill
\end{figure}

The primary limitation on W-AM's trigger performance arises from its two-jet-per-event constraint. This is demonstrated in Figures \ref{fig:mult2} and \ref{fig:mult1}, which show events with three and four jet clusters, respectively. Although WOMBAT identifies two jet centers satisfying the $\Delta$R $< 0.4$ matching criterion in both cases, overall efficiency remains low due to the presence of additional unmatched jets. As detailed in Chapter V, Section 3, an analysis of jet multiplicity based on leading-order offline jet $p_T$ reveals a theoretical upper limit to W-AM's trigger efficiency, most pronounced in the high-$p_T$ regime.

\subsection{WOMBAT Rate Analysis}

WOMBAT performance is evaluated by comparing the rates of W-AM and W-MM to the Single Jet 180 trigger, as well as the JEDI algorithm (see Chapter~V, Section 5). Optimal L1T design aims to minimize rate while maximizing efficiency. Rates were derived from 2023 ZB data corresponding to an integrated luminosity of $0.64$ fb$^{-1}$. Using CRAB, events were processed through the WOMBAT trigger paths, and rates for W-AM, W-MM, Single Jet 180, and JEDI were recorded.

\begin{table}[ht]
    \centering
    \begin{tabular}{|l|c|}\hline
        \textbf{L1T Algorithm} & \boldmath{$p_T$ at $1$ kHz} \\
        \hline\hline
        Single Jet 180 & $187.4\pm 5.50$ GeV \\
        W-MM         & $146.8\pm 5.50$ GeV \\
        W-AM         & $140.4\pm 5.50$ GeV \\
        \hline
    \end{tabular}
    \caption{Summary of $p_T$ Values Associated with a $1$ kHz Trigger Rate}
    \label{tab:r}
\end{table}

\begin{figure}[ht]
  \centering
    \begin{minipage}[b]{0.49\textwidth}
    \centering
   \includegraphics[width=\linewidth]{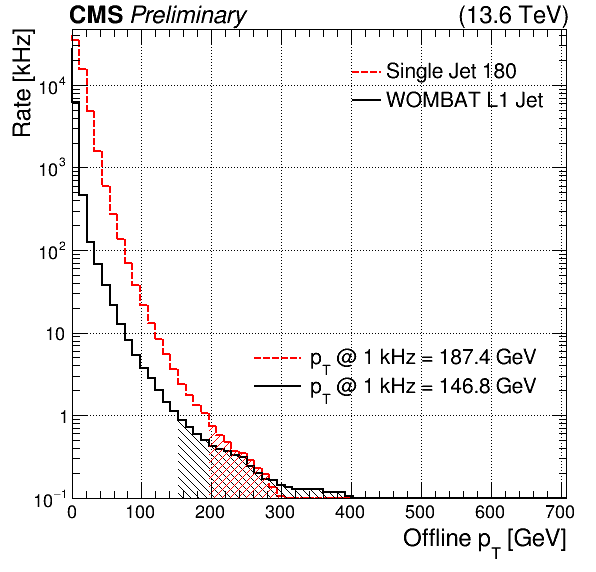}
    \caption{W-MM and Single Jet 180 Trigger Rate vs. Offline $p_T$ With $R(p_T)=1$ kHz Threshold}
    \label{fig:wmmrate}
  \end{minipage}
  \hfill
  \begin{minipage}[b]{0.49\textwidth}
    \centering
    \includegraphics[width=\linewidth]{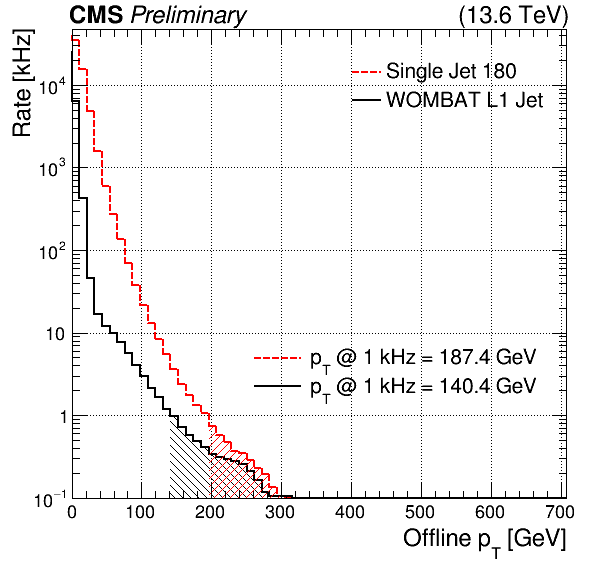}
    \caption{W-AM and Single Jet 180 Trigger Rate vs. Offline $p_T$ With $R(p_T)=1$ kHz Threshold}
    \label{fig:wamrate}
  \end{minipage}
  \hfill
\end{figure}

Figures \ref{fig:wmmrate} and \ref{fig:wamrate} present the trigger rates, $R(p_T)$, of W-MM and W-AM, respectively, in comparison to the Single Jet 180 algorithm. The shaded regions encompass events that fall below the comparison threshold of $1$ kHz. Both WOMBAT models demonstrate lower trigger rates than Single Jet 180 at this threshold, indicating improved background suppression. This reduction is particularly significant in the high-rate regime, where efficient rejection of less physics-relevant jets is essential for maintaining L1T (and DAQ) system performance.

The $1$ kHz threshold is chosen to reflect realistic per-trigger rate constraints within the CMS L1T architecture. While the total L1 bandwidth is on the order of $100$ kHz, individual trigger paths typically operate in the $1-10$ kHz to accommodate bandwidth sharing among multiple physics triggers and to preserve headroom for calibration and control paths. Evaluating WOMBAT against a $1$ kHz benchmark provides a practical and stringent test of its suitability for deployment in real-time systems constrained by latency, FPGA resource limits, and global rate ceilings.

The numerical results corresponding to the trigger rates of W-MM, W-AM, and Single Jet 180 are summarized in Table \ref{tab:r}. The table reports the jet $p_T$ at which each algorithm reaches a trigger rate of $1$ kHz.

WOMBAT models achieve the 1 kHz trigger rate at significantly lower jet $p_T$ thresholds compared to the Single Jet 180 algorithm. The Single Jet 180 requires a jet $p_T$ of $187.4$ GeV to stay within the imposed 1 kHz rate limit, whereas the WOMBAT variants W-MM and W-AM reach this rate at just $146.8$~GeV and $140.4$ GeV, respectively. This represents a reduction of $40.6$ GeV for W-MM and $47.0$ GeV for W-AM. These improvements highlight the enhanced background rejection capabilities of WOMBAT, allowing effective operation at lower $p_T$ while meeting the rate constraint.


As shown in Figure \ref{fig:wmmrate}, above $p_T\approx 300$ GeV, the W-MM rate exceeds that of the Single Jet 180 trigger. While Single Jet 180 drops below $10^{-1}$ kHz near $p_T= 300$ GeV, W-MM reaches this level around $400$ GeV. This discrepancy is partially due to W-MM's capacity to tag up to three boosted $H\rightarrow b\bar{b}$ candidates. In events with jet multiplicity mismatches, where W-MM predicts three jets while offline reconstruction identifies less (see Figure \ref{fig:i1}), the rate increases. Such over-predictions are more pronounced above $200$ GeV, where energetic jets generate complex TP patterns that may be misinterpreted by the ML model as additional Higgs-like jets. Unlike traditional algorithms, ML-based triggers are more sensitive to subtle features in the input, making them more prone to these classification ambiguities.

Since W-AM uses a fixed jet multiplicity of $2$, it yields a lower trigger rate than Single Jet 180 across all $p_T$ values. While reduced jet multiplicity is not inherently required to lower trigger rates, in this case, it is correlated with rate suppression for both W-AM and W-MM. While there are multiple strategies for reducing trigger rates, such as adjusting selection thresholds, applying tighter isolation, or incorporating refined object definitions, constraining multiplicity proves effective in the WOMBAT models. Unlike W-MM, W-AM does not overpredict jet counts in the $p_T>200$ GeV regime, leading to fewer false positives. In terms of rate alone, W-AM is the most selective. However, the trigger rate does not directly reflect signal acceptance. In this respect, W-MM and Single Jet 180 outperform W-AM, as the fixed multiplicity in W-AM can lead to underprediction of boosted $H\rightarrow b\bar{b}$ decays.


\subsection{WOMBAT Efficiency Analysis and Jet Multiplicity Distribution}\label{wameff}

While rate comparisons provide insight into background suppression, they do not fully capture a trigger's physics performance. Signal efficiency, $\epsilon(p_T)$, is a critical complement to rate in evaluating L1T algorithms. For $H\rightarrow b\bar{b}$ tagging, the efficiency curve quantifies a trigger's ability to correctly identify jets as a function of jet $p_T$, while the rate reflects the frequency at which events are accepted in a realistic collision environment. Given the low production cross-section of $H\rightarrow b\bar{b}$ relative to QCD multijet backgrounds, efficiency is evaluated using MC signal samples, as detailed in Chapter III, Section 1.

{
\begin{table}[ht]
    \centering \footnotesize
    \begin{tabular}{|l|c|c|c|}\hline
        \textbf{L1T Algorithm} & $p_T$ Threshold & \boldmath{$\epsilon(p_T)$ at $R(p_T)=1$ kHz} & \boldmath{$\epsilon(p_T)$ at $R(p_T)=1$ kHz} \\
      &  & \boldmath{Condition: $\Delta$R $<0.4$} & \boldmath{Condition: $\Delta$R $<0.8$}\\
        \hline\hline
        Single Jet 180 & $187.4\pm5.50$ GeV &$0.91^{+0.03}_{-0.04}$ & $0.95^{+0.02}_{-0.03}$\\
        W-MM      &  $146.8\pm5.50$ GeV  & $0.71^{+0.05}_{-0.05}$& $0.89^{+0.03}_{-0.04}$\\
        W-AM     & $140.4\pm5.50$ GeV  & $0.53^{+0.06}_{-0.06}$& $0.73_{-0.05}^{+0.05}$\\
        \hline
    \end{tabular}
    \caption{\centering Summary of $p_T$ Values Associated with a $1$ kHz Trigger Rate on Full Evaluation Dataset}
    \label{tab:e}
\end{table}

}

\begin{figure}[ht]
    \centering
    \includegraphics[width=0.6\linewidth]{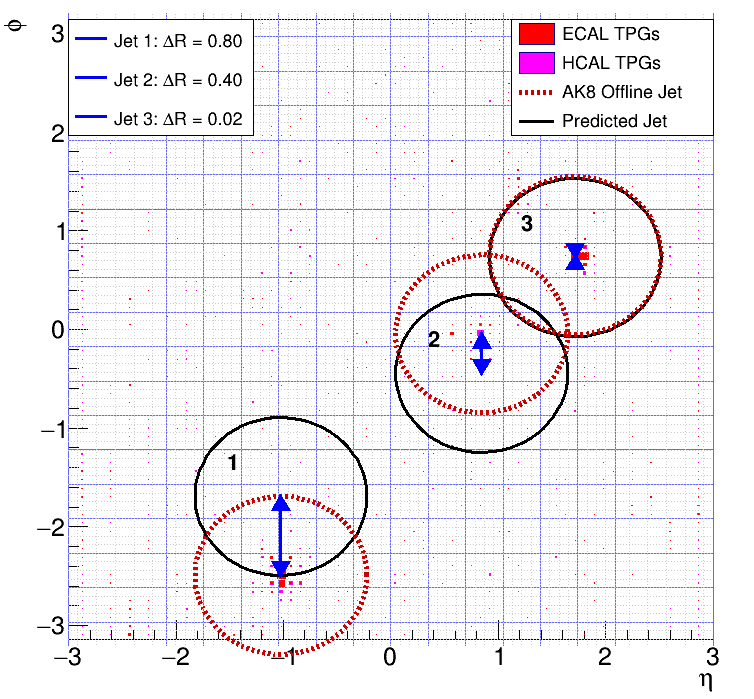}
    \caption{$\Delta$R Matching Condition Visualization for $\Delta$R Separations of 0.80, 0.40, and 0.02}
    \label{fig:dr}
\end{figure}

Table \ref{tab:e} shows a summary of the efficiency, $\epsilon(p_T)$, at the $p_T$ threshold associated with a rate of $1$ kHz for each algorithm. To demonstrate the effect of the $\Delta$R matching condition, the results when imposing $\Delta$R $<0.8$, in addition to the more stringent condition of $\Delta$R $<0.4$, are presented. Figure \ref{fig:dr} visualizes the $\Delta$R condition using three offline reconstructed jets with associated predictions at $\Delta$R $= {0.02, 0.40, 0.80}$. For this particular event, W-MM and W-AM predict all jets within $\Delta$R $<0.40$ (see Appendix A for Figures \ref{fig:wam99} and \ref{fig:wmm99}). For illustration purposes, the predicted jets shown were manually placed and do not reflect actual WOMBAT outputs.

As shown in Table \ref{tab:e}, both W-AM and W-MM demonstrate the capacity to accept lower-$p_T$ events compared to the baseline Single Jet 180 trigger when operating under the $R(p_T) \leq 1$ kHz rate constraint. While their absolute efficiencies near the threshold are lower, the presence of a non-zero $\epsilon(p_T)$ at reduced $p_T$ allows for extended coverage into kinematic regions that remain inaccessible to the Single Jet 180 trigger. This characteristic is particularly advantageous for capturing a wider spectrum of boosted $H\rightarrow b\bar{b}$ processes, especially those occurring below the $p_T$ threshold enforced by Single Jet 180.

The W-MM algorithm exhibits performance comparable to the Single Jet 180 trigger, due to its EDA architecture, which efficiently encodes global event-level TP features and jet substructure. Its capacity to predict up to three jets enables high correspondence with AK8 offline reconstructed jets. For a target rate of $1$ kHz, W-MM achieves a $p_T$ threshold of $146.8$ GeV, granting access to the $146.8-187.4$ GeV region, populated by hadronic W/Z decays, moderate-$p_T$ QCD jets, and boosted $H \to b\bar{b}$ events. This region remains inaccessible to Single Jet 180 under the same rate constraint.

\begin{figure}[H]
  \centering

    \begin{minipage}[b]{0.49\textwidth}
    \centering
    \includegraphics[width=\linewidth]{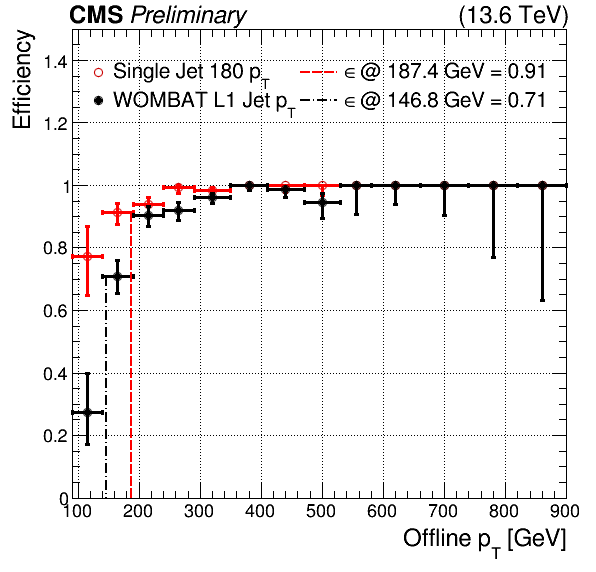}
    \caption{W-MM and Single Jet 180 Trigger Efficiency vs. Offline $p_T$ With $\epsilon(p_T)$ Threshold for $\Delta$R $<0.4$}
    \label{fig:wmmeff}
  \end{minipage}
  \hfill
  \begin{minipage}[b]{0.49\textwidth}
    \centering
    \includegraphics[width=\linewidth]{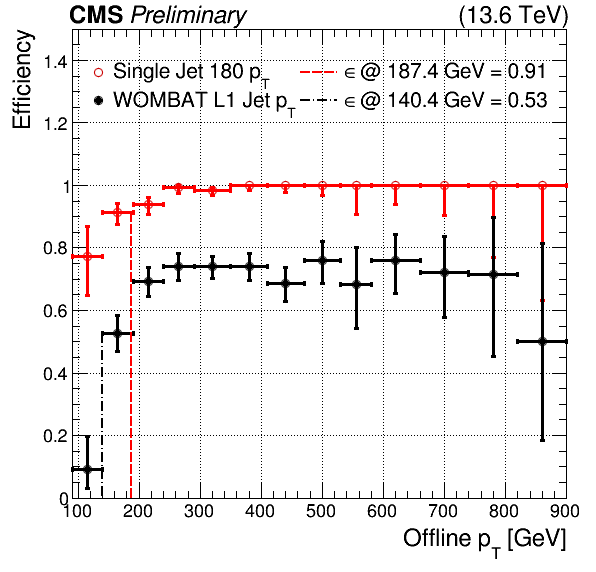}
    \caption{W-AM and Single Jet 180 Trigger Efficiency vs. Offline $p_T$ With $\epsilon(p_T)$ Threshold for $\Delta$R $<0.4$}
    \label{fig:wameff}
  \end{minipage}

  \vspace{0.5cm} 

    \begin{minipage}[b]{0.49\textwidth}
    \centering
    \includegraphics[width=\linewidth]{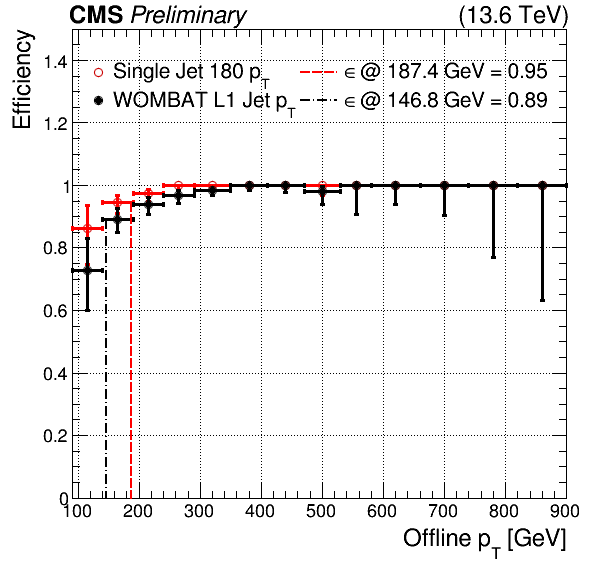}
    \caption{W-MM and Single Jet 180 Trigger Efficiency vs. Offline $p_T$ With $\epsilon(p_T)$ Threshold for $\Delta$R $<0.8$}
    \label{fig:wmmeff8}
  \end{minipage}
  \hfill
    \begin{minipage}[b]{0.49\textwidth}
    \centering
    \includegraphics[width=\linewidth]{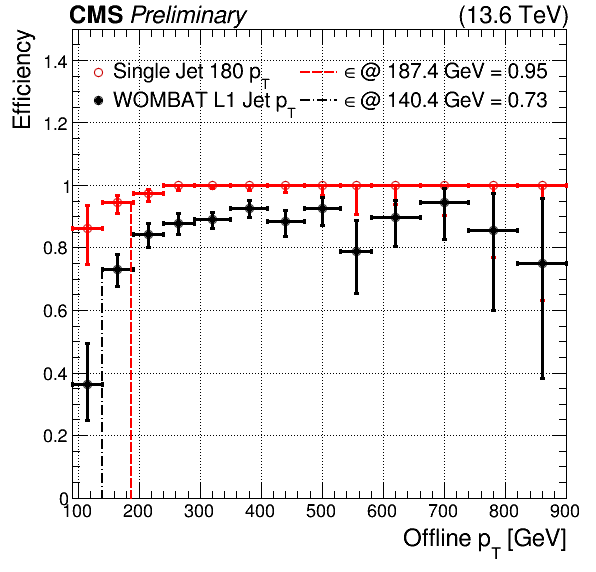}
    \caption{W-AM and Single Jet 180 Trigger Efficiency vs. Offline $p_T$ With $\epsilon(p_T)$ Threshold for $\Delta$R $<0.8$}
    \label{fig:wameff8}
  \end{minipage}

\end{figure}

\begin{figure}[ht]
    \centering
    \includegraphics[width=0.7\linewidth]{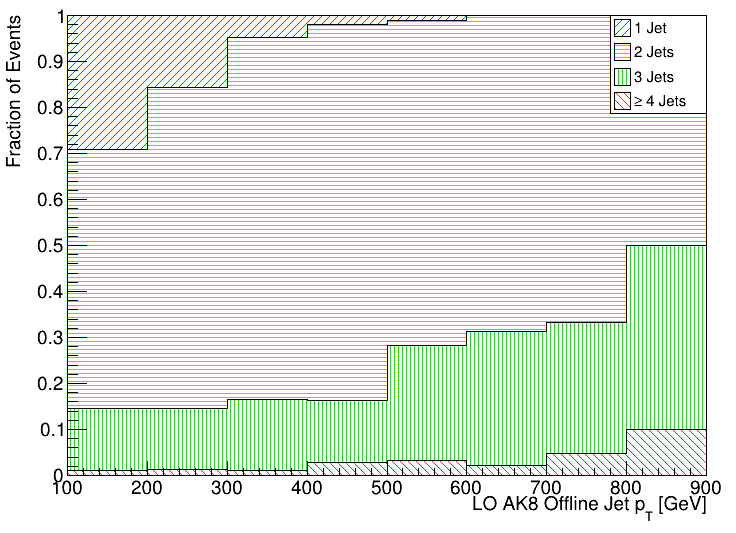}
    \caption{MC Evaluation Dataset Jet Multiplicity per Event Leading Order (LO) Jet $p_T$}
    \label{fig:jmult}
\end{figure}

The efficiency profile of the W-AM model, illustrated in Figure \ref{fig:wameff}, exhibits a peak at $\epsilon(p_T) \approx 0.75$, followed by a pronounced decline in the high transverse momentum regime ($p_T > 650$ GeV). This drop is mainly attributed to the rising jet multiplicity in the MC efficiency evaluation dataset, as demonstrated in Figure \ref{fig:jmult}. Specifically, for events containing leading order jets with $p_T > 600$~GeV, over $34.3$\% of events have a jet multiplicity $>2$. Since W-AM is architecturally constrained to predict exactly two jets per event, its performance deteriorates in scenarios where the reconstructed jet multiplicity exceeds this fixed topology. In such cases, the model can at best recover $\frac{2}{3}$ of the event content for three-jet topologies and only $\frac{1}{2}$ for events with four jets, thereby imposing a theoretical upper bound on efficiency. This limitation results in a systematic underperformance at high $p_T$, where multi-jet configurations become increasingly prevalent, causing the efficiency curve to decline rather than plateau. The effect is not due to a failure in inference per se, but rather a structural mismatch between the model's output dimensionality and the true event complexity in this kinematic regime.

Using the values from Figure \ref{fig:jmult}, the maximal efficiency in each $p_T$ bin is given by: 
\begin{gather}
    \epsilon_{max}(p_T) = \chi_1(p_T) + \chi_2(p_T) + \frac{2}{3}\chi_3(p_T) + \frac{1}{2}\chi_4(p_T),
\end{gather}
where $\chi_i(p_T)$ denotes the fraction of events with jet multiplicity $i$ in a given $p_T$ bin.

\begin{figure}[ht]
    \centering
    \includegraphics[width=0.65\linewidth]{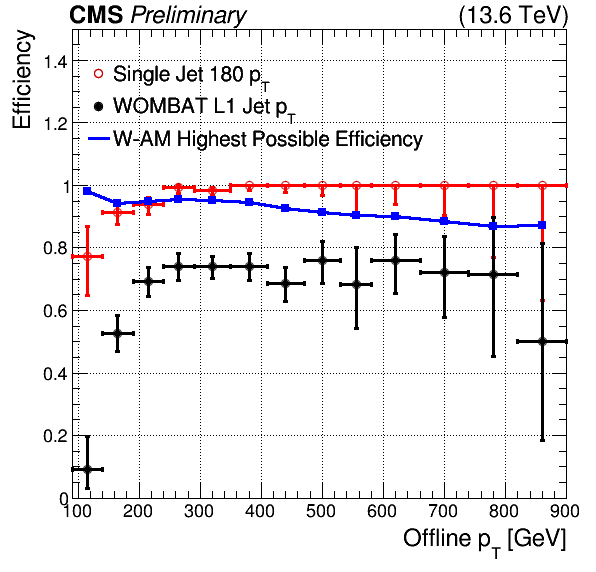}
    \caption{Efficiency Curve of W-AM and Single Jet 180 Compared to the Maximal Theoretical Efficiency for W-AM}
    \label{fig:jmulttheory}
\end{figure}

This expression represents the theoretical upper bound for W-AM efficiency, accounting for its dependence on jet multiplicity. This constraint explains the concave-down shape observed in the W-AM efficiency curve (Figure \ref{fig:wameff}). To illustrate this, Figure \ref{fig:jmulttheory} plots the corresponding upper bound. As this limit declines with increasing $p_T$, so too does W-AM's efficiency. Notably, for $p_T > 300$ GeV, even ideal W-AM predictions cannot match the efficiency of Single Jet 180, which is not subject to this multiplicity constraint. While W-MM also has a theoretical upper bound resulting from events with jet multiplicities of $4$, it proves insignificant given the fraction of $4$-jet events ranges from $0.004$ for $p_T = [100.0,140.0)$ GeV to $0.069$ for $p_T = [820.0,900.0)$ GeV.\footnote{Quantitatively, the most restrictive upper bound on W-MM arises in the $p_T = [820.0, 900.0)$ GeV bin, where the maximal efficiency is limited to $\epsilon_{\text{max}}(p_T) = 0.98$ due to the contribution of $4$-jet events. However, this constraint is not statistically significant, as it lies well within the uncertainty associated with the efficiency axis.}

Consequently, evaluating W-AM efficiency over the full dataset can obscure its true performance. To address this, Chapter V, Section 4, re-evaluates all algorithms using only events with exactly two jets.

\begin{figure}[ht]
    \centering
    \includegraphics[width=0.65\linewidth]{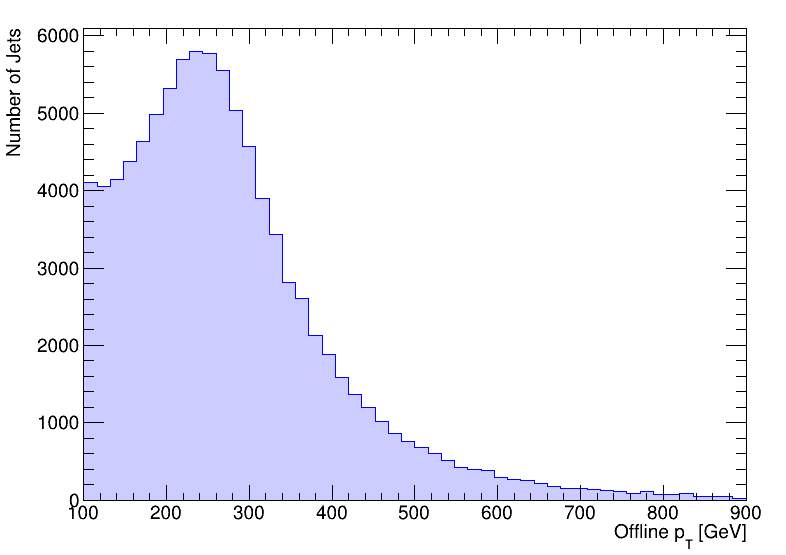}
    \caption{Training $H\rightarrow b\bar{b}$ MC Dataset Jet $p_T$ Distribution}
    \label{fig:enter-labe}
\end{figure}

Moreover, W-AM's high-$p_T$ efficiency degradation is compounded by the training and evaluation strategy, which intentionally prioritizes low-to-moderate $p_T$ regions to reflect the dominant phase space of LHC collisions. As shown in Figure \ref{fig:enter-labe}, the MC datasets are densely populated in the $150$-$350$ GeV range, aligning with regions where efficient trigger rate control is most crucial. While this ensures optimal performance in the most statistically relevant regions, it limits W-AM's exposure to, and generalization in, high-$p_T$ regimes with nontrivial jet substructure and multiplicity. Although both W-MM and W-AM were trained on this distribution, W-MM's larger parameter space enables it to capture high-$p_T$ features despite their rarity.


When the matching criterion is relaxed from $\Delta$R $< 0.4$ to $\Delta$R $< 0.8$, the W-AM model exhibits a similar overall efficiency trend, but with higher $\epsilon(p_T)$ values, as illustrated in Figure \ref{fig:wameff8}. This increase is expected, as the looser matching condition results in a greater number of WOMBAT-tagged jets being considered correctly matched. Similarly, W-MM and Single Jet 180 retain their characteristic efficiency profiles under the relaxed condition, but with slightly elevated $\epsilon(p_T)$. In practical terms, a $\Delta$R $< 0.8$ condition permits matches within approximately two CaloLayer1 TP regions from the offline-reconstructed jet center (see Figure \ref{fig:dr}), effectively broadening the spatial tolerance of the matching process.

While the $\Delta$R $< 0.8$ condition yields higher $\epsilon(p_T)$ values by allowing more WOMBAT tagged jets to be considered matched, it reflects a looser spatial association and is therefore less suitable for rigorous performance evaluation. In contrast, the original $\Delta$R $< 0.4$ criterion imposes a stricter correspondence, roughly aligning with the size of a single CaloLayer1 TP region around the offline jet center. This tighter matching enhances the sensitivity of the efficiency curve to spatial and energetic biases, providing a more accurate view of trigger behavior. Nevertheless, the $\Delta$R $< 0.8$ results remain informative for understanding model performance in contexts where broader spatial resolution or relaxed deployment conditions are relevant.

\subsection{WOMBAT Efficiency Analysis on Events with Fixed Jet Multiplicity of 2}\label{wameffjet2}

Given the theoretical efficiency constraints of W-AM (and of W-MM in the presence of $4$-jet topologies), performance was reevaluated using a subset of the original MC efficiency dataset restricted to events with exactly $2$ jets. This isolates the models' behavior under controlled conditions, eliminating the impact of mismatched jet multiplicities.

\begin{table}[ht]
    \centering\footnotesize
    \begin{tabular}{|l|c|c|c|}\hline
        \textbf{L1T Algorithm} & $p_T$ Threshold & \boldmath{$\epsilon(p_T)$ at $R(p_T)=1$ kHz} & \boldmath{$\epsilon(p_T)$ at $R(p_T)=1$ kHz} \\
      &  & \boldmath{Condition: $\Delta$R $<0.4$} & \boldmath{Condition: $\Delta$R $<0.8$}\\
        \hline\hline
        Single Jet 180 & $187.4\pm5.50$ GeV &$0.96_{-0.05}^{+0.06}$ & $1.00_{-0.02}^{+0.00}$\\
        W-MM      &  $146.8\pm5.50$ GeV  & $0.81_{-0.06}^{+0.05}$& $0.98^{+0.02}_{-0.03}$\\
        W-AM     & $140.4\pm5.50   $ GeV  & $0.53_{-0.07}^{+0.07}$& $0.85^{+0.03}_{-0.04}$\\
        \hline
    \end{tabular}
    \caption{\centering Summary of $p_T$ Values Associated with a $1$ kHz Trigger Rate on Subset of the Evaluation Dataset Containing Only Events with Jet Multiplicity of $2$}
    \label{tab:e2jet}
\end{table}

\begin{figure}[H]
  \centering

  \begin{minipage}[b]{0.49\textwidth}
    \centering
    \includegraphics[width=\linewidth]{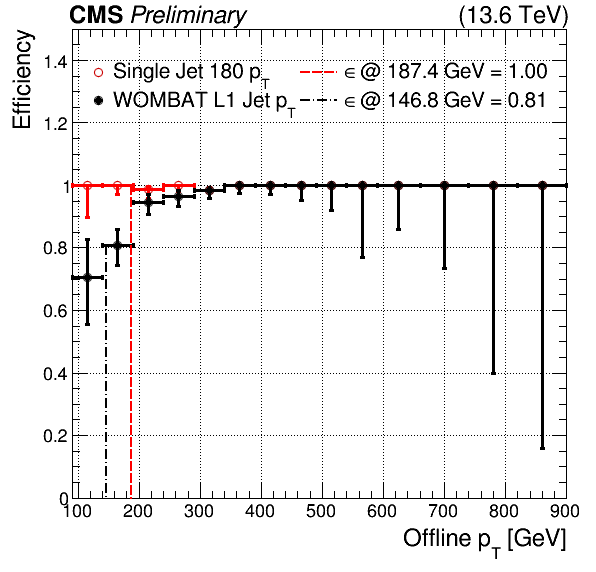}
    \caption{W-MM Trigger Efficiency vs. Offline $p_T$ on Events with Jet Multiplicity of 2}
    \label{fig:wmmeffjet2}
  \end{minipage}
  \hfill
  \begin{minipage}[b]{0.49\textwidth}
    \centering
    \includegraphics[width=\linewidth]{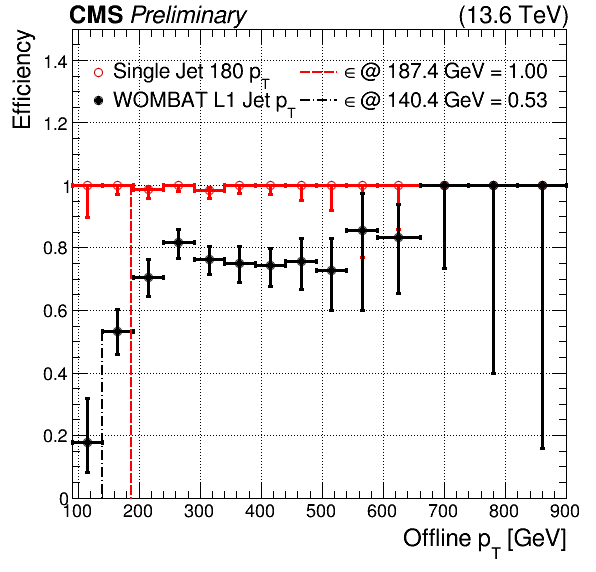}
    \caption{W-AM Trigger Efficiency vs. Offline $p_T$ on Events with Jet Multiplicity of 2}
    \label{fig:wameffjet2}
  \end{minipage}

  \vspace{0.5cm} 

  \begin{minipage}[b]{0.49\textwidth}
    \centering
    \includegraphics[width=\linewidth]{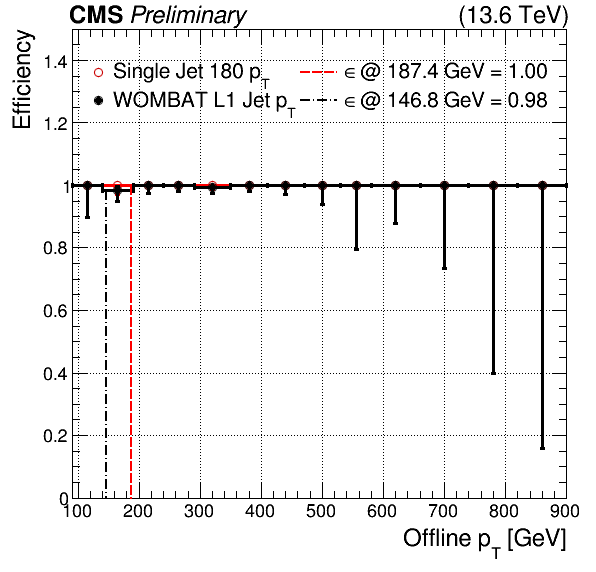}
    \caption{W-MM Trigger Efficiency vs. Offline $p_T$ on Events with Jet Multiplicity of 2 for $\Delta R < 0.8$}
    \label{fig:wmmeff8jet2}
  \end{minipage}
  \hfill
  \begin{minipage}[b]{0.49\textwidth}
    \centering
    \includegraphics[width=\linewidth]{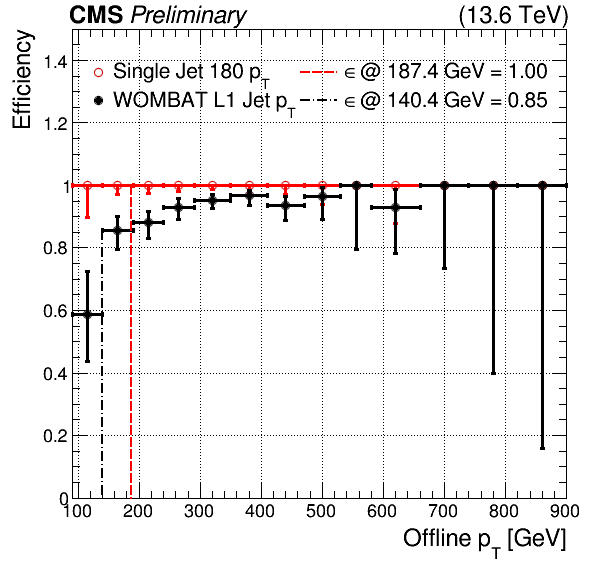}
    \caption{W-AM Trigger Efficiency vs. Offline $p_T$ on Events with Jet Multiplicity of 2 for $\Delta R < 0.8$}
    \label{fig:wameff8jet2}
  \end{minipage}

\end{figure}

The results in Table \ref{tab:e2jet} demonstrate improved performance of both W-AM and W-MM when evaluated on events containing only two $H\rightarrow b\bar{b}$ jets rather than the entire MC dataset. As with the previous efficiency study, the analysis was done following the formula outlined in Equation \ref{effformula}. Given the high efficiency of W-MM and Single Jet 180 when tested on the entire dataset, with $\epsilon(p_T>300\text{ GeV}) > 0.9$, improved performance is expected under reduced event complexity. This is confirmed in Figure \ref{fig:wameffjet2}, where Single Jet 180 maintains $\epsilon(p_T) \approx 1.0$ across the entire $p_T$ range, while W-MM reaches this efficiency at approximately $300$ GeV. W-MM's slightly lower efficiency at $p_T<300$ GeV leads to a suppressed rate, thereby decreasing the $p_T$ threshold required to remain within the $1$ kHz limit.

Figure \ref{fig:wameffjet2} illustrates an overall improvement in W-AM trigger efficiency, most notably at high $p_T$. Instead of the concave-down behavior seen in Figure~\ref{fig:wameff}, the W-AM trigger efficiency generally increases with an increase in $p_T$. This is because the high-$p_T$ events ($p_T>600$ GeV) were populated with $3$-jet and $4$-jet events, with $\approx 30\%-50\%$ of events being in this category, as shown in Figure \ref{fig:jmult}. While eliminating these events increases statistical uncertainty, it yields a more accurate performance metric for W-AM, which is constrained to two jets per event. 

Notably, no significant performance improvement is observed in the $p_T$ range of $300-500$ GeV, as the evaluation dataset in this region is dominated by $2$-jet events. Consequently, removing $3$- and $4$-jet events has minimal impact here, unlike in the $p_T > 600$ GeV range, which contains a higher proportion of high jet multiplicity events. The primary difference from the analysis in Figure \ref{fig:wameff} is the altered curve shape: rather than concave-down, it now rises at high $p_T$. This indicates that W-AM effectively identifies more well-defined high-$p_T$ jets --- an insight obscured in the full dataset analysis.

While W-AM does not surpass the Single Jet 180 trigger in overall efficiency, it offers complementary advantages that make it valuable in a combined trigger strategy. Its significantly lower trigger rate permits a reduced $p_T$ threshold, allowing it to capture low-to-moderate $p_T$ events that Single Jet 180 cannot within the $1$ kHz rate constraint. This makes W-AM particularly effective in extending coverage to regions otherwise excluded due to rate limitations. For high $p_T$ events, where W-AM's performance is limited by increasing jet multiplicity, a logical OR with Single Jet 180 would ensure efficient selection across a broader $p_T$ range without violating rate constraints.


Alternative evaluation strategies, such as restricting the $\eta$ or $\phi$ phase space, were investigated but introduced additional statistical uncertainty without substantially altering the efficiency curve. Example plots with constraints of $|\eta| < 2.4$ (W-AM's TP input boundary) and $|\phi| < 0.349$ radians (excluding the 0$^\text{th}$ and 17$^\text{th}$ CaloLayer1 TP regions) are shown in Appendix A. The negligible impact of these constraints suggests that W-AM detects jets at the edges of the TP grid with comparable accuracy to those located centrally. This is a favorable outcome, as it indicates no significant location-based bias influencing the trigger's efficiency.


To complement the analysis in Chapter V, Section 3, Figures \ref{fig:wmmeff8jet2} and \ref{fig:wameff8jet2} show the efficiencies of all algorithms evaluated on the 2-jet subset using a relaxed matching condition of $\Delta$R $< 0.8$. As before, algorithm efficiency increases with a looser matching threshold. Under this condition, W-AM achieves an efficiency above 0.90 for $p_T > 250$ GeV and exhibits a smoother efficiency curve. Notably, W-AM with $\Delta$R $< 0.8$ performs comparably to W-MM with the stricter $\Delta$R $< 0.4$ condition shown in Figure \ref{fig:wmmeffjet2}.

With the expanded matching criterion, both W-MM and Single Jet 180 reach near-unity efficiency across the full kinematic range. Although $\Delta$R $< 0.8$ still enforces close spatial proximity (approximately within two CaloLayer1 TP regions), stricter matching thresholds offer more discriminating insight into trigger behavior and inter-algorithm differences.

Therefore, while W-AM appears to perform well under these relaxed matching conditions, this alone does not provide definitive evidence of its competitiveness relative to other triggers. The looser matching threshold tends to elevate efficiency across all algorithms, reducing the ability to distinguish nuanced performance differences. For instance, the near-unity efficiency observed for both W-MM and Single Jet 180 in this setting offers limited insight into their sensitivity to jet $p_T$ or spatial resolution.

\subsection{Comparative Analysis of Trigger Rate and Efficiency for WOMBAT and JEDI}

Originally developed as a predecessor to WOMBAT, the JEDI algorithm served as a baseline for exploring whether ML-based L1T systems can surpass traditional rule-based designs when implemented on FPGAs. This section presents a comparative analysis of the trigger rates and selection efficiencies for both algorithms, evaluated over the full test dataset and a subset restricted to two-jet events. Since the W-MM model exceeds the resource constraints of the target FPGA, it is excluded from comparative analysis. Consequently, only the W-AM model is evaluated against JEDI for performance benchmarking.

\subsubsection{W-AM and JEDI Rate Analysis}

As summarized in Table \ref{tab:rJEDI}, W-AM continues to maintain the lowest $p_T$ threshold at a fixed rate of $R(p_T) = 1$ kHz among the FPGA-implemented algorithms. This is advantageous, as W-AM exhibits greater pileup resilience and enables the selection of lower $p_T$ jets under a set rate constraint. The new component of this analysis, the JEDI algorithm, achieves a lower $p_T$ threshold than Single Jet 180, but fails to meet W-AM's threshold of $140.4$ GeV. W-AM's low rate stems from its fixed low jet multiplicity, unlike JEDI, which allows up to 6 jets, or Single Jet 180, which has variable multiplicity. In the kinematic region of $p_T>300$ GeV, JEDI exhibits a higher rate than Single Jet 180, similar to the behavior of W-MM observed in Figure \ref{fig:wmmrate}. The elevated rate arises from JEDI and W-MM's tendency to over-predict jet multiplicities or tag higher-energy ZB events with multiple jets.

\begin{table}[ht]
    \centering
    \begin{tabular}{|l|c|}\hline
        \textbf{L1T Algorithm} & \boldmath{$p_T$ at $1$ kHz} \\
        \hline\hline
        Single Jet 180 & $187.4\pm 5.50$ GeV \\
        JEDI         & $150.2\pm 5.50$ GeV \\
        W-AM         & $140.4\pm 5.50$ GeV \\
        \hline
    \end{tabular}
    \caption{Summary of $p_T$ Values Associated with a $1$ kHz Trigger Rate for FPGA Implemented Algorithms}
    \label{tab:rJEDI}
\end{table}

\begin{figure}[ht]
    \centering
    \includegraphics[width=0.65\linewidth]{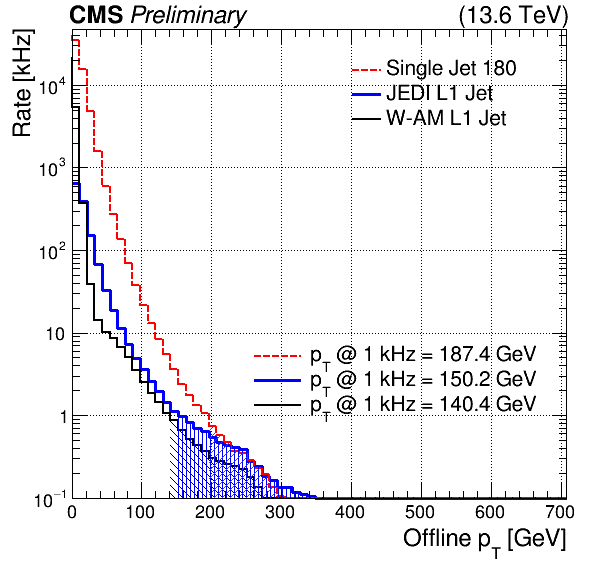}
    \caption{Rate vs Offline $p_T$ for W-AM, JEDI, and Single Jet 180 With Threshold At $R(p_T)=1$ kHz}
    \label{fig:wamjedirate}
\end{figure}

As shown in Figure \ref{fig:wamjedirate}, W-AM maintains a consistently lower rate than JEDI across all $p_T$ values. JEDI relies on fixed selection rules, making it sensitive to category definitions and unable to adapt to unmodeled data features. This rigidity can introduce systematic biases and reduce robustness to variations in jet topology. In contrast, W-AM's (as well as W-MM's) learned representations enable broader generalization. The sharp rate suppression in W-AM reflects its capacity to reject background without overfitting to specific patterns. This distinction is particularly relevant in high-pileup environments, where static thresholds are less effective. Furthermore, the adaptive nature of W-AM could support improved long-term stability under evolving detector conditions.

\subsubsection{W-AM and JEDI Efficiency Analysis}

As previously discussed, the theoretical constraint on W-AM's efficiency prevents it from achieving $\epsilon(p_T)\approx1$ at high jet $p_T$, even as jets become more collimated and background levels decrease. Given JEDI's fixed output of $6$ jets per event, it is able to efficiently capture all $H\rightarrow b\bar{b}$ decays in TPs with jet multiplicities between $3$ and $6$, which are dominant in the high-$p_T$ regime. As a result, full-dataset evaluations introduce a bias in favor of JEDI when assessing pure algorithmic performance. However, this evaluation remains essential, as it reflects realistic LHC conditions where high jet multiplicity events can occur and are beyond the capture capability of W-AM.

\begin{table}[ht]
    \centering
    \begin{tabular}{|l|c|c|}\hline
        \textbf{L1T Algorithm} & $p_T$ Threshold & \boldmath{$\epsilon(p_T)$ at $R(p_T)=1$ kHz}  \\
      &  & \boldmath{Matching Condition} $\Delta$R $<0.4$ \\
        \hline\hline
        Single Jet 180 & $187.4\pm5.50$ GeV &$0.91^{+0.03}_{-0.04}$ \\
        JEDI      &  $150.2\pm5.50$ GeV  & $0.23^{+0.05}_{-0.04}$\\
        W-AM     & $140.4\pm5.50$ GeV  & $0.53^{+0.06}_{-0.06}$\\
        \hline
    \end{tabular}
    \caption{\centering Summary of $p_T$ Values Associated with a $1$ kHz Trigger Rate on Full Evaluation Dataset for W-AM, JEDI, and Single Jet 180}
    \label{tab:ejedi}
\end{table}

As shown in Table \ref{tab:ejedi}, the W-AM algorithm achieves a trigger rate of $R(p_T)=1$ kHz at a lower transverse momentum threshold compared to JEDI. Furthermore, in the vicinity of this threshold, W-AM demonstrates a $0.3$ higher efficiency in the low-$p_T$ regime. These characteristics make W-AM more effective for tagging lower-$p_T$, boosted $H\rightarrow b\bar{b}$ jets. This performance advantage persists up to $p_T \lesssim 300$ GeV, within which W-AM consistently yields higher efficiency and lower trigger rates than JEDI. A contributing factor to JEDI's reduced efficiency is the super-region activity veto condition, as detailed in Chapter IV, Section 4, and summarized in Table \ref{bitp}. This veto, combined with stringent pileup mitigation cuts on $E_T$, imposes tight constraints that exclude lower-energy events. While these criteria limit the algorithm's sensitivity to less energetic signatures, they were deliberately designed to suppress high-rate QCD background processes, which dominate in this kinematic regime.

\begin{figure}[ht]
  \centering
    \begin{minipage}[b]{0.49\textwidth}
    \centering
    \includegraphics[width=\linewidth]{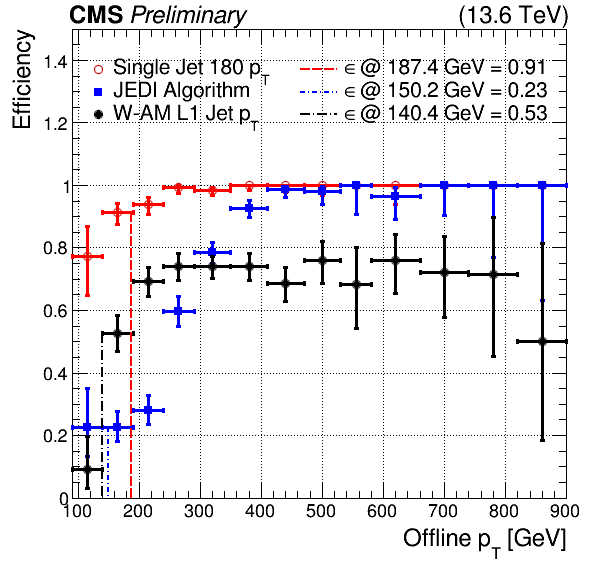}
    \caption{Trigger Efficiency vs. Offline $p_T$ for W-AM, JEDI, and Single Jet 180 Evaluated on Full Dataset $(\Delta$R $<0.4)$}
    \label{fig:wamjedieff}
  \end{minipage}
  \hfill
  \begin{minipage}[b]{0.49\textwidth}
    \centering
    \includegraphics[width=\linewidth]{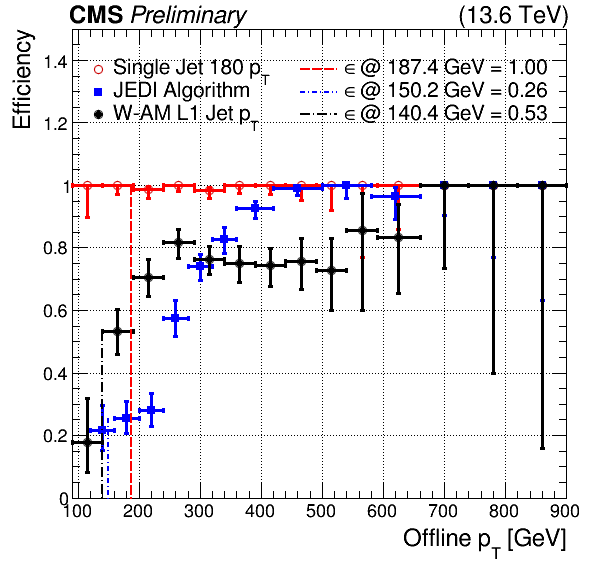}
    \caption{Trigger Efficiency vs. Offline $p_T$ for W-AM, JEDI, and Single Jet 180 Evaluated on Jet Multiplicity of 2 Events $(\Delta$R $<0.4)$}
    \label{fig:wamjedieff2}
  \end{minipage}
  \hfill
\end{figure}


In contrast, for $p_T > 300$ GeV, JEDI outperforms W-AM. This is evident in Figures \ref{fig:wamjedieff} and \ref{fig:wamjedieff2}, where both algorithms are evaluated on the full efficiency dataset and a subset constrained to events with exactly two jets. In the full dataset, JEDI asymptotically reaches an efficiency of $\epsilon(p_T) \approx 1.0$, whereas W-AM underperforms due to limitations in handling multi-jet topologies.

When the dataset is restricted to events with exactly 2 jets, W-AM exhibits improved efficiency in the high-$p_T$ regime. In contrast, JEDI's performance remains largely unaffected by this constraint, reflecting the stability of its rule-based, deterministic design. JEDI consistently evaluates the 6 leading jet candidates, regardless of total jet multiplicity. Its robustness stems from iterating over the full event space using a fixed $3\times3$ grid structure, computing energy sums, and applying predefined veto conditions on a per-candidate basis. As a result, the algorithm's response is minimally influenced by event-level complexity, handling both single-jet and multi-jet topologies in a uniform manner.



\subsection{FPGA Timing and Resource Usage Analysis}

All timing and utilization figures in this section are obtained after HLS synthesis but before placement-and-route
(P\&R) for the designated Xilinx Virtex-7 device. At this stage, Vitis HLS provides cycle-true latency and
initiation interval (II) reports, which remain unchanged after P\&R, as well as an estimated clock period and resource count that do not yet include routing delays or clock-tree overhead. For 7-series devices the HLS estimates are usually pessimistic: post-route clock periods are typically $10-30$\% shorter than the HLS report, and LUT utilization falls by roughly $20-40$\% once logic-level optimizations are applied during implementation \cite{ug}. These synthesis-only reports are therefore adequate for algorithmic comparison, as they provide consistent and conservative estimates that preserve relative performance and resource trends across design variants.

For online of a boosted $H\rightarrow b\bar{b}$ algorithm in the CMS L1T, the total processing latency must remain below $14$ clock cycles (CCs). Given that the designated L1T FPGA operates on a $160$ MHz clock, each CC corresponds to $6.25$ ns, resulting in a total allowable processing time of $14\times 6.25\text{ ns} = 87.5$ ns. For the trigger system to be consistent with the $40$ MHz bunch crossing rate at the LHC, a new event must be accepted into the processing pipeline every $25$ ns, which corresponds to an II of $4$ CCs. Moreover, the CMS L1T hardware ensures that the full set of input data for a single event, comprising calorimetric information, is available to the algorithm after $4$ CCs from the bunch crossing. This structure allows the algorithm to operate on complete event information with a fixed latency budget while maintaining alignment with the continuous event stream produced by the LHC.

As discussed in Chapter IV, Sections 5 and 6, the W-AM and JEDI algorithms were synthesized onto FPGA using HLS. Two implementations of W-AM were evaluated: one utilizing the \texttt{DATAFLOW} directive for parallelism through task-level pipelining, and another employing \texttt{PIPELINE} and \texttt{INLINE} pragmas to optimize function-level latency and resource reuse. As shown in Table \ref{tab:fpgaSUMMARYWAM}, none of the algorithms meet the L1T FPGA processing time requirement of $<87.5$ ns. Through the \texttt{DATAFLOW} implementation, W-AM manages to achieve the lowest latency of $22$ CCs, which results in a processing time of $137.5$, with respect to the target time per CC of $6.25$ ns. In contrast, the optimized \texttt{PIPELINE}+\texttt{INLINE} implementation achieves a minimal latency of $24$ CCs, indicating that the \texttt{DATAFLOW} approach provides better task parallelism and execution speed.

Compared to W-AM, the JEDI algorithm has a latency of $56$ CCs, which translates to a total processing time of $350.0$ ns. This is significantly higher than the target latency of $14$ CCs, making this algorithm much less optimal than W-AM for online L1T FPGA deployment. JEDI's computational complexity, which stems from dynamic pileup estimation, energy summing over sliding windows, and bitonic sorting of jet candidates, significantly increases execution latency. The extensive reliance on LUTs for pileup correction and veto logic further strains FPGA resources, particularly evident in higher LUT consumption compared to W-AM implementations. While the JEDI system has higher trigger efficiency, its high resource demands limit its suitability for real-time applications under strict L1T constraints.

While the upper bound of the timing uncertainty for all algorithms exceeds the $6.25$ ns target, the average time per clock cycle remains well below the threshold at $5.79$ ns (or $5.76$ ns for the \texttt{PIPELINE}+\texttt{INLINE} implementation) and $4.56$ ns for W-AM and JEDI, respectively. This indicates that the designs meet performance expectations under nominal conditions. Moreover, as noted earlier, post-route synthesis typically reduces clock periods by an additional $10$-$30$\% compared to the initial HLS estimates, further improving timing margins \cite{ug}. Taken together, these results indicate that, although the upper bounds of the timing per CC estimates exceed the $6.25$ ns target, the implementations still exhibit sufficiently low average CC times to support potential online deployment.

\begin{table}[ht]
\centering\footnotesize
\begin{tabular}{|l|c|c|c|p{3cm}|}
\hline 
\textbf{Algorithm} & Latency (CC) & II (CC)&
CC Estimate & Total Processing Time
For Target CC of $6.25$ ns\\
\hline\hline
W-AM (\texttt{PIPE+INLINE}) & 24 & {4} &  $5.76\pm 1.69$ ns & \parbox{\linewidth}{\centering $150$ ns} \\
W-AM (\texttt{DATAFLOW}) & 22 & {4} & $5.79\pm 1.69$ ns & \parbox{\linewidth}{\centering$137.5$ ns }\\
JEDI                     & 56 & {4} & $4.56\pm 1.69$ ns & \parbox{\linewidth}{\centering$350$ ns} \\
\hline
\end{tabular}
\caption{Synthesis-Level Timing Summary}
\label{tab:fpgaSUMMARYWAM}
\end{table}

Table \ref{tab:fpgaSUMMARYRESOURCE} presents the FPGA resource utilization for each implementation. The key hardware resources reported include \cite{handbook}:
\begin{itemize}
    \item \textbf{Block Random Access Memory (BRAM)}: On-chip memory blocks embedded within FPGAs that typically provide a storage capacity of $18,432$ bits per block. These blocks offer configurable data widths and depths, support dual-port access, and are optimized for low-latency, high-bandwidth operations. They serve as local memory for storing intermediate computation results, buffering data streams, and facilitating efficient data exchange between logic modules. W-AM and JEDI are highly pipelined algorithms that use directives to enable function inlining and dataflow, thus avoiding temporary storage in dedicated memory blocks.
    \item \textbf{Digital Signal Processing (DSP) Blocks}: Specialized hardware units optimized for high-speed arithmetic operations such as multiplication, addition, and multiply-accumulate (MAC). In CNN implementations, DSP blocks are critical for executing convolutional kernels and matrix multiplications with high throughput, making use of pipelined architectures for efficient fixed-point or floating-point computations that accelerate both filtering and feature extraction processes. 
    \item \textbf{Flip-Flops (FF)}: Fundamental sequential logic elements that capture and store single-bit information on clock edges. They are critical for implementing registers, synchronizers, and pipeline stages in digital circuits. Key parameters such as setup time, hold time, and propagation delay determine the maximum operational frequency and reliability of timing in synchronous digital designs.
    \item \textbf{Look-Up Table (LUT)}: Configurable combinatorial logic components that implement arbitrary Boolean functions by mapping a set number of input values to predetermined outputs. They form the backbone of FPGA logic synthesis, enabling the implementation of efficient digital circuits. In addition to general-purpose logic, LUTs can be repurposed as fixed-function look-up tables for storing constants and precomputed values, such as those used in the JEDI algorithm for pileup mitigation.
    \item \textbf{Ultra Random Access Memory (URAM)}: High-density memory blocks provided in some FPGA architectures, designed for scenarios that demand large volumes of on-chip storage with high throughput. URAM offers a greater storage capacity per block compared to BRAM, making it suitable for applications requiring extensive data buffering and processing. In both W-AM and JEDI, URAM remains unused for the same reasons as BRAM: the algorithms rely on pipelining and dataflow optimizations to minimize the need for on-chip memory storage.
\end{itemize}

\begin{table}[ht]
    \centering\footnotesize
   \begin{tabular}{|l|c|c|c|c|c|}
\hline
      \textbf{L1T Algorithm} & {BRAM} & {DSP} & {FF} & LUT & URAM  \\
        \hline\hline
        W-AM (\texttt{PIPE}+\texttt{INLINE}) &  $0$\%&$10$\%&$4$\%&$19$\%&$0$\%\\
        W-AM (\texttt{DATAFLOW})      &  $0$\% &$11$\% & $4$\% & $20$\% & $0$\%\\
        JEDI    & $0$\%& $1$\% &$14$\%&$121$\%&$0$\%\\
        \hline
    \end{tabular}
    \caption{\centering Summary FPGA Resource Usage For W-AM and JEDI}
    \label{tab:fpgaSUMMARYRESOURCE}
\end{table}

The key difference between the JEDI and W-AM algorithms is in the utilization of DSP blocks and LUTs. W-AM maps quantized convolutions and dense layers to DSPs, maintaining low LUT usage ($\leq20$\%, Table \ref{tab:fpgaSUMMARYRESOURCE}) and achieving low latency ($22-24$ cycles) with sub-$6$ ns clocks (Table \ref{tab:fpgaSUMMARYWAM}). JEDI, by contrast, uses minimal DSPs ($1$\%) but extremely LUT-heavy control logic, leading to an estimated $121$\% LUT utilization during Vivado HLS synthesis. This value exceeds the device's physical capacity due to conservative overestimation by HLS, which does not account for optimizations applied during placement and routing. In practice, post-implementation resource usage is often reduced by $20-40$\%, making the design fit feasible. The inflated estimate reflects high logic density, not an unimplementable design, and highlights the trade-off: JEDI achieves a shorter clock period ($4.56$ ns) but incurs high latency ($56$ cycles), leading to a total processing time $\approx2.5$ times higher than W-AM's \texttt{DATAFLOW} implementation.


\subsection{Analysis Discussion: Comparative Assessment of L1T Algorithms}

The overall evaluation of the L1T trigger algorithms reveals a series of trade-offs when comparing physics performance with FPGA implementability. Table \ref{tab:physicsSummary} summarizes key physics performance metrics, while Table \ref{tab:fpgaSummary} focuses on the synthesis-level FPGA implementation results. In both tables, checkmarks indicate that the criteria are met.

\begin{table}[ht]
\centering \footnotesize
\begin{tabular}{|l|p{2.1cm}|p{1.5cm}|p{1.5cm}|p{1.5cm}|p{2.2cm}|}
\hline
\textbf{L1T Algorithm} &
Lowest $p_T$ at $1$ kHz &
Highest $\epsilon(p_T)$ for $p_T < 300$ GeV &
Highest $\epsilon(p_T)$ for $p_T > 300$ GeV &
Handles $\geq 3$ Jets &
FPGA Implemented \\
\hline\hline
Single Jet 180 &   \xmark  & tie & tie & \cmark & \cmark \\
W-MM           & \xmark  & tie & tie & \cmark &   \xmark          \\
W-AM           & \cmark  & \xmark &  \xmark   &    \xmark        & \cmark \\
JEDI           &   \xmark  &  \xmark  & \xmark & \cmark & \cmark \\
\hline
\end{tabular}
\caption{Trigger Physics Summary}
\label{tab:physicsSummary}
\end{table}

From a physics standpoint (Table \ref{tab:physicsSummary}), the Single Jet 180 trigger consistently achieves high efficiency both for low and high $p_T$ jets and also accommodates events with $\geq 3$ jets. However, its relatively higher $p_T$ threshold at $R(p_T)=1$~kHz limits the ability to capture lower energy jets.

It is important to note that $\epsilon(p_T)$ is defined as the fraction of correctly identified signal jets to the total number of true signal jets at each $p_T$ bin --- that is, it quantifies the fraction of true positive predictions. However, this metric does not capture false positives; an algorithm can achieve a high efficiency by accepting a large number of events, which include both true and false positives. In practice, although Single Jet 180 achieves one of the highest efficiencies, it does so at the expense of a comparatively high trigger rate. This implies that while it captures many true signal events, it also selects an inflated number of events that do not correspond to boosted $H\rightarrow b\bar{b}$ jets when compared to the WOMBAT and JEDI algorithms.

\begin{table}[ht]
\centering \footnotesize
\begin{tabular}{|l|p{1.1cm}|p{1.5cm}|p{1.6cm}|p{1.9cm}|p{1.9cm}|}
\hline
\textbf{L1T Algorithm} & II$= 4$ & Lowest Latency & Clock Period $\leq 6.25$ ns & LUT Usage $< 50$\% & DSP Usage $<20$\% \\
\hline\hline
W-AM (\texttt{PIPE+INLINE}) & \cmark &      \cmark      &  \cmark    & \cmark & \cmark \\
W-AM (\texttt{DATAFLOW})    & \cmark & \xmark &  \cmark     & \cmark & \cmark \\
JEDI                  & \cmark &   \xmark   & \cmark &    \xmark        & \cmark \\
\hline
\end{tabular}
\caption{Synthesis-level FPGA Implementation Summary}
\label{tab:fpgaSummary}
\end{table}

The goal of an optimal L1T system is to minimize the trigger rate (thus minimizing false positive jets tagged with L1A) while maximizing true positive efficiency. This balance is crucial for ensuring that the downstream processing and data acquisition systems are not overwhelmed while still retaining the maximum number of target physics events. 

WOMBAT's student-teacher framework is part of a two-pronged approach. On one hand, W-MM employs an EDA architecture to better extract jet substructure and distinguish true boosted  $H\rightarrow b\bar{b}$ jets from QCD background. This complexity enables W-MM to achieve efficiency nearly equivalent to that of the Single Jet 180 algorithm, yet at a significantly lower $p_T$ threshold, approximately $40.6$ GeV lower. Consequently, W-MM can operate at a notably lower rate, reflecting its enhanced discrimination ability and reduced false positive L1A tagging.

\begin{table}[ht]
\centering \footnotesize
\begin{tabular}{|p{2cm}|p{5cm}|p{6cm}|}
\hline
\textbf{Aspect} & \textbf{Physics Performance} & \textbf{FPGA Implementation} \\
\hline\hline
Efficiency &
High efficiency indicates a high true positive jet tagging rate. 
&
Simple, FPGA-compatible trigger systems may inflate efficiency by over-tagging, while resource-constrained CNNs can underperform due to limited substructure resolution.
 \\
\hline
$p_T$ Threshold (Rate) &
A low trigger rate implies fewer false positives, enabling a lower jet $p_T$ selection threshold.
&
Simple, FPGA-compatible designs with fixed multiplicity or strict cuts reduce rate but limit signal acceptance; permissive models raise efficiency but increase false positives.
\\
\hline
Model Complexity & Complex architectures improve jet discrimination and efficiency. & Higher complexity increases resource usage and latency, potentially violating L1T online processing constraints and making models unfeasible for FPGA implementation.  \\
\hline
Jet-Multiplicity Handling & Flexible jet multiplicity permits efficient capture of events with multiple jets. & Fixed jet multiplicity simplifies design, making ML models FPGA-compatible, and reduces rate but can miss genuine multi-jet events.  \\
\hline
Latency & Extremely low latency ($<14$ cycles) is critical for real-time triggering. & More complex algorithms typically incur higher latencies, often exceeding online limits. \\
\hline
Resource Utilization & Advanced detection schemes may require extensive logical resources to achieve high performance & Keeping resource usage (LUTs, DSPs) low often necessitates algorithm simplifications, limiting efficiency .  \\
\hline
\end{tabular}
\caption{Summary of Trade-offs in L1T Trigger Evaluation}
\label{tab:tradeoffsSummaryy}
\end{table}

On the other hand, W-AM was developed to meet the real-time constraints of FPGA implementation. This system represents a deliberate tradeoff between model complexity and hardware feasibility. To meet strict FPGA resource and latency constraints, W-AM employs a simplified architecture with a fixed jet multiplicity. This limits its ability to match the efficiency of Single Jet 180, especially in high-multiplicity or high-$p_T$ kinematic regions. However, this same simplicity results in a lower overall rate. Because W-AM is structurally constrained to predict only $2$ jets per event, it inherently selects fewer jets, leading to the lowest trigger rate of all systems evaluated, even if it occasionally misses legitimate signal jets.

In contrast to WOMBAT, the JEDI algorithm represents a more traditional, rule-based approach to jet tagging, relying on deterministic selection logic, super-region vetoes, and pileup mitigation through fixed energy thresholds. JEDI achieves higher efficiency than W-AM for $p_T > 300$ GeV and can handle high jet multiplicity events due to its fixed $6$-jet output structure. However, this same structure results in an elevated trigger rate, as the algorithm tends to tag background jets with L1A. From an implementation standpoint, JEDI meets FPGA initiation interval requirements and achieves the shortest estimated clock period among the algorithms tested. Yet, its high logical complexity, resulting from extensive LUT-based control logic, translates to excessive resource usage and a latency of $56$ cycles, more than twice that of W-AM and far exceeding the $14$-cycle maximum needed for online CMS L1T deployment. This places JEDI even further than W-AM from being a viable candidate for real-time tagging.

Table \ref{tab:tradeoffsSummaryy} distills the algorithm-specific trade-offs identified in this study, evaluated relative to the baseline Single Jet 180 trigger:

\begin{itemize}
    \item \textbf{W-MM}: Achieves high signal efficiency and reduced trigger rate through an expressive EDA-based architecture capable of capturing complex jet substructure. However, the design exceeds available FPGA logic and timing constraints, rendering it infeasible for real-time deployment on the designated FPGA device.\\
    {Trade-off: Performance vs.\ Hardware Feasibility}
    
    \item \textbf{W-AM}: Yields the lowest trigger rate due to a fixed low-multiplicity output and a compact CNN architecture compatible with FPGA resources. This simplification, however, results in reduced efficiency, particularly for high-$p_T$ and multi-jet events.\\
   {Trade-off: FPGA Compatibility vs.\ Physics Reach}
    
    \item \textbf{JEDI}: Offers moderate efficiency and rate via a rule-based design that is tunable and deterministic. While synthetically implementable on FPGA, its logic-heavy structure incurs high latency and lacks the capacity to adapt to unmodeled features in the input or changing detector conditions.\\
    {Trade-off: Tunability and Performance vs.\ Latency and Adaptability}
\end{itemize}

\newpage

\stepcounter{section}
\section*{Chapter {VI}: Conclusion, Future Prospects, and Acknowledgments}
\addcontentsline{toc}{section}{Chapter {VI}: Conclusion, Future Prospects, and Acknowledgments}
\setcounter{figure}{0}
\markboth{Chapter {VI}: Conclusion, Future Prospects, and Acknowledgments}{}

This thesis presents a systematic evaluation of ML-based L1T algorithms for the CMS detector, with a specific focus on boosted $H\rightarrow b\bar{b}$ jet tagging under Run 3 conditions. Two neural network-based models were developed for this study: W-AM, a quantized convolutional neural network designed to meet current FPGA resource and timing constraints, and W-MM, a more expressive EDA architecture aimed at improved jet substructure resolution. Both were benchmarked against a traditional rule-based JEDI algorithm and the standard Single Jet 180 trigger. W-AM and JEDI were implemented in Vitis HLS for the Virtex-7 XC7VX690T-2FFG1927I FPGA to assess real-time hardware feasibility.

The designated FPGA device, used in the present (Run 3) CMS L1T, imposes strict latency and resource constraints that limit the complexity of ML models suitable for real-time jet tagging. Within these limitations, WOMBAT demonstrates superior FPGA performance relative to traditional rule-based systems, such as JEDI, underscoring the potential of ML models to achieve higher accuracy with reduced computational overhead. While the lightweight W-AM model does not surpass the Single Jet 180 trigger in efficiency, it operates at a significantly lower trigger rate and approaches the required L1T processing timing. In contrast, W-MM achieves a lower rate with comparable efficiency across the full $p_T$ range, outperforming Single Jet 180 in terms of physics performance, although it exceeds current hardware resource limits.

With the CMS Phase-2 upgrade on the horizon, these hardware limitations are expected to be significantly relaxed. The adoption of next-generation FPGA platforms, featuring expanded logic, memory, and DSP capabilities, will accommodate more sophisticated models at reduced timing costs. Under such conditions, architectures like W-MM are likely to become viable for online deployment. As a prototype system developed within the constraints of Phase-1, WOMBAT offers a forward-compatible foundation: its strong physics performance and FPGA compatibility position it as a compelling candidate for future L1T systems operating in the high pileup environment of the HL-LHC.

In particular, maintaining low trigger rates under high pileup conditions will require enhanced substructure discrimination, as the increased number of simultaneous collisions will exacerbate background contamination. The W-MM model demonstrates precisely this capability, offering both low rate and high efficiency, indicative of effective background rejection through learned substructure features. In contrast, Single Jet 180 incurs a high rate due to its broad selection logic, while JEDI, by design, maintains low efficiency in the densely populated low-$p_T$ region to suppress the rate. Its rule-based architecture lacks the adaptability to learn substructure features, limiting performance in complex jet environments.

Although WOMBAT was not tested in a live CMS trigger environment as the development timeline did not align with the operational schedule of the experiment, the simulation-based evaluation presented in this thesis strongly supports its future viability. The demonstrated ability of WOMBAT to exploit low-level calorimeter inputs for real-time jet substructure tagging lays the groundwork for ML-based L1T systems during Phase 2. Furthermore, with the forthcoming HGCAL upgrade providing increased spatial granularity, ML-based triggers like WOMBAT will have even greater access to fine-grained features, enabling improved substructure resolution and enhanced discrimination power within the stringent latency requirements of the L1T.

Future research will focus on refining the WOMBAT architecture, enhancing W-AM's physics performance within FPGA constraints, and investigating the feasibility of deploying W-MM through alternative hardware pipelines beyond the HLS4ML framework. Although the current WOMBAT architecture was developed in the context of the Run 3 detector and trigger system, it is envisioned as a prototype for deployment during Phase 2 of the CMS experiment. The upcoming LS3 period offers a critical window to adapt and extend WOMBAT, along with similar ML-based jet tagging algorithms, to align with the upgraded Phase 2 architecture.

Regardless of WOMBAT's future trajectory toward online implementation, the present study establishes a crucial foundation for ML-based jet tagging within the CMS L1T framework. At this stage, the methods, architectures, and evaluation strategies developed provide essential groundwork, establishing a technical foundation for the development and integration of sophisticated trigger algorithms anticipated in Phase 2 of the CMS experiment.

\newpage

{
\section*{Acknowledgments}


First and foremost, I extend my deepest gratitude to my advisor, Professor Isobel Ojalvo, and her postdoctoral researcher, Doctor Pallabi Das. Their guidance has been immeasurable, pushing me beyond every perceived boundary, inspiring my passion for physics, and laying the foundation for my academic future. 

I am profoundly thankful to Professors Olsen, Visnjic, and Abanin, whose firm belief in my potential has been a cornerstone of my academic growth. From my earliest experiences in physics under Professor Visnjic's mentorship, to Professor Olsen's continued support as my initial research mentor and thesis second reader, and Professor Abanin's invaluable guidance in both advanced coursework and research discussions, each has significantly shaped my trajectory and confidence as a physicist.

To my friends --- Deniz, Rafael, Bel, and David --- I owe you immense thanks for filling these past four years with laughter, support, and unforgettable memories. Your presence transformed challenges into joy, making my undergraduate years truly remarkable and cherished.

Above all, I express my heartfelt gratitude to my partner, Alex, whose love, encouragement, and support have profoundly impacted my life and academic pursuits. Your quiet strength and brilliance inspire me daily, continually renewing my love for physics, learning, and living.

My sincerest appreciation goes to my family, particularly my parents, whose endless love, sacrifices, and encouragement have been my greatest source of strength. Mom and Dad, your steadfast belief in me and unconditional support have guided me through every step of my journey --- I am endlessly grateful to have you.

Lastly, to my grandmother Mirjana, to whom I dedicate the deepest acknowledgment. You have been my guiding star, the inspiring force behind every ambition, and the unwavering believer in my potential. Everything I am and every milestone I have achieved is because of your profound influence and the boundless love with which you raised me. You are the heart of my journey, and all my work is forever dedicated to you.


}

\newpage
\appendix

\stepcounter{section}
\section*{Appendix {A}: Supplemental Figures}
\addcontentsline{toc}{section}{Appendix {A}: Supplemental Figures}
\setcounter{figure}{0}
\markboth{Appendix {A}: Supplemental Figures}{}
\renewcommand{\thefigure}{\Alph{section}.\arabic{figure}}

\subsection{Common Production Mechanisms of Higgs Bosons}\label{supap}

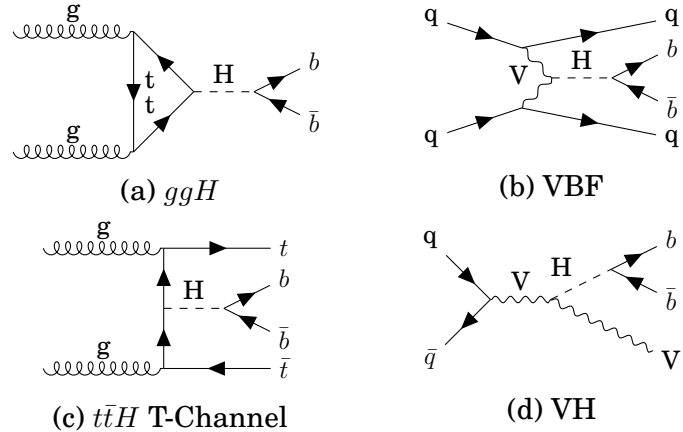
\begin{figure}[H]
  \centering

  \begin{minipage}{.3\textwidth}
    \centering
    \begin{tikzpicture}[scale=0.8, transform shape]
             \begin{feynman}
    \vertex (gluon1) at (0,0);
    \vertex (g1) at (2,0);
    \vertex (gluon2) at (0,-2);
    \vertex (g2) at (2,-2);
    \vertex (higgs) at (3,-1);
    \vertex (h) at (4,-1);
    \vertex (b) at (5,-0.5) {$b$};
    \vertex (antib) at (5,-1.5) {$\bar{b}$};

    \diagram* {
     (gluon1) -- [gluon, edge label=g] (g1),
     (gluon2) -- [gluon, edge label=g] (g2),
    (g1) -- [fermion] (g2),
    (g2) -- [fermion, edge label=t] (higgs),
    (higgs) -- [fermion, edge label=t] (g1),
    (higgs) -- [scalar, edge label=H] (h),
    (h) -- [fermion] (b),
    (antib) -- [fermion] (h),
    };
  \end{feynman}
    \end{tikzpicture}
    \subcaption{$ggH$}\label{ghh}
  \end{minipage} 
   \begin{minipage}{.3\textwidth}
    \centering
    \begin{tikzpicture}[scale=0.8, transform shape]
       \begin{feynman}
    \vertex (q1) at (0,0){q};
    \vertex (mid1) at (1.5,-0.5);
    \vertex (q2) at (4,0){q};
    \vertex (q11) at (0,-2){q};
    \vertex (mid11) at (1.5,-1.5);
    \vertex (q21) at (4,-2){q};
    \vertex (comb) at (2,-1);
    \vertex (h) at (3,-1);
    \vertex (b) at (4, -0.5) {$b$};
    \vertex (antib) at (4, -1.5) {$\bar{b}$};

    \diagram* {
     (q1) -- [fermion] (mid1);
     (mid1) -- [fermion] (q2);
     (q11) -- [fermion] (mid11);
     (mid11) -- [fermion] (q21);
     (mid11) -- [boson, edge label=V] (comb);
     (comb) --[boson] (mid1);
     (comb) -- [scalar, edge label = H] (h);
     (h) -- [fermion] (b);
     (antib) -- [fermion] (h);
    };
  \end{feynman}
    \end{tikzpicture}
    \subcaption{VBF}\label{vbf}
  \end{minipage}%

\begin{minipage}{.3\textwidth}
    \centering
    \begin{tikzpicture}[scale=0.8, transform shape]
             \begin{feynman}
    \vertex (gluon1) at (0,0);
    \vertex (g1) at (2,0);
    \vertex (gluon2) at (0,-2);
    \vertex (g2) at (2,-2);
    \vertex (higgs) at (2,-1);
    \vertex (t) at (4,0) {$t$};
    \vertex (hbranch) at (3,-1);
    \vertex (antit) at (4,-2) {$\bar{t}$};
    \vertex (b) at (4,-0.5) {$b$};
    \vertex (antib) at (4, -1.5) {$\bar{b}$};

    \diagram* {
     (gluon1) -- [gluon, edge label=g] (g1),
     (gluon2) -- [gluon, edge label=g] (g2),
     (g2) -- [fermion] (higgs),
     (higgs) -- [fermion] (g1),
     (g1) -- [fermion, particle=] (t),
     (higgs) -- [scalar, edge label=H] (hbranch),
     (antit) -- [fermion] (g2),
     (hbranch) -- [fermion] (b),
     (antib) -- [fermion] (hbranch),
    };
  \end{feynman}
    \end{tikzpicture}
    \subcaption{$t\bar{t} H$ T-Channel}\label{htt}
  \end{minipage}
    \begin{minipage}{.3\textwidth}
    \centering
    \begin{tikzpicture}[scale=0.8, transform shape]
       \begin{feynman}
    \vertex (q1) at (0,0){q};
    \vertex (int) at (1,-1);
    \vertex (q2) at (0,-2){$\bar{q}$};
    \vertex (v) at (2,-1);
    \vertex (h) at (3,-0.5);
    \vertex (v2) at (4,-2) {V};
    \vertex (b) at (4,0){$b$};
    \vertex (antib) at (4,-1){$\bar{b}$};

    \diagram* {
     (q1) -- [fermion] (int);
     (int) -- [fermion] (q2);
     (int) -- [boson, edge label=V] (v);
     (v) -- [scalar, edge label=H] (h);
     (v) -- [boson] (v2);
     (h) -- [fermion] (b);
     (antib) -- [fermion] (h);
    };
  \end{feynman}
    \end{tikzpicture}
    \subcaption{VH}\label{vh}
  \end{minipage}%
  \small
  \caption{ Common Production Mechanisms of $H\rightarrow b\bar{b}$}{
  \small
  {Figure \ref{ghh}}
  illustrates the $ggH$ production mode. Figure \ref{vbf} refers to the Vector Boson Fusion production mode, whereas Figure \ref{vh} depicts the Higgs production in association with a Vector boson. Figure \ref{htt} shows Higgs production in association with a top quark-antiquark pair.}\label{supp}
\end{figure}

\subsection{Additional Event Displays}\label{appendisp}

\begin{figure}[H]
  \centering
    \begin{minipage}[b]{0.49\textwidth}
    \centering
    \includegraphics[width=\linewidth]{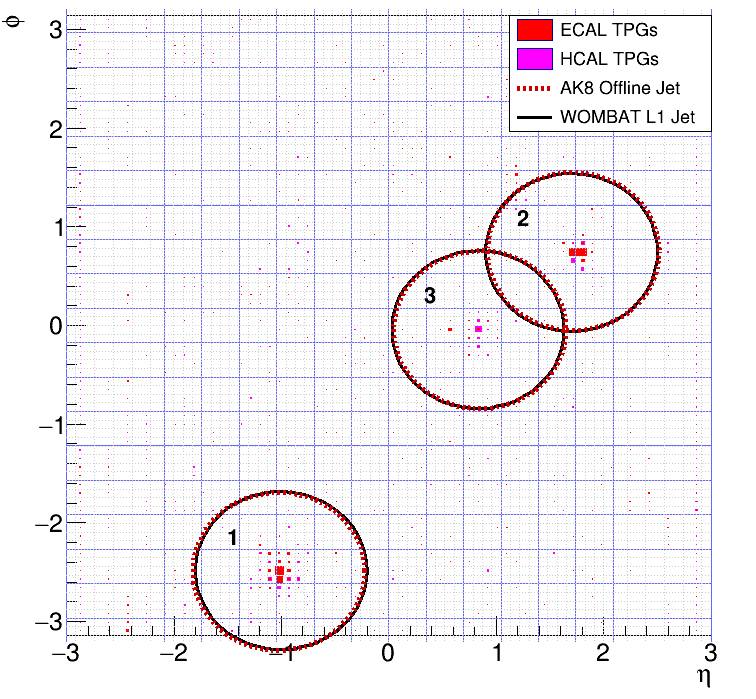}
    \caption{W-MM TP Display - Event 829}
    \label{fig:wmm99}
  \end{minipage}
  \hfill
      \begin{minipage}[b]{0.49\textwidth}
    \centering
    \includegraphics[width=\linewidth]{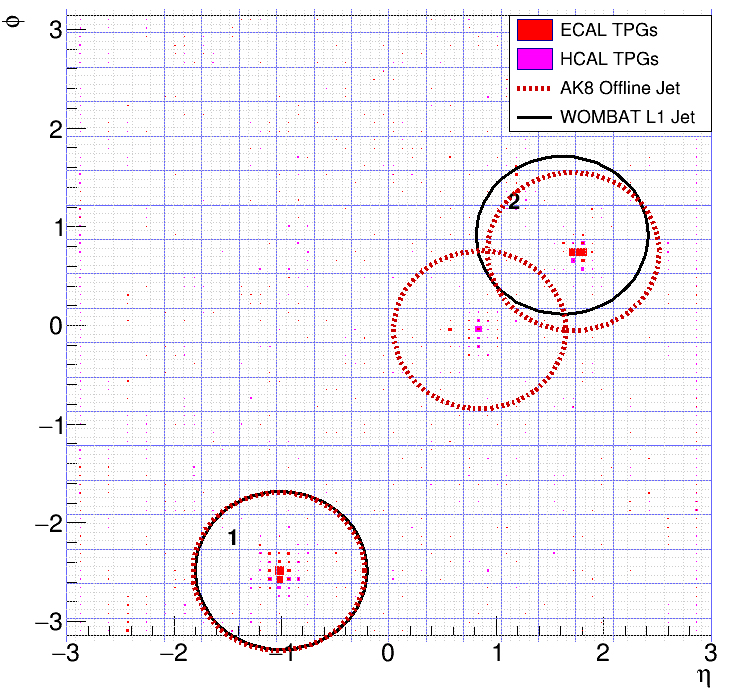}
    \caption{W-AM TP Display - Event 829}
    \label{fig:wam99}
  \end{minipage}
\end{figure}

\begin{figure}[ht]
  \centering
    \begin{minipage}[b]{0.49\textwidth}
    \centering
    \includegraphics[width=\linewidth]{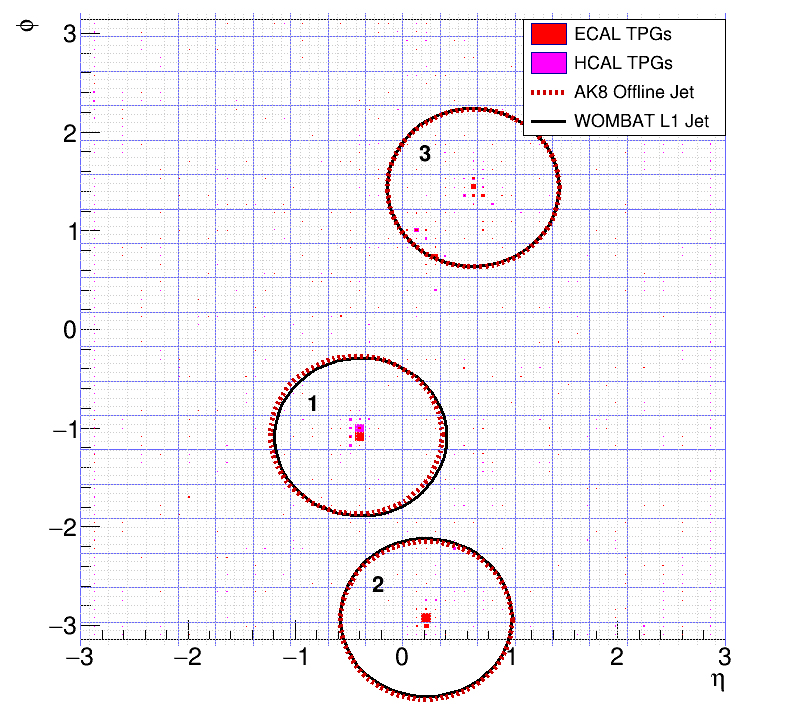}
    \caption{W-MM TP Display - Event 2549}
    \label{fig:wam99EXX}
  \end{minipage}
  \begin{minipage}[b]{0.49\textwidth}
    \centering
    \includegraphics[width=\linewidth]{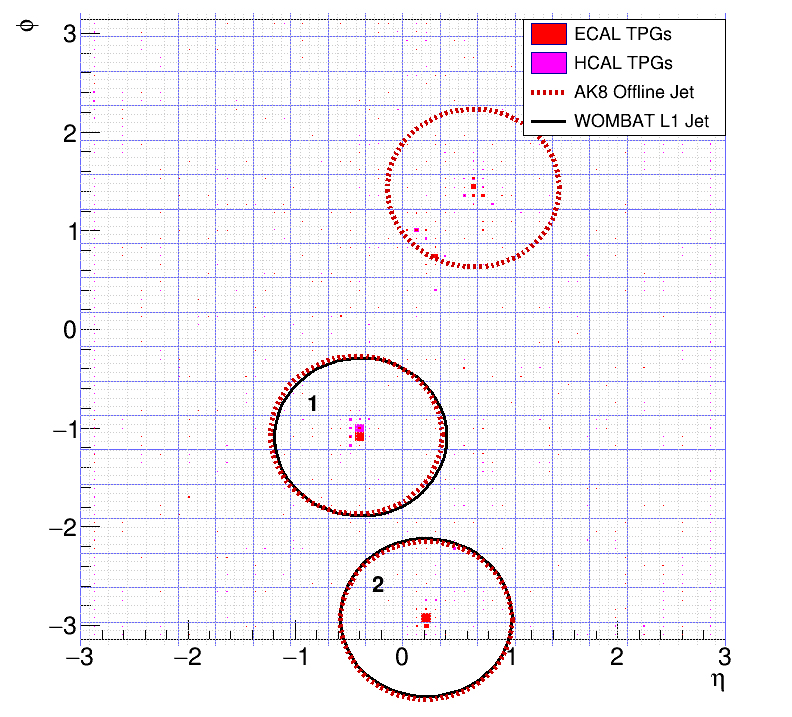}
    \caption{W-AM TP Display - Event 2549}
    \label{fig:wmm99EXX}
  \end{minipage}
  \hfill
\end{figure}

\subsection{Efficiency Analysis Implementing Space Constraints}\label{appeneffpe}

\begin{figure}[H]
  \centering
    \begin{minipage}[b]{0.49\textwidth}
    \centering
    \includegraphics[width=\linewidth]{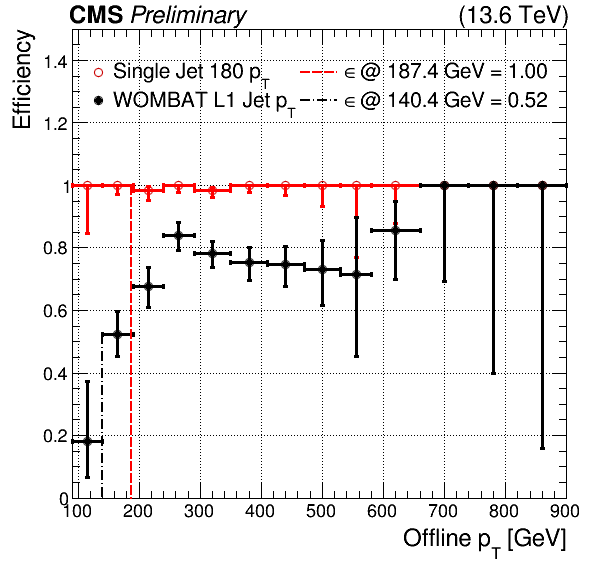}
    \caption{W-AM and Single Jet 180 $\epsilon(p_T)$ for $|\eta|<2.4$}
    \label{fig:wameff8jet2ETA}
  \end{minipage}
  \hfill
  \begin{minipage}[b]{0.49\textwidth}
    \centering
    \includegraphics[width=\linewidth]{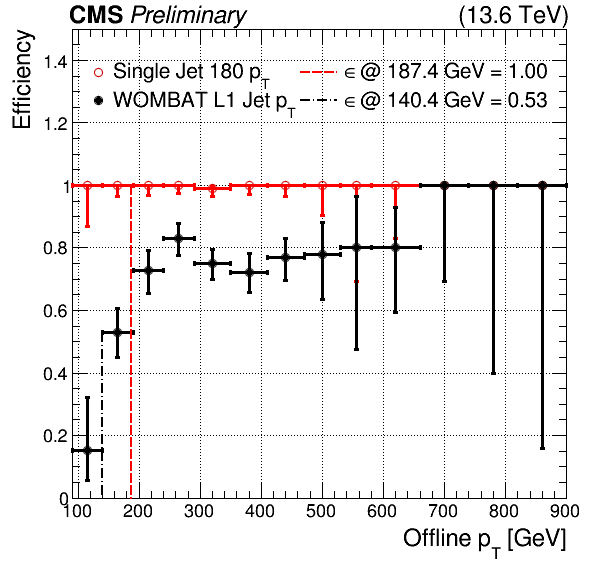}
    \caption{W-AM and Single Jet 180 $\epsilon(p_T)$ for $|\phi|<0.349$ Radians}
    \label{fig:wameff8jet2PHI}
  \end{minipage}
  \hfill
\end{figure}

\stepcounter{section}

\newpage

{
\singlespacing
\section*{Appendix \textbf{B}: Z Boson Mass Derivation: Higgs Mechanism Continuation}\label{zboson}
\addcontentsline{toc}{section}{Appendix \textbf{B}: Z Boson Mass Derivation: Higgs Mechanism Continuation}
\setcounter{figure}{0}
\markboth{Appendix \textbf{B}: Z Boson Mass Derivation: Higgs Mechanism Continuation}{}
\renewcommand{\thefigure}{\Alph{section}.\arabic{figure}}

The coupling of neutral gauge fields to the Higgs doublet can be described as \cite{chris}:
\begin{equation}
    \mathcal{L} = \frac{1}{4}\{(g'B_\mu Y_\Phi+gW_\mu^3\tau_3)\Phi  \}^\dagger (g'B_\mu Y_\Phi+g W^{3\mu}\tau_3)\Phi.
\end{equation}
Evaluating this at the vacuum expectation value of $\Phi$, $\Phi_{min}$, gives:
\begin{gather}
    \mathcal{L} = \frac{v^2}{8}\begin{bmatrix}
        W_\mu^{3\dagger}&B_\mu^\dagger
    \end{bmatrix}\begin{bmatrix}
        g^2(\tau_3^{\langle\Phi\rangle})^2 & gg'Y_\Phi\tau_3^{\langle\Phi\rangle}\\
        gg'Y_\Phi \tau_3^{\langle\Phi\rangle}& g'^{2}Y_\Phi^2
    \end{bmatrix}\begin{bmatrix}
        W^{3\mu}\\
        B^\mu
    \end{bmatrix}
\end{gather}

Where the mass-squared matrix can be identified as:
\begin{equation}
    M^2 = \frac{v^2}{4}\begin{bmatrix}
        g^2(\tau_3^{\langle\Phi\rangle})^2 & gg'Y_\Phi\tau_3^{\langle\Phi\rangle}\\
        gg'Y_\Phi \tau_3^{\langle\Phi\rangle}& g'^{2}Y_\Phi^2
    \end{bmatrix}
\end{equation}

Defining a unitary transformation of the sort:
\begin{gather}\label{uni}
    U = \frac{1}{\sqrt{g^2(\tau_3^{\langle\Phi\rangle})^2+g'^{2}Y_\Phi^2}}\begin{bmatrix}
        g'Y_\Phi&-g\tau_3^{\langle\Phi\rangle}\\
        g\tau_3^{\langle\Phi\rangle}& g'Y_\Phi
    \end{bmatrix},
\end{gather}
and setting the lower component of the Higgs field to be electrically neutral, i.e. $\frac{1}{2}\tau_3^{\langle\Phi\rangle}=-\frac{1}{2}$ and $Y_\Phi=1$, the diagonalized matrix becomes \cite{chris}:
\begin{equation}\label{mass}
    M_{Diag}^2 = UM^2U^{-1}=\begin{bmatrix}
        0&0\\
        0& \frac{v^2}{4}(g^2+g'^{2})
    \end{bmatrix}.
\end{equation}

The zero-mass eigenvalue of this matrix corresponds to the four-potential that results from \textit{Eq.}\ref{uni}. The gauge interaction is given by:
\begin{equation}
    eQ =  \frac{1}{\sqrt{g^2(\tau_3^{\langle\Phi\rangle})^2+g'^{2}Y_\Phi^2}} \Big(  Y_\Phi\frac{\tau_3}{2}-\frac{\tau_3^{\langle\Phi\rangle}}{2}Y\Big).
\end{equation}
In this interaction, by definition, the coupling to the Higgs field is 0. $Q$ is commonly known as the electric charge operator. $A^\mu$ is defined as the massless four-potential, known as the photon, which is a consequence of the choice for Higgs VEV (which was taken to be minimal). From \textit{Eq.}\ref{mass}, the non-zero eigenvalue is the squared mass of the $Z$ boson, $M_Z$:
\begin{equation}
    M_Z^2 = \frac{v^2}{4}(g^2+g'^{2}) \equiv \frac{M_{W^\pm}^2}{\cos^2\theta_W}
\end{equation}

}

\stepcounter{section}

\newpage

\section*{Appendix {C}: Detector Geometry}\label{geom}
\addcontentsline{toc}{section}{Appendix {C}: Detector Geometry}
\setcounter{figure}{0}
\markboth{Appendix {C}: Detector Geometry}{}
\renewcommand{\thefigure}{\Alph{section}.\arabic{figure}}

This section will briefly cover some basic definitions, as well as a schematic representation of the Phase 2 ECAL barrel region, which is closely related to the formatting of the TP output used as a sample in the study. From the center of the LHC ring, the CMS detector is located North. The coordinate system used is defined as follows:
\begin{itemize}
    \item \textbf{x-axis:} horizontal, pointing towards the center of the LHC;
    \item \textbf{y-axis:} vertical, pointing upwards;
    \item \textbf{z-axis:} horizontal, pointing in the beam direction;
    \item $\phi$\textbf{:} defined as $0^{\circ}$ in the x-axis and $90^{\circ}$ in the y-axis;
    \item $\eta$\textbf{:} $0^\circ$ in the x-y plane, positive in $+z$ and negative in $-z$.
\end{itemize}

\begin{figure}[ht]
    \centering
    \includegraphics[width=\linewidth]{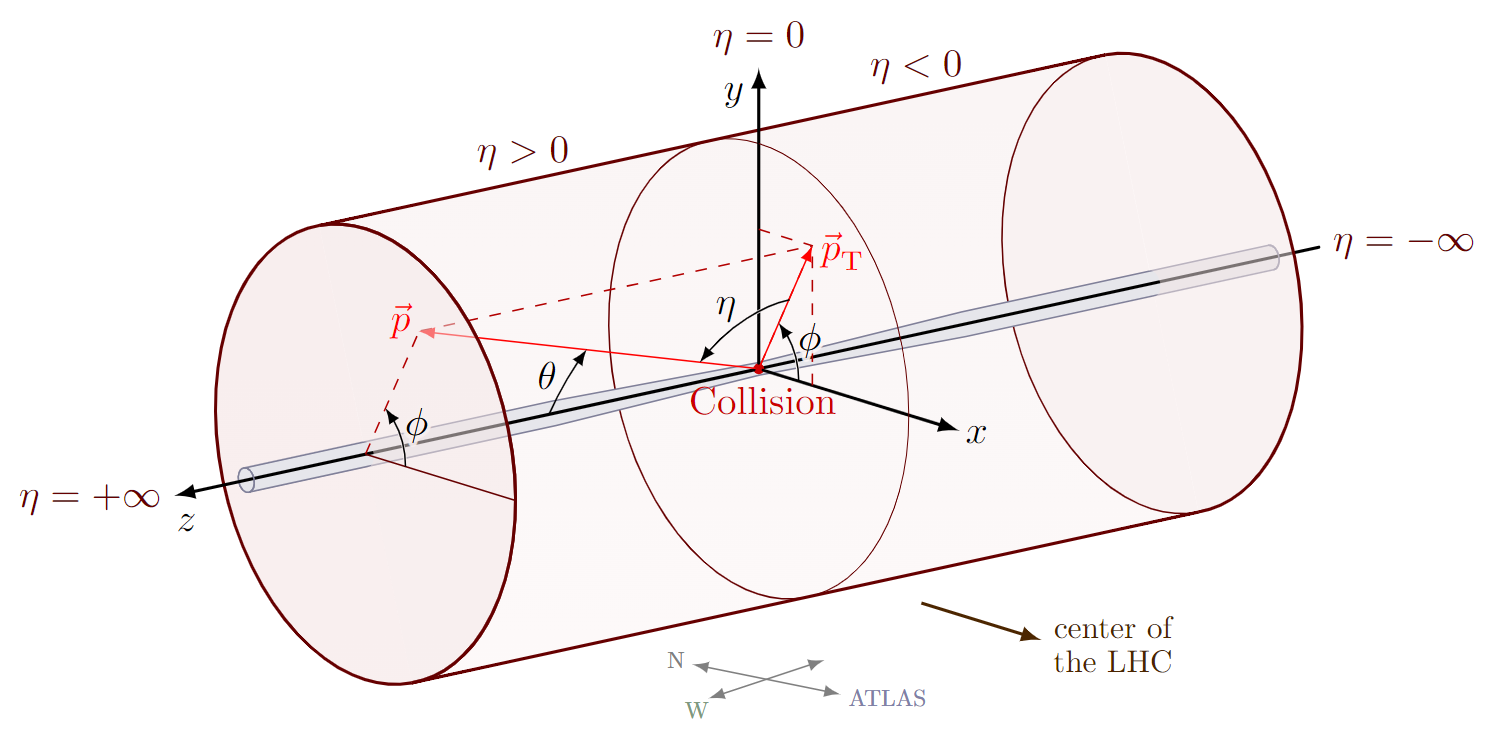}
    \caption{Geometric View of the CMS Detector With Coordinate Axis \cite{geom}}
    \label{fig:geom}
\end{figure}

An illustrative graphic of the aforementioned definitions can be seen in \textit{Fig.}\ref{fig:geom} alongside vectors indicating the direction of the transverse momentum.

\stepcounter{section}

\newpage
\section*{Appendix {D}: Schematic View of WOMBAT Models}\label{models}
\addcontentsline{toc}{section}{Appendix {D}: Schematic View of WOMBAT Models}
\setcounter{figure}{0}
\markboth{Appendix {D}: Schematic View of WOMBAT Models}{}

\begin{figure}[H]
    \centering
    \includegraphics[width=\linewidth]{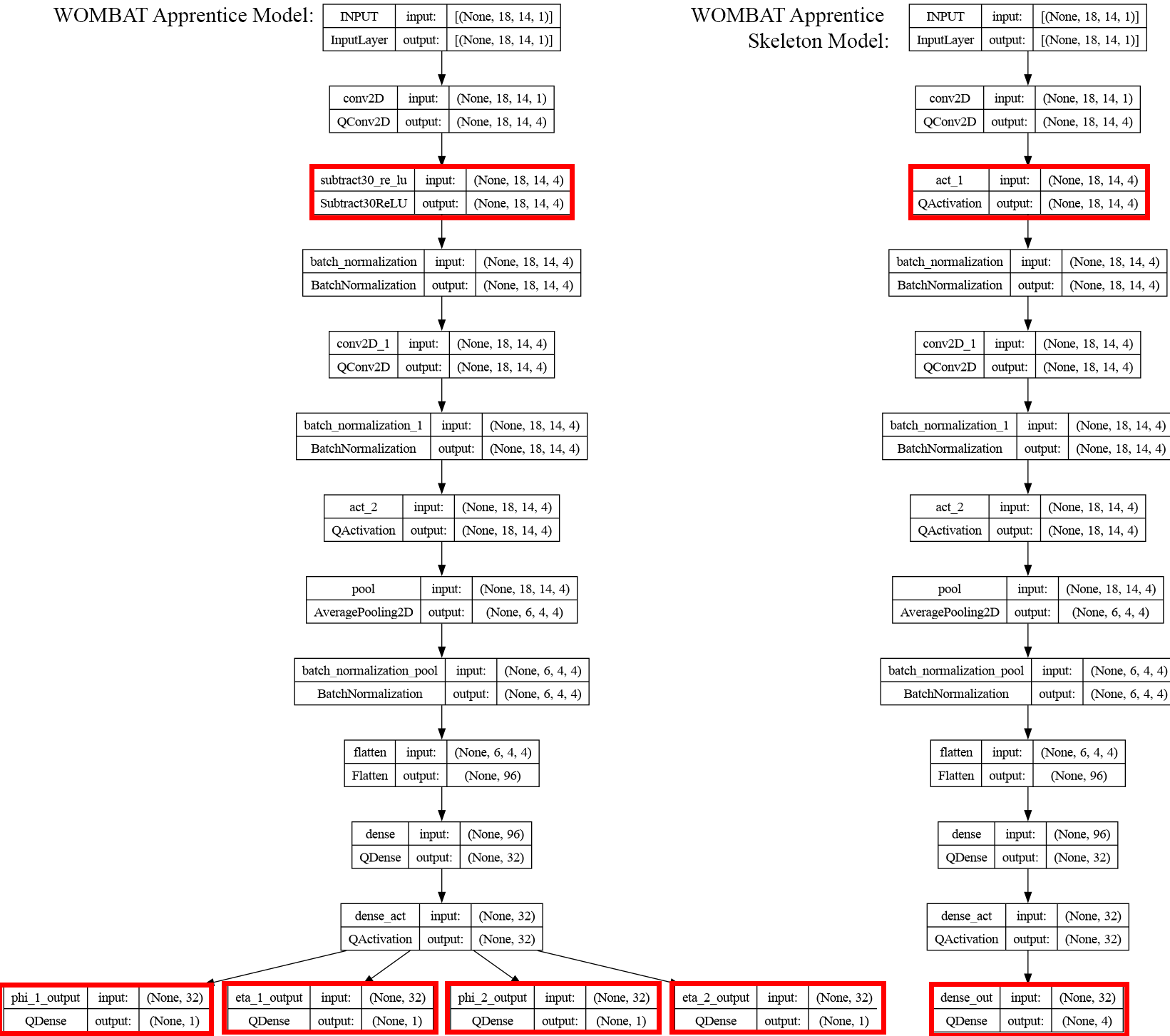}
    \caption{Schematic Architecture of WOMBAT Apprentice Model}
    \label{fig:apprentice}
    {\small
    Schematic view of the WOMBAT Apprentice Model and WOMBAT Apprentice Skeleton Model. Differences are highlighted in red.
    }
\end{figure}

\vspace{5cm}

\begin{figure}[H]
    \centering
    \includegraphics[angle=90,width=0.81\linewidth]{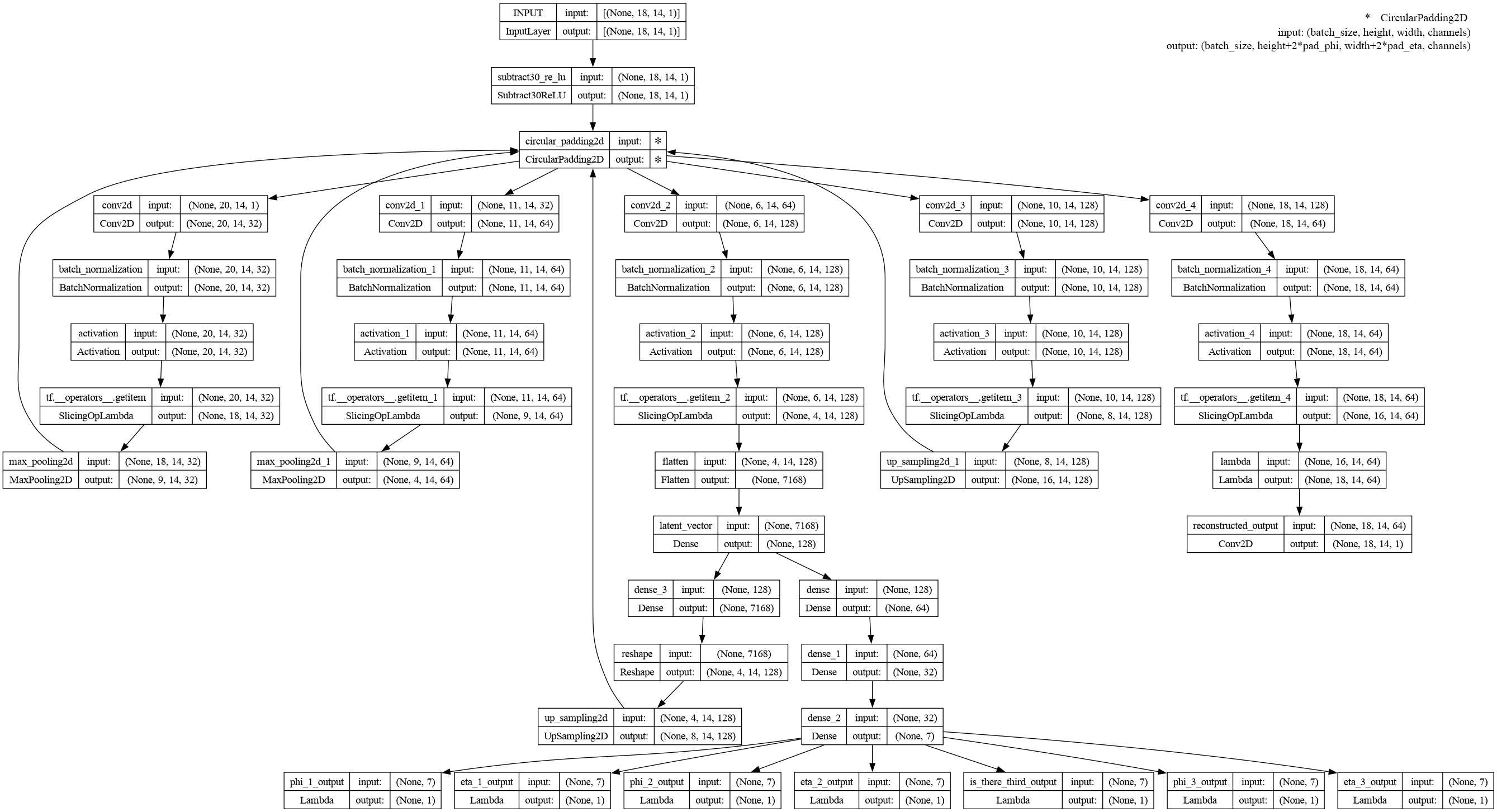}
    \caption{Schematic Architecture of WOMBAT Master Model}
    \label{fig:master}
\end{figure}

\clearpage

\stepcounter{section}

\newpage
\section*{Appendix {E}: Control Plots}
\addcontentsline{toc}{section}{Appendix {E}: Control Plots}
\setcounter{figure}{0}
\markboth{Appendix {E}: Control Plots}{}

\subsection{$p_T$ Resolution}

The $p_T$ resolution is computed through:
\begin{gather}
    p_T^{\text{resolution}} = \frac{p_T^{\text{trig}}-p_T^{\text{reco}}}{p_T^{\text{reco}}},
\end{gather}
where $p_T^{\text{trig}}$ is the jet $p_T$ reported by the trigger system, and $p_T^{\text{reco}}$ is the fully reconstructed offline jet, which serves as the ground truth.

To ensure a fair shape-based comparison between trigger algorithms with differing numbers of accepted events, the resolution histograms are normalized:
\begin{gather} 
\text{Histogram}(x) \rightarrow \frac{1}{\text{N}} \cdot \text{Histogram}(x), 
\end{gather}
where N is the total number of entries in the histogram. This normalization allows direct visual comparison of the resolution distributions without being biased by absolute event counts.

Evaluating the $p_T$ resolution is essential in trigger performance studies because it directly impacts the sharpness and stability of the trigger response. Poor $p_T$ resolution leads to broader turn-on curves, which represent the efficiency as a function of offline $p_T$. This is equivalent to the efficiency plots in Chapter V. A wide turn-on indicates that the trigger's response is smeared, making it difficult to define a precise threshold. This smearing causes efficiency losses near the threshold and increases the inclusion of lower-energy background jets, degrading the system's background rejection. Furthermore, poor resolution results in rate instability, as small fluctuations in input can cause significant changes in trigger rates. High-resolution performance ensures that the trigger accurately reflects the true kinematics of jets, enabling tighter thresholds and more reliable rate control under high-luminosity conditions.

Benchmarking the resolution of new algorithms against the existing Single Jet 180 trigger is critical to ensure that improvements in rate or acceptance are not achieved at the cost of degraded $p_T$ fidelity. Good resolution indicates a tight correlation with offline jets, enabling sharp efficiency turn-ons and reliable threshold tuning.

\begin{figure}[h]
    \centering
    \includegraphics[width=0.5\linewidth]{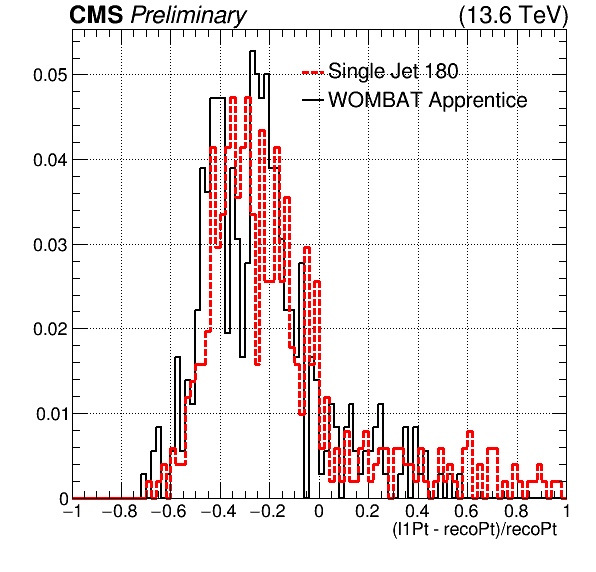}
    \caption{W-AM $p_T$ Resolution Benchmarked Against Single Jet 180}
    \label{fig:wamres}
\end{figure}

Additionally, the resolution distribution provides a diagnostic tool for identifying potential biases in the scale of the new algorithm. By comparing it directly with the baseline, one can determine if scale factors are needed to calibrate the trigger output, ensuring consistency across algorithms and physics analyses.

\begin{figure}[H]
    \centering
    \includegraphics[width=0.5\linewidth]{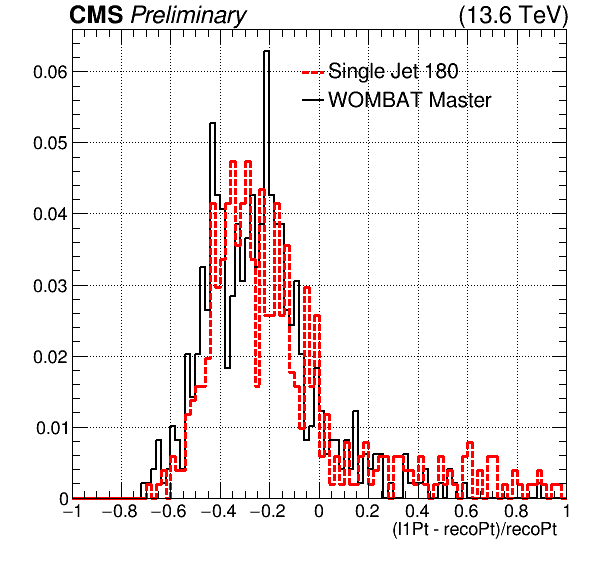}
    \caption{W-MM $p_T$ Resolution Benchmarked Against Single Jet 180}
    \label{fig:wmmres}
\end{figure}

Both W-AM and W-MM exhibit $p_T$ resolution distributions broadly consistent with Single Jet 180, with similar spread and peaks in the range $[-0.6, -0.4]$. JEDI, driven by rule-based logic and hard veto conditions, produces a narrower distribution centered around $-0.2$. Notably, W-MM also shows a secondary peak near $-0.2$, suggesting it effectively captures jets targeted by JEDI's selection logic.

\begin{figure}[H]
    \centering
    \includegraphics[width=0.5\linewidth]{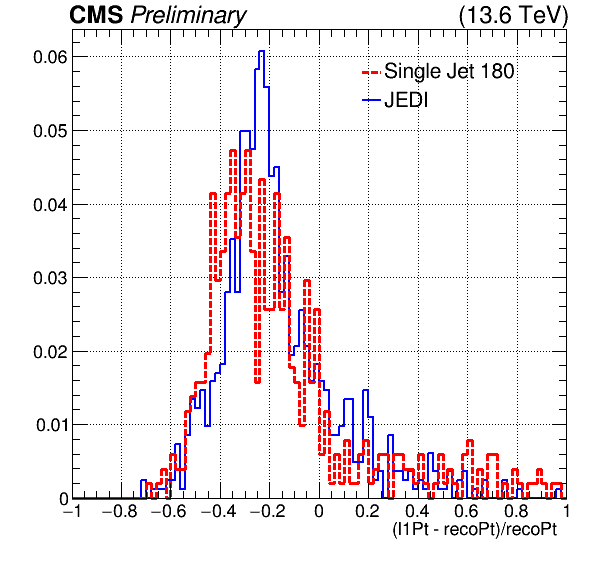}
    \caption{JEDI $p_T$ Resolution Benchmarked Against Single Jet 180}
    \label{fig:jedires}
\end{figure}

Although an ideal trigger would peak at zero resolution, the W-series algorithms were tuned to reproduce the behavior of Single Jet 180. This ensures compatibility with current CMS trigger thresholds and maintains continuity in downstream selection performance.

\subsection{Zero Bias Jet $p_T$ Distribution}

The ZB count vs. $p_T$ control plots provide a direct, unweighted view of the raw event distributions as observed in ZB data. Unlike the normalized rate computation presented in Chapter V, these plots are not scaled to reflect a physical rate but instead represent the absolute number of jets identified by each algorithm per $p_T$ bin. This distinction is important: while Chapter V focuses on the trigger rate prediction under pileup and luminosity scaling, the current plots offer a baseline diagnostic of trigger behavior, free from external scaling factors.

The JEDI algorithm demonstrates a sharp turn-on near the bin $p_T \approx 11-22$ GeV, which reflects its use of a rule-based pileup mitigation threshold that effectively suppresses low-$p_T$ jets. This thresholding behavior is clearly visible as a near-absence of counts in the lowest bins. In contrast, W-AM and W-MM display broader low-$p_T$ activity, indicating less robust suppression. This is expected, as the thresholding behavior in the WOMBAT models is not hard-coded but rather learned during training, resulting in greater flexibility but also reduced sharpness at the low end. All algorithms are benchmarked against Single Jet 180, which serves as the reference in both rate and resolution performance.

\begin{figure}[H]
  \centering
  \begin{subfigure}[b]{0.49\linewidth}
    \includegraphics[width=\linewidth]{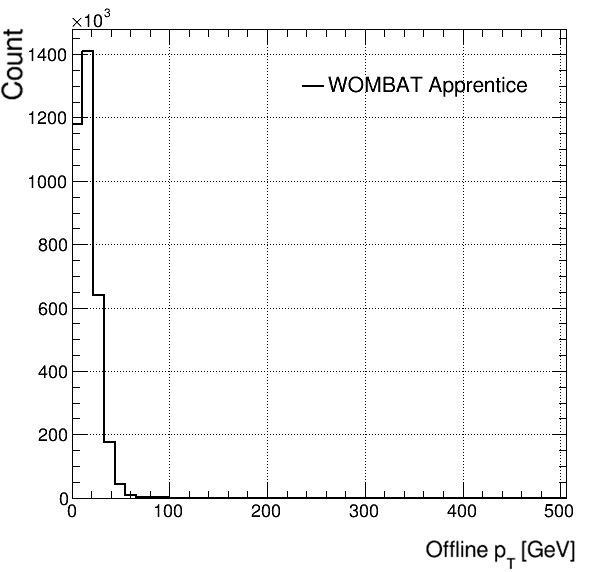}
  \end{subfigure}
  \hfill
  \begin{subfigure}[b]{0.49\linewidth}
    \includegraphics[width=\linewidth]{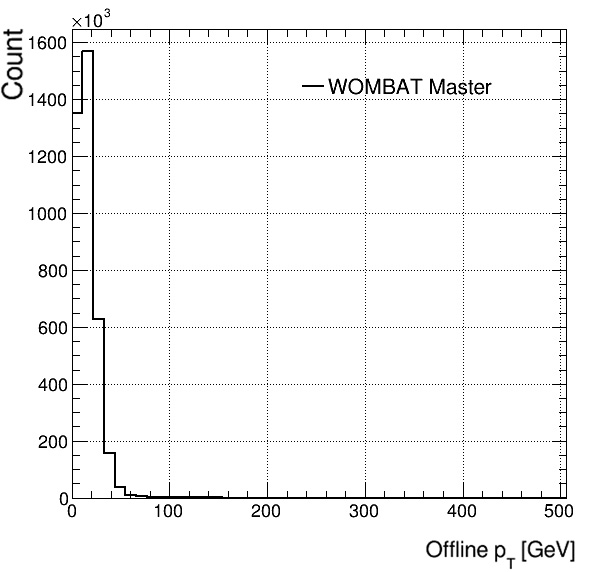}
  \end{subfigure}
  
  \vspace{0.5em}
  
  \begin{subfigure}[b]{0.49\linewidth}
    \includegraphics[width=\linewidth]{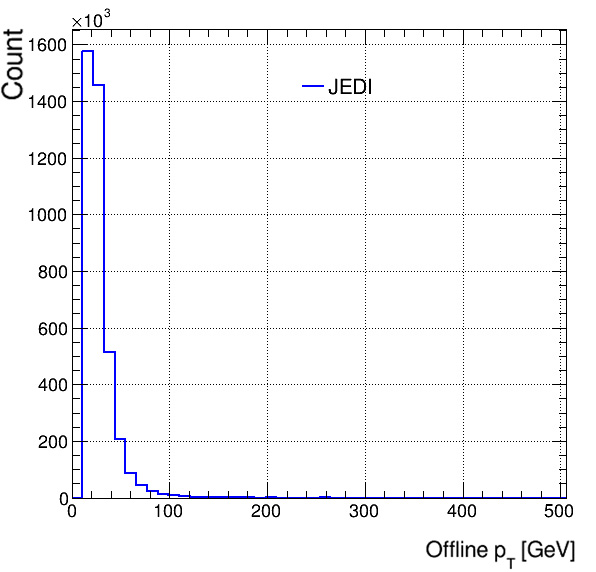}
  \end{subfigure}
  \hfill
  \begin{subfigure}[b]{0.49\linewidth}
    \includegraphics[width=\linewidth]{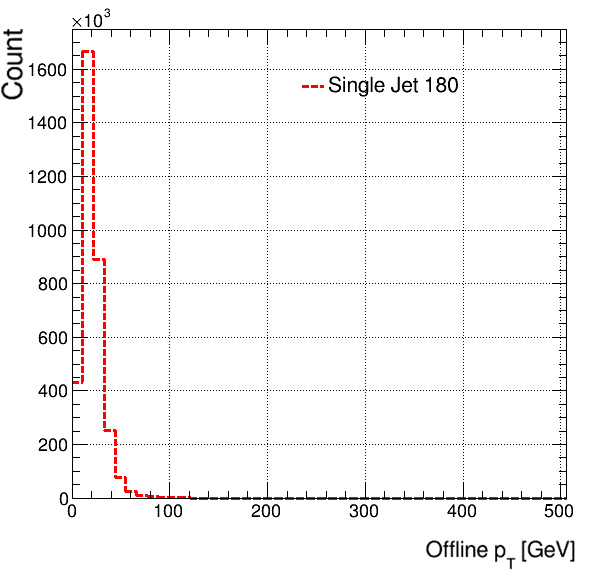}
  \end{subfigure}
  
  \caption{Raw ZB $p_T$ Distribution for WAM, WMM, JEDI, and Single Jet 180}
  \label{fig:zbcontrol}
\end{figure}

\stepcounter{section}

\newpage
\section*{Appendix {F}: Documentation and Repositories}\label{repo}
\addcontentsline{toc}{section}{Appendix {F}: Documentation and Repositories}
\setcounter{figure}{0}
\markboth{Appendix {F}: Documentation and Repositories}{}
\renewcommand{\thefigure}{\Alph{section}.\arabic{figure}}

\begin{table}[h!]
\centering
\begin{tabular}{|c|c|}
\hline
\textbf{Repository} & \textbf{Link} \\
\hline\hline
TP Displays & \href{https://github.com/mbileska/WOMBAT_TP_Displays}{github.com/mbileska/WOMBAT\_TP\_Displays} \\
Main WOMBAT Repository & \href{https://github.com/mbileska/WOMBAT_Preview}{github.com/mbileska/WOMBAT\_Preview}\\ 
\hline
\end{tabular}
\caption{GitHub Repositories Related to the WOMBAT Project}
\end{table}

\newpage
\newpage

\footnotesize



\end{document}